\documentclass[fleqn,usenatbib]{mnras}

\usepackage{newtxtext,newtxmath}
\usepackage[T1]{fontenc}
\usepackage{ae,aecompl}

\usepackage{graphicx,color}	
\usepackage{amsmath}	
\usepackage{amssymb}	
\usepackage{natbib}
\usepackage{xspace}
\usepackage{hyperref}
\usepackage[dvipsnames]{xcolor}
\usepackage{pdflscape}
\usepackage{tikz}
\usepackage{subcaption}
\usepackage{floatrow}
\usepackage{enumitem}
\usepackage{siunitx}
\usepackage{multirow}
\usepackage{setspace}
\usepackage{hhline}
\usepackage{longtable}
\usepackage{colortbl}
\usepackage{makecell}
\setcellgapes{5pt}

\setlist[enumerate]{leftmargin=3em, labelwidth=30em, listparindent=30em}
\setlist[itemize]{leftmargin=3em, labelwidth=30em, listparindent=30em}

\graphicspath{{./images/}}

\newcommand{\hi}{\text{H\,\sc{i}}\xspace}
\newcommand{\unit}[2][]{\ensuremath{\text{#2}^{#1}}\xspace}
\newcommand{\units}[3][]{\ensuremath{\text{#2}\cdot\text{#3}^{#1}}\xspace}
\newcommand{\sunit}[2]{\ensuremath{#1\text{#2}}\xspace}
\newcommand{\numunit}[2]{\mbox{\ensuremath{#1\,#2}\xspace}}
\newcommand{\z}{\ensuremath{z}\xspace}
\newcommand{\mstar}{\ensuremath{\text{M}_\star}\xspace}
\newcommand{\mhi}{\ensuremath{\text{M}_\text{\sc{Hi}}}\xspace}
\newcommand{\stacker}{HISS\xspace}
\newcommand{\stackerfull}{\hi Stacking Software\xspace}
\newcommand{\nancay}{Nan\c{c}ay\xspace}
\newcommand{\kms}{\ensuremath{\text{km}\,\text{s}^{-1}}\xspace}
\newcommand{\msol}{\ensuremath{\text{M}_\odot}\xspace}
\newcommand{\omegahi}{\ensuremath{\Omega_\text{}}\xspace}
\newcommand{\ave}[1]{\ensuremath{\langle #1 \rangle}\xspace}
\newcommand{\fhi}{\ensuremath{f_\text{\sc{Hi}}}\xspace}
\newcommand{\wfifty}{\ensuremath{\text{W}_{50}}\xspace}
\newcommand{\scube}{\ensuremath{\text{S}^3}\xspace}

\newcommand{\astropy}{AstroPy\xspace}
\newcommand{\numpy}{NumPy\xspace}
\newcommand{\scipy}{SciPy\xspace}

\newcommand{\m}{\ensuremath{\text{m}}\xspace}
\newcommand{\ur}{\ensuremath{u - r}\xspace}

\newcommand{\wim}{\citet{VanDriel2016}\xspace}

\newcommand{\appendref}[1]{Appendix~\ref{#1}\xspace}

\newcommand{\resultmhi}[5]{\ensuremath{ \num[round-mode=places, round-precision=#1, group-digits=false, scientific-notation=false]{#2} ^{ \pm \num[round-mode=places, round-precision=#1, group-digits=false, scientific-notation=false]{#3}\,\text{(sys)} } _{ \pm \num[round-mode=places, round-precision=#1, group-digits=false, scientific-notation=false]{#4}\,\text{(stat)} } \times 10^{#5} }\xspace}
\newcommand{\resultfhi}[4]{\ensuremath{ \num[round-mode=places, round-precision=#1, group-digits=false, scientific-notation=false]{#2} ^{ \pm \num[round-mode=places, round-precision=#1, group-digits=false, scientific-notation=false]{#3}\,\text{(sys)} } _{ \pm \num[round-mode=places, round-precision=#1, group-digits=false, scientific-notation=false]{#4}\,\text{(stat)} } }\xspace}

\usetikzlibrary{shapes.arrows, arrows.meta,shapes.misc, shapes,arrows,backgrounds,positioning, calc}
\tikzset{
block/.style={
  rectangle,
  draw,
  text width=5.5em,
  text centered,
  rounded corners,
  minimum height=4em},
line/.style={draw, -latex'},
edge/.style={draw}
}
\tikzset{
block1/.style={
  rectangle,
  draw,
  text centered,
  rounded corners},
line/.style={draw, -latex'},
edge/.style={draw}
}
\tikzset{cross/.style={cross out, draw=black, minimum size=2*(#1-\pgflinewidth), inner sep=0pt, outer sep=0pt}, cross/.default={1pt}}
\usetikzlibrary{automata,positioning}

\title[NIBLES III: Stacking]{HISS, a new tool for \hi stacking: application to NIBLES spectra}%

\author[J. Healy et al.]
{\parbox{\textwidth}{
J. Healy,$^{1,2}$\thanks{E-mail: julia@ast.uct.ac.za} 
S-L. Blyth,$^{1}$ 
E. Elson,$^{1,3}$ 
W. van Driel,$^{4,5}$ 
Z. Butcher,$^{6}$ 
S. Schneider,$^{6}$ 
M.D. Lehnert,$^{7}$ and 
R. Minchin$^{8,9}$}
\vspace{0.4cm}
\\
\parbox{\textwidth}{
$^{1}$Department of Astronomy, University of Cape Town, Private Bag X3, Rondebosch 7701, South Africa\\
$^{2}$Kapteyn Astronomical Institute, University of Groningen, Landleven 12, 9747 AV Groningen, The Netherlands\\
$^{3}$Department of Physics and Astronomy, University of the Western Cape, Robert Sobukwe Road, Bellville, 7535, South Africa\\
$^{4}$GEPI, Observatoire de Paris, PSL Research University, CNRS, 5 place Jules Janssen, 92190 Meudon, France
\\
$^{5}$Station de Radioastronomie de Nan\c{c}ay, Observatoire de Paris, CNRS/INSU USR 704, Universit\'{e} d'Orl\'{e}ans OSUC, route de Souesmes, 18330 Nan\c{c}ay, France\\
$^{6}$University of Massachusetts, Astronomy Program, 619E LGRT-B, Amherst, MA 01003, U.S.A.\\
$^{7}$Sorbonne Universit\'{e}, CNRS UMR 7095, Institut d'Astrophysique de Paris, 98bis bd Arago, 75014 Paris, France\\
$^{8}$Arecibo Observatory, National Astronomy and Ionosphere Center, Arecibo, PR 00612, USA\\
$^{9}$SOFIA-USRA, NASA Ames Research Center, MS 232-12, Moffett Field, CA 94035, USA
}}

\date{Accepted XXX. Received YYY; in original form ZZZ}

\pubyear{2017}
\begin{document}
\label{firstpage}
\pagerange{\pageref{firstpage}--\pageref{lastpage}}
\maketitle

\begin{abstract}
\hi stacking has proven to be a highly effective tool to statistically analyse average \hi properties for samples of galaxies which may or may not be directly detected. With the plethora of \hi data expected from the various upcoming \hi surveys with the SKA Precursor and Pathfinder telescopes, it will be helpful to standardize the way in which stacking analyses are conducted. In this work we present a new python-based package, \stacker, designed to stack \hi (emission and absorption) spectra in a consistent and reliable manner. 
As an example, we use \stacker to study the \hi content in various galaxy sub-samples from the NIBLES survey of SDSS galaxies which were selected to represent their entire range in total stellar mass without a prior colour selection. This allowed us to compare the galaxy colour to average \hi content in both detected and non-detected galaxies. Our sample, with a stellar mass range of $10^8 < \mstar\, (\msol) < 10^{12}$, has enabled us to probe the \hi-to-stellar mass gas fraction relationship more than half an order of magnitude lower than in previous stacking studies.
\end{abstract}

\begin{keywords}
galaxies: evolution -- methods: data analysis -- galaxies: ISM
\end{keywords}


\section{Introduction}
	\label{sec:introduction}
	
    How galaxies evolve over cosmic time is currently a key area of research in astrophysics. A great deal is already known about the evolution of the stellar content of galaxies due to  recent infrared (\textit{Spitzer}, \citealt{Werner2004}), optical (Sloan Digital Sky Survey, \citealt{York2000}), and ultraviolet (Galaxy Evolution Explorer, \citealt{Bianchi2000}) surveys. However, comparatively little is known about the evolution of the gas in galaxies. Understanding how the cold gas evolves is important as it is the raw fuel for the formation of stars and thus galaxies. \\

Neutral atomic hydrogen (\hi) forms the most significant reservoir of neutral gas in galaxies. Studies have shown that blue, star-forming galaxies have a higher fraction of \hi gas compared to red, quiescent galaxies \citep[e.g.][]{Roberts1994a, Gavazzi1996, McGaugh1997, Cortese2011, Fabello2011, Brown2015}, which suggests that \hi plays an important role in star formation. \hi is difficult to detect in most galaxies beyond the local universe ($z \sim 0.06$) with existing radio telescopes due to the weak nature of the emission. Researchers have had to exploit different techniques in order to measure very weak \hi signals from galaxies. One such technique is \hi stacking; with this technique, an average \hi mass per galaxy can be estimated by co-adding the \hi spectra of a sample of galaxies in which \hi is not necessarily directly detected. \\

	The idea of co-adding the undetected \hi spectra in studies of the gas content in galaxies was first presented by \citet{Zwaan2001} and \citet{Chengalur2001}. Both groups were studying the \hi within galaxies located in and around clusters. With low detection counts in their samples, both groups independently co-added the non-detections in an effort to obtain a statistically meaningful averaged detection for their samples.\\
    
    Stacking analyses have become commonplace in the last $15+$ years. The technique has been applied in various areas: \hi content of galaxies in dense environments and how the gas content relates to other observables \citep{Chengalur2001,Verheijen2007,Lah2009,Jaffe2016}; gas content of active galaxies \citep{Gereb2013,Gereb2015}; measurement of \omegahi at low to intermediate ($z < 0.4$) redshifts \citep{Lah2007, Delhaize2013, Rhee2013, Rhee2016}; and using stacking to study the relations between \hi and various stellar mass/star formation indicators \citep{Fabello2011,Fabello2012,Brown2015,Brown2016c,Gereb2015}.\\

    The gas scaling relations for galaxies are average trends which relate the \hi mass (\mhi) or \hi mass to stellar mass (\mstar) ratio (gas fraction, \fhi) to various other galaxy properties. Stacking a sample of $\sim$5000 galaxies with \numunit{\mstar > 10^{10}}{\msol} that had both ultraviolet and optical imaging, \citet{Fabello2011} found using \hi stacking that the gas fraction correlates better with $\text{NUV} - r$ colour than stellar mass. This result was later confirmed and extended to a sample with \numunit{\mstar > 10^{9}}{\msol} by \citet{Brown2015}. To date, studies of the \hi gas scaling relations have been limited to samples with \numunit{\mstar > 10^9}{\msol}, a mass which has been identified as a turning point around which the \mhi vs \mstar slope changes \citep{Huang2012,Maddox2015}.\\

    Directly detecting \hi with current radio telescopes beyond $z$~$\sim$~$0.1$ is challenging due to the long observing times required (e.g. HIGHz \citep{Catinella2014} on Arecibo using \numunit{>300}{\text{hours}} and CHILES \citep{Fernandez2016,Hess2018} on the JVLA requiring \numunit{\sim 1000}{\text{hours}} of observing time). The upcoming surveys with the SKA Precursors and Pathfinders (e.g. MeerKAT, \citealt{Booth2009}; ASKAP, \citealt{Johnston2008}; APERTIF, \citealt{Verheijen2008a}) will greatly extend the redshift range over which \hi in galaxies is studied, either directly or indirectly. Techniques such as \hi stacking will have an important role to play in order to study the average \hi properties of different galaxy samples, particularly at higher redshifts ($z \gtrsim 0.6$). Deep SKA Precursor surveys such as LADUMA\footnote{Looking At the Distant Universe with the MeerKAT Array} \citep{Holwerda2011} on MeerKAT have identified stacking as an important tool to probe higher redshifts, and push to lower \hi mass limits.\\
   
    As we rapidly approach the start of the Precursor Surveys, work is under-way on developing a data analysis toolkit that will enable consistent and comparable studies of the survey data. Already available is the versatile source finder, \textsc{SoFiA} \citep{Serra2015}. With the important role that stacking will play in the analysis of high redshift and low mass samples, it is imperative that a tool capable of reliably and consistently stacking \hi spectra is developed.\\
    
    In this work, we present our new software package, \stacker, that has been designed for the astronomy community in response to the need for a stacking software package. \stacker can be downloaded from \url{https://github.com/healytwin1/HISS}. We use \stacker  to revisit the gas scaling relations with a sample of 1000 galaxies from the \nancay Interstellar Baryon Legacy Extragalactic Survey \citep[NIBLES][]{VanDriel2016}. NIBLES is an SDSS-selected targeted \hi survey with the \nancay Radio Telescope which aims to study the \hi properties of galaxies as a function stellar mass, covering a representable \mstar range of the ensemble of galaxies in the nearby Universe.\\

	The outline for this paper is as follows: the first half of the paper (Section 2) describes the design of the \hi Stacking Software (\stacker). In the second half of the paper (Section 3 and Section 4), we give an introduction to the NIBLES Survey and the ancillary data we use in our stacking analysis and describe the classification of \hi non-detections. We use the NIBLES sample to explore the well-known \hi mass to stellar mass scaling relations in Section 4, and finally, summarise our findings in Section 5.

\section{The H {\sc{i}} Stacking Software (HISS)}
	\label{sec:software}
	
    The stacking method that the software needs to implement can be summarized by the following steps:
	\begin{enumerate}
		\item Ingest 1D \numunit{21}{\text{cm}} \hi spectra (radio data) and list of associated redshifts for the sample of galaxies to be stacked;
		\item Spectrally shift and re-scale the \hi spectra to align the expected line emission at a common frequency (usually the \hi rest frequency: \numunit{1420.4058}{\text{MHz}});
		\item Weight the spectra (according to a preferred weighting scheme) and co-add the spectra.
	\end{enumerate}

We have designed and created the \hi Stacking Software (\stacker) package, after discussion with colleagues and with the following guiding principles in mind:

	\begin{itemize} 
    	\item be freely available in most operating systems
		\item to be open source
		\item easy to modify (extensible)
		\item stack hundreds or thousands of galaxy spectra in an efficient and reliable manner.
	\end{itemize}

	The Python\footnote{Developed in Python~2.7, but compatible with Python~3} programming language was chosen for development of \stacker because it is freely available, can be used on any operating system, and is also one of the most commonly used languages within the astronomy community. Where appropriate, \stacker makes use of the publicly available Python modules (such as \astropy, \numpy, \scipy, etc.) which have been optimized for data input, manipulation, and display.\\
    
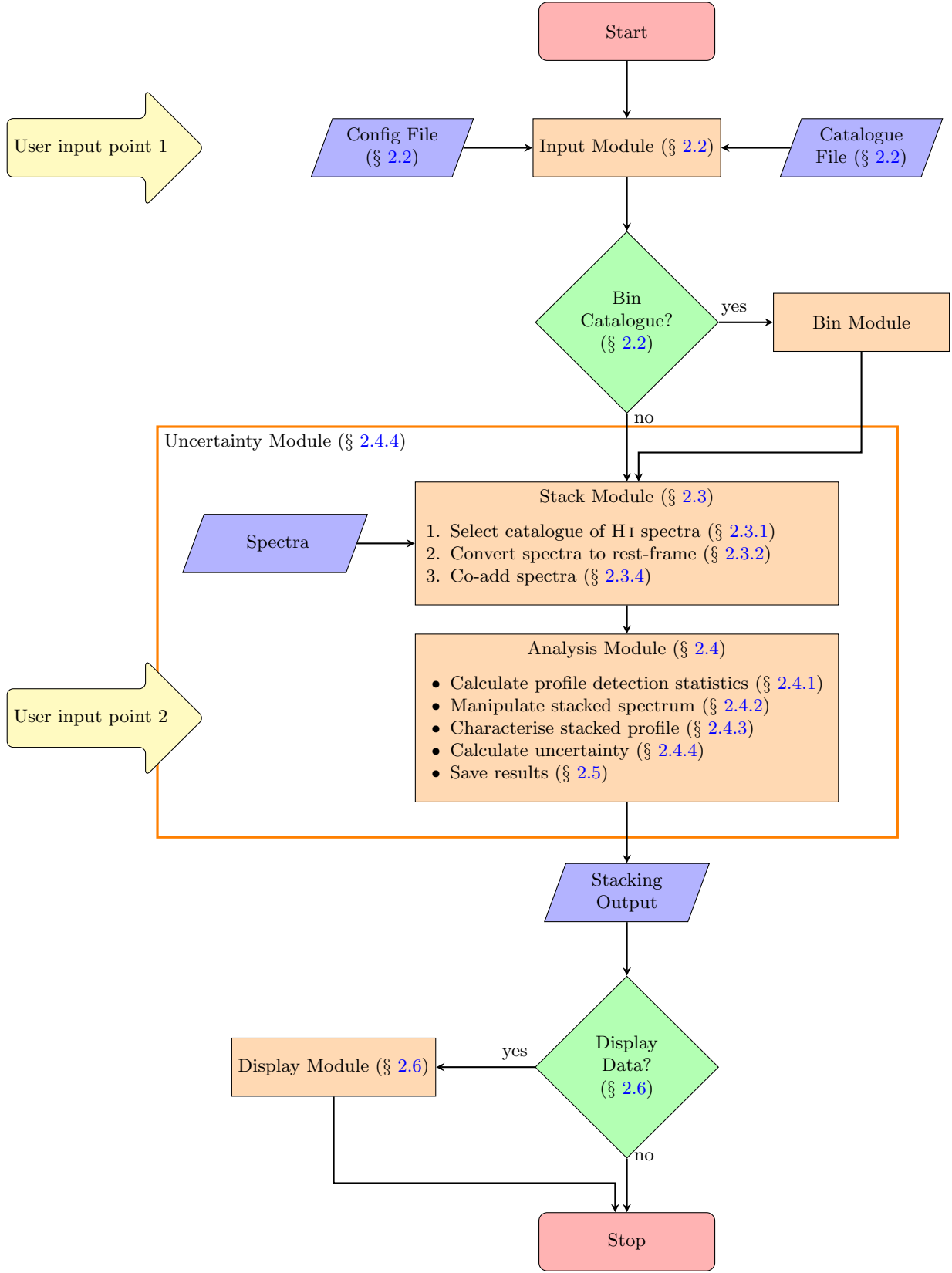
\begin{figure*}
		\centering
		\tikzstyle{startstop} = [rectangle, rounded corners, minimum width=3cm, minimum height=1cm,text centered, draw=black, fill=red!30]
\tikzstyle{io} = [trapezium, trapezium left angle=70, trapezium right angle=110, minimum width=1cm, minimum height=1cm, text width=5em, align=center, draw=black, fill=blue!30]
\tikzstyle{process} = [rectangle, minimum width=3cm, minimum height=1cm, text centered, draw=black, fill=orange!30] 
\tikzstyle{decision} = [diamond, minimum width=3cm, minimum height=1cm, text width=5em, text centered, draw=black, fill=green!30]
\tikzstyle{arrow} = [thick,->,>=stealth]

\begin{tikzpicture}[node distance=2cm]
	\node (start) [startstop] {Start};

	\node (inputmod) [process, below of=start] {Input Module (\S~\ref{sec:inputmodule})};
	\node (userin) [io, left of=inputmod, xshift=-2cm] {{Config File (\S~\ref{sec:inputmodule})}};
	\node (catin) [io, right of=inputmod, xshift=2cm] {{Catalogue File (\S~\ref{sec:inputmodule})}};
    \node[single arrow, rounded corners=3pt, fill=yellow!30, draw, align=center, left of=userin, xshift=-1.2cm, minimum height=1cm, minimum width=2cm, anchor=east]{User input point 1};

	\node (checkbin) [decision, below of=inputmod, yshift=-1cm] {Bin Catalogue? (\S~\ref{sec:inputmodule})};
	\node (bin) [process, right of=checkbin, xshift=2cm] {Bin Module};

	\node (stack) [process, below of=checkbin, yshift=-1.8cm, text width=7cm] {\vbox {Stack Module (\S~\ref{sec:stackmodule})
          {\begin{itemize}
              \itemindent=-1.5em
              \item [1.] Select catalogue of \hi spectra (\S~\ref{sec:readspectra})
              \item [2.] Convert spectra to rest-frame (\S~\ref{sec:restframe})
              \item [3.] Co-add spectra (\S~\ref{sec:coaddspectra})
          \end{itemize}}
																				}};
	\node (spectra) [io, text width=3em, left of=stack, xshift=-4cm] {Spectra};

	\node (analysis) [process, below of=stack, yshift=-1cm, text width=7cm] {\vbox {Analysis Module (\S~\ref{sec:analysismodule})
              {\begin{itemize}
                  \itemindent=-1.5em
                  \item Calculate profile detection statistics (\S~\ref{sec:profilestats})
                  \item Manipulate stacked spectrum (\S~\ref{sec:specmanip})
                  \item Characterise stacked profile (\S~\ref{sec:fitprofile})
                  \item Calculate uncertainty (\S~\ref{sec:uncertcalc})
                  \item Save results (\S~\ref{sec:outputmethods})
              \end{itemize}}
																}};
	\draw [very thick, color=orange] ([xshift=-1.5cm, yshift=1.5cm]spectra.north west) rectangle ([xshift=1cm, yshift=-0.6cm]analysis.south east);
	\node [anchor=north west] at ([xshift=-1.5cm, yshift=1.5cm]spectra.north west) {Uncertainty Module (\S~\ref{sec:uncertcalc})};
	\node (savedata) [io, below of=analysis, yshift=-1cm] {Stacking Output};
	\node (dispyn) [decision, below of=savedata, yshift=-1cm] {Display Data? (\S~\ref{sec:displaymodule})};
     \node[single arrow, rounded corners=3pt, fill=yellow!30, draw, align=center, left of=analysis, xshift=-5.2cm, minimum height=1cm, minimum width=2cm, anchor=east] {User input point 2};

	\node (display) [process, left of=dispyn, xshift=-3cm] {Display Module (\S~\ref{sec:displaymodule})};

	\node (stop) [startstop, below of=dispyn, yshift=-1cm] {Stop};

	\draw [arrow] (start) -- (inputmod);
	\draw [arrow] (userin) -- (inputmod);
	\draw [arrow] (catin) -- (inputmod);

	\draw [arrow] (inputmod) -- (checkbin);
	\draw [arrow] (checkbin.south) -- node[anchor=west, yshift=+0.5cm] {no} (stack.north);
	\draw [arrow] (checkbin) -- node[anchor=south, xshift=-0.2cm] {yes} (bin);
	\draw [arrow] (bin.south) |- ([xshift=+0.2cm, yshift=+0.5cm]stack.north) -- ([xshift=+0.2cm]stack.north);
	
	\draw [arrow] (spectra) -- (stack);
	\draw [arrow] (stack) -- (analysis);

	\draw [arrow] (analysis) -- (savedata);
	\draw [arrow] (savedata) -- (dispyn);
	\draw [arrow] (dispyn.south) -- node[anchor=west, yshift=+0.5cm] {no} (stop);
	\draw [arrow] (dispyn.west) -- node[anchor=south, xshift=+0.5cm] {yes} (display.east);


	\draw [arrow] (display.south) |- ([xshift=-0.2cm, yshift=+0.5cm]stop.north) -- ([xshift=-0.2cm]stop.north);

\end{tikzpicture}
		\caption[\stacker Flow Diagram]{This flow diagram shows how the individual spectra and user information are taken by \stacker to produce a stacked spectrum from which average galaxy properties (such as total \hi mass, \mhi, and the \hi-to-stellar mass ratio, \fhi) may be extracted. The orange rectangles show the six modules of the package, the blue parallelograms show the points of input or output, and the green diamonds show where the user may choose to incorporate the optional modules.}
		\label{fig:outlineflowuncert}
	\end{figure*}

    \autoref{fig:outlineflowuncert} shows a flow diagram of how the processes required to create a stacked spectrum are incorporated in the different modules of \stacker. In the sections that follow we will illustrate how each of the processes highlighted in \autoref{fig:outlineflowuncert} are implemented. Basic instructions on how to run \stacker can be found in \appendref{sec:runhiss}.

	\subsection{Simulated Data}
		\label{sec:artificialdata}

		When developing new data analysis tools, one of the most important steps is to quantify the accuracy and reliability of any generated results. For this purpose, simulated spectra are used instead of real spectra because properties of the input data are then well known. In this work, we use a data set of 1000 simulated \hi profiles of galaxies to illustrate and test the capabilities of \stacker. The simulated \hi spectra are created using the formulation outlined in \citet[Appendix A]{Obreschkow2009c}, and are based on the evaluated properties extracted from the \scube-SAX catalogue \citep{Obreschkow2009a,Obreschkow2009b,Obreschkow2009c} for galaxies in a redshift range of $ 0.1 < z < 0.2$ with a channel width of \numunit{7}{\kms} and Gaussian noise of \numunit{16}{\mu\text{Jy}} per channel. 

	\subsection{Catalogue of H~{\sc{i}} spectra}
		\label{sec:inputmodule}

        \begin{figure*}
			\centering
			\includegraphics[width=\textwidth]{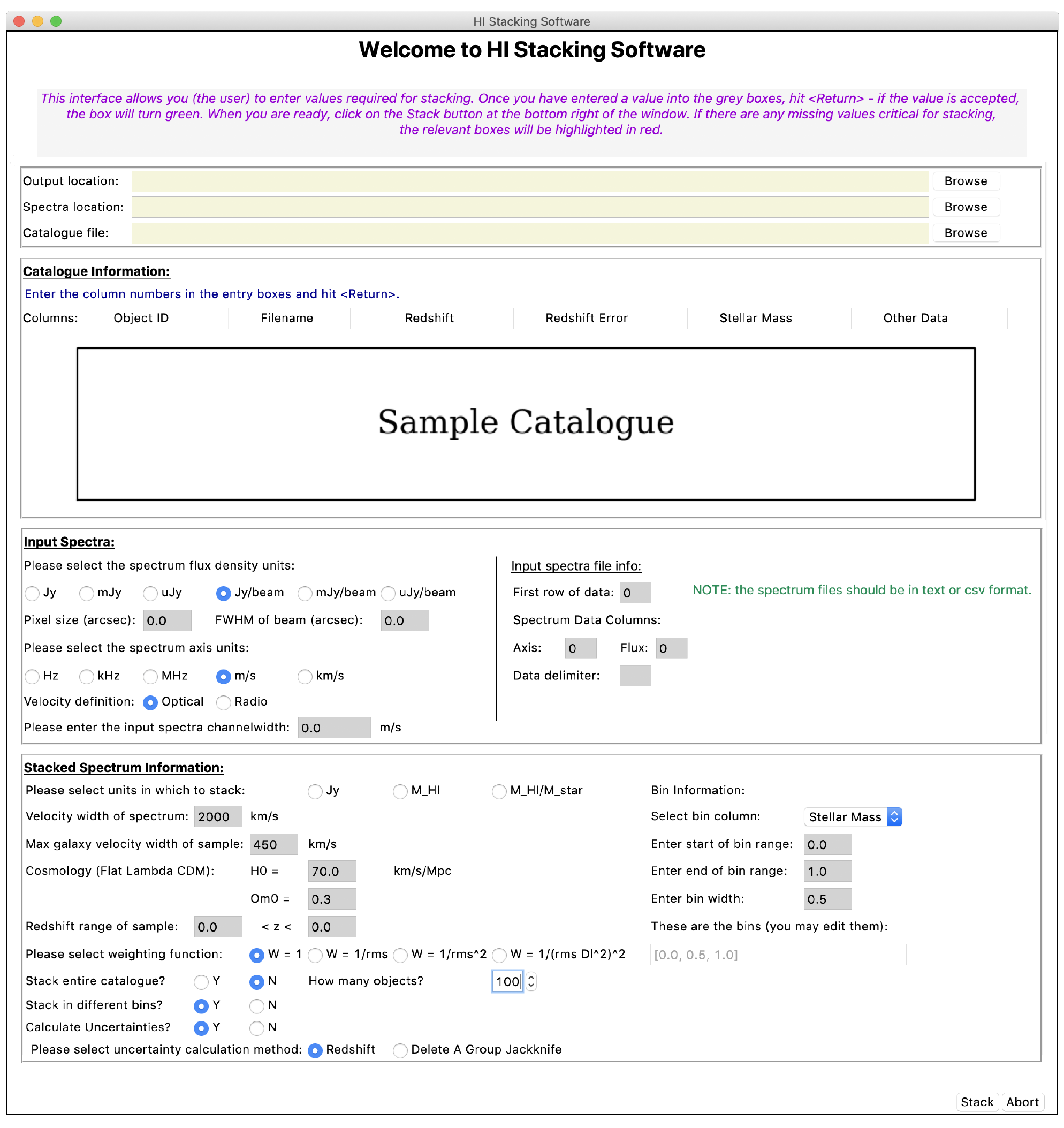}
			\caption{Screen shot from GUI main window, useful for first-time \stacker users. It shows an overview of all initial parameters required by \stacker. This graphical interface fills in a configuration file whose parameters are explained in \autoref{sec:configfile}.}
			\label{fig:guimainwindow}
		\end{figure*}
        
		When initiated, \stacker requires some information about the sample of \hi spectra -- see \autoref{fig:guimainwindow} for an overview. The Input Module requires two input files: a configuration file and a catalogue file. There are two ways to provide \stacker with the required information: the user can provide a text file in JSON\footnote{JSON stands for JavaScript Object Notation, a JSON formatted file has the extension \texttt{.json}} format or use the graphical user interface shown in \autoref{fig:guimainwindow} to generate a configuration file (also in JSON format). The JSON format is used for the configuration file as it is easy to read/write, and when imported into Python, the information is dumped into an easy~-~to~-~access Python dictionary. The configuration file includes options such as different weighting functions for the stacking procedure, and options to bin the stacking sample based on additional data provided in the catalogue file. \\
        
		The catalogue file is a user-created text file in CSV\footnote{comma separated values} format that contains at least the following columns, 
		\begin{itemize}
			\item Object ID
			\item Spectrum filename
			\item Redshift
			\item Redshift uncertainty
			\item Other data (optional)
		\end{itemize}
		for each spectrum that the user would like to include in the stack. Although the catalogue file may contain any number of columns, the non-optional columns mentioned above are those that are required to run \stacker. Additional columns containing numerical information such as stellar mass or optical colour may be used to refine the catalogue into a number of different sub-samples to be stacked separately. This refining is done through the Bin Module (\autoref{fig:outlineflowuncert}). In this first version of HISS, bins can only be created using one quantity, e.g. stellar mass or colour, and the stacked results are stored for each sub-sample\footnote{In the current implementation this process is sequential i.e. each binned sub-sample of spectra is stacked one after the other. This process will be parallelised in future versions.}. \\
       
       During the processing of the configuration file, the code checks that the catalogue file and location of the spectra exist. If these checks are passed, \stacker will continue.

	\subsection{Stacking the spectra}
		\label{sec:stackmodule}

		The stacking procedure is the heart of \stacker, and is controlled by the Stack Module which is responsible for reading in and preparing each spectrum for stacking, as well as maintaining the stacked spectrum and associated information (e.g.,number of objects in the stack, stacked noise). One of the features of this module is the option that allows the user to watch the progress of the stacking in a window such as the one in \autoref{fig:progressplot}. The progress window shows each of the spectra in the observed frame as they are read in, as well as the individual spectra as they are converted to the rest-frame and then added to the total stacked spectrum.

		\subsubsection{Reading in the spectra}
          \label{sec:readspectra}

			Regardless of how the \hi spectra were created, \stacker requires that the spectra be in a text file type format containing a column for the spectral axis (either frequency or velocity) and another for the flux density (in units of $\mu$Jy, mJy, Jy or flux density/beam). \\

			\autoref{fig:inputspectra} shows a selection of ten of the 1000 simulated \hi profiles that will be used to illustrate the different parts of \stacker. Each of these have had \numunit{16}{\sunit{\mu}{Jy}} Gaussian noise added, so as to simulate the spectra that are anticipated from the deep LADUMA survey.\\ 

			As each spectrum is read into \stacker, it is put through a series of quality checks:
			\begin{itemize}
				\item The spectrum's channel width is checked to ensure it is within $5\%$ of the entered channel width. A warning is raised if the spectrum channel width is $5\% - 10\%$ different. If the channel width is more than $10\%$ different the spectrum is discarded.
				\item The input object contains the minimum and maximum redshift values for the sample of galaxies. This check makes sure that the redshift associated with the spectrum falls within these bounds.
				\item If the user has decided to stack the spectra in units of gas fraction, the catalogue will be checked for an associated stellar mass value.
				\item The final check is for spectrum length. A galaxy mask is created based on the expected maximum width galaxy width for the sample. Every spectrum is checked to ensure that it has more channels than the length of mask. An example of a spectrum that would fail such a check is shown in the panel (j) of \autoref{fig:inputspectra} and \autoref{fig:extendedspectra}.
			\end{itemize}
			
            Due to the discrete nature of the input spectra, the channel width check is important to avoid co-adding arrays where the rest-frame size of the channel (i.e velocity width in \kms) changes significantly ($>10\%$) between different arrays. The spectra that do not pass at least one of the quality checks are excluded from the stack. The object IDs for the excluded spectra and the reason for exclusion are recorded in the log file.

		\subsubsection{Convert spectra to rest-frame}
			\label{sec:restframe}

			If the spectrum passes all the quality checks, it moves to the next step which is to align the spectrum such that the galaxy signal is centred about the central channel of the spectrum. There are two steps to aligning the spectra. The first step is to convert the observed frequencies to a set of rest-frame frequencies corresponding to the systemic velocity of the galaxy being placed at $z=0$. The spectral axes are converted to the rest frame by
		    \begin{equation} \label{eqn:shiftfunc}
		      \mathrm{v}_\mathrm{emit} = \mathrm{v}_\mathrm{obs} - cz \qquad \text{or} \qquad \nu_\mathrm{emit} = \nu_\mathrm{obs} (1 + z)
		    \end{equation}
		    where $\mathrm{v}_\mathrm{obs}$ and $\nu_\mathrm{obs}$ are the spectrum's spectral axis in either velocity or frequency, respectively. The galaxy's redshift is \z, and $c$ is the speed of light given in kilometres per second.\\  

		    An input spectrum with a velocity spectral axis is assumed to have rest frame channel widths; this means that the conversion from the observed frame to the rest frame only requires a shift from the recessional velocity to \numunit{0}{\kms} as given by the velocity relation in \autoref{eqn:shiftfunc}. For a spectrum with a frequency spectral axis, the rest frame channel width is larger than the observed frame by a factor of $(1 + z)$; thus in order to convert the spectrum to the rest frame, both the spectral axis and flux density must be appropriately scaled such that the integrated line flux density is conserved. The frequency axis is converted to the rest frame using the frequency relation in \autoref{eqn:shiftfunc}, and the flux is scaled according to: $S_{\nu,emit} = S_{\nu,obs}/(1+z)$. \\

		\subsubsection{Spectrum wrapping}
			\label{sec:spectrumwrapping}

		    The second step is to shift the galaxy emission to the centre of the spectrum. Since the spectra are now in the rest frame, the target emission is located at either \numunit{0}{\kms} or \numunit{1420}{\text{MHz}} (depending on the units of the spectral axis) and is easily shifted to the centre of the spectrum. In this step, some authors \citep[e.g.,][]{Zwaan2001} will concatenate the spectra such that every channel has the same number of measurements (i.e. all of the spectra are the same length). Other authors \citep[e.g.,][]{Lah2007,Lah2009,Rhee2013} do not concatenate the spectra, instead choosing to weight each channel in the stacked spectrum relative to the number of measurements in that channel.\\

		    Instead, \stacker wraps the flux that would be shifted out of the now centred spectrum to append to the other end of the spectrum so that the original spectrum length is maintained; this method also maintains the spectrum noise properties. This is illustrated in \autoref{fig:inputspectra}: the green part of the spectrum is the main part that is shifted to the centre and purple highlights the section of the spectrum that is appended to the other side of the spectrum array. The centred spectra are shown in blue in \autoref{fig:extendedspectra}. It is clear from each of the panels of input spectra in \autoref{fig:inputspectra} that the spectra need not all be the same length. Each spectrum is thus converted to some appropriate length specified by the user as an input parameter, (see bottom block of \autoref{fig:guimainwindow} -- \textit{Velocity width of spectrum}\footnote{The default length is \numunit{2000}{\kms} which is $4\times$ width of a galaxy rotating at \numunit{500}{\kms}. A wider spectral axis is not advised due to possible noise variations across the bandwidth.})\\

		    In order to avoid issues with small number statistics in the spectrum noise calculations, the length of the stacked spectrum should be at least three times the width of the broadest galaxy profile (i.e., the largest \wfifty). The spectra that are longer than necessary are simply truncated at the edges while keeping the galaxy emission at the centre of the spectrum. The shorter spectra are extended by using the so-called ``noisy'' channels from the outer channels that are not expected to contain any galaxy emission to fill up the new empty channels at either edge of the spectrum. \\
            
            In order to determine which channels contain only noise, the galaxy emission is masked. The galaxy mask is centred at the spectral location of the galaxy and has some pre-determined width (this is usually a conservative estimate on the maximum possible velocity width of galaxies in the sample). In \autoref{fig:inputspectra}, for illustrative purposes, the galaxy emission is masked using the pale blue band, all other channels are considered to contain only noise, and so can be used in the spectrum extension process. There are some caveats associated with this process: if there are too few noisy channels in the spectrum, then a ringing phenomenon may appear in the extended spectrum. \autoref{fig:extendedspectra} shows the original spectra centred in blue, while black indicates the noise that has been used to fill up the channels of the extended spectrum. \autoref{fig:extendedspectra}~(j) highlights the ringing phenomenon.

		\subsubsection{Co-adding the spectra}
        	\label{sec:coaddspectra}

			Before adding each rest-frame spectrum to the stack, the spectra are converted to the units (Jy, \msol, \fhi) that the user specified as the ``stacking units'' during the configuration process. \\

			To convert the \hi flux density to \hi mass, we use the following relation from \citet{Wieringa1992}:
		    \begin{equation}\label{eqn:mhi}
		      \frac{\mhi}{\msol} = \frac{2.36 \times 10^5}{1+z} \left( \frac{D_L(z)}{\text{Mpc}} \right)^2 \left( \frac{\int S_v \text{d}v}{\units{Jy}{\units[-1]{km}{s}}} \right)
		    \end{equation}
		    where \z is the galaxy redshift and $D_L(z)$ the associated luminosity distance. $\int S_v \text{d}v$ is the galaxy rest frame integrated line flux density for the profile in \units{Jy}{\units[-1]{km}{s}}. This relation between the flux density and the \hi mass assumes a spherical \hi cloud that is optically thin with a uniform internal velocity distribution.  \\

		    The \hi gas fraction (or more accurately, the \hi to stellar mass ratio) is simply
			\begin{equation}\label{eqn:fhi}
				\fhi = \frac{\mhi}{\mstar},
			\end{equation}
			where \mstar is the total stellar mass and \mhi is the total \hi mass. The right $y$-axes of the spectra in \autoref{fig:extendedspectra} are in units of \msol per channel; this choice of units allows us to simply add the values in each channel together when measuring the integrated value. \\

\begin{figure}
				\centering
				\includegraphics[width=\textwidth]{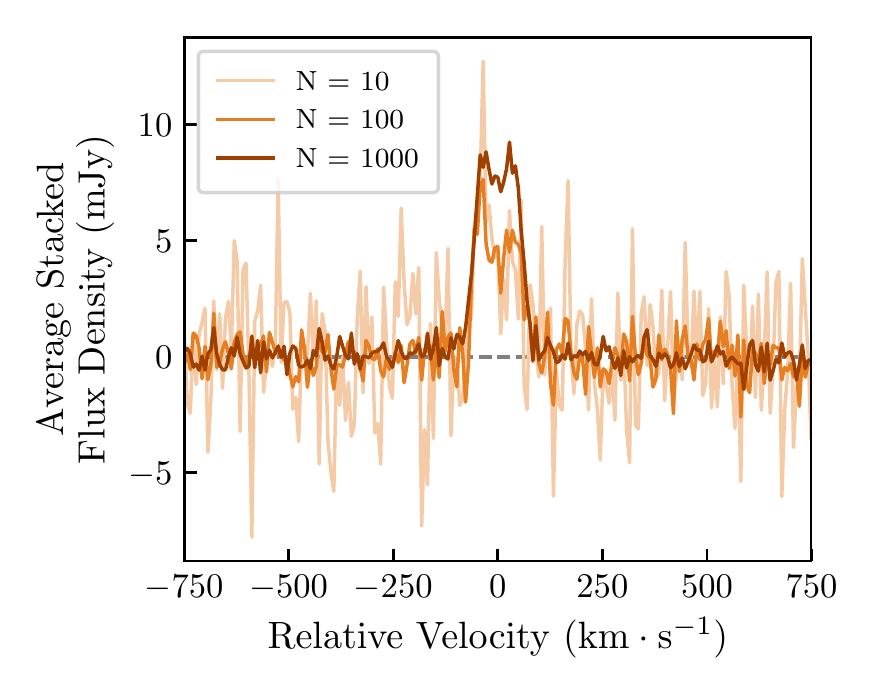}
				\caption[The power of stacking.]{This plot shows how the S/N of the average spectrum increases with increasing number of profiles. The average noise decreases as $1/\sqrt{N}$.}
				\label{fig:totalstacks}
			\end{figure}

			The final step required to produce a stacked spectrum is to co-add the spectra. This is typically done as a weighted average. The benefit of using the average is that Gaussian noise decreases proportional to $1/\sqrt{N}$ ($N$ is the number of profiles included in the stacked spectrum), which is favourable to the creation of a spectrum with a higher signal-to-noise ratio, S/N. \autoref{fig:totalstacks} shows how the average noise decreases as the number of profiles (in our simulated sample) included in the stack increase. The weighted average is defined by
			\begin{equation}\label{eqn:stack}
				S_\mathrm{stack} = \frac{\sum_{i=1}^N w_i S_i}{\sum_{i=1}^N w_i}
			\end{equation}
			where $i$ is the number of the spectrum to be included in the stack and $N$ is the total number of spectra in the sample. Each spectrum $S_i$ has an associated weighting factor $w_i$. By default, the weighting factor is 1, but the user may choose an alternate option such as 
			\begin{itemize}\label{eqn:weight}
            	\setlength\itemsep{0.5em}
				\item $w_i = {\sigma_{\mathrm{rms},i}}^{-2} $ \citep{Fabello2011}
                \item $w_i = {\sigma_{\mathrm{rms},i}}^{-1} $ \citep{Lah2007}
                \item $w_i = (\sigma_{\mathrm{rms},i} D_{L(z), i}^2 )^{-2} $ \citep{Delhaize2013}
			\end{itemize}
			where $\sigma_{\mathrm{rms},i}$ is the RMS of the channels in a spectrum that contains no galaxy emission (i.e. the noisy channels), and $D_{L(z),i}$ is the luminosity distance to the galaxy. We use the interquartile range of each spectrum as the noise estimate over the conventional standard deviation in order to deal with outlying high or low values for example from RFI spikes. \\

		\subsubsection{Reference/Control Spectrum}
			\label{sec:refspec}

			\begin{figure*}
				\centering
				\includegraphics[width=\textwidth]{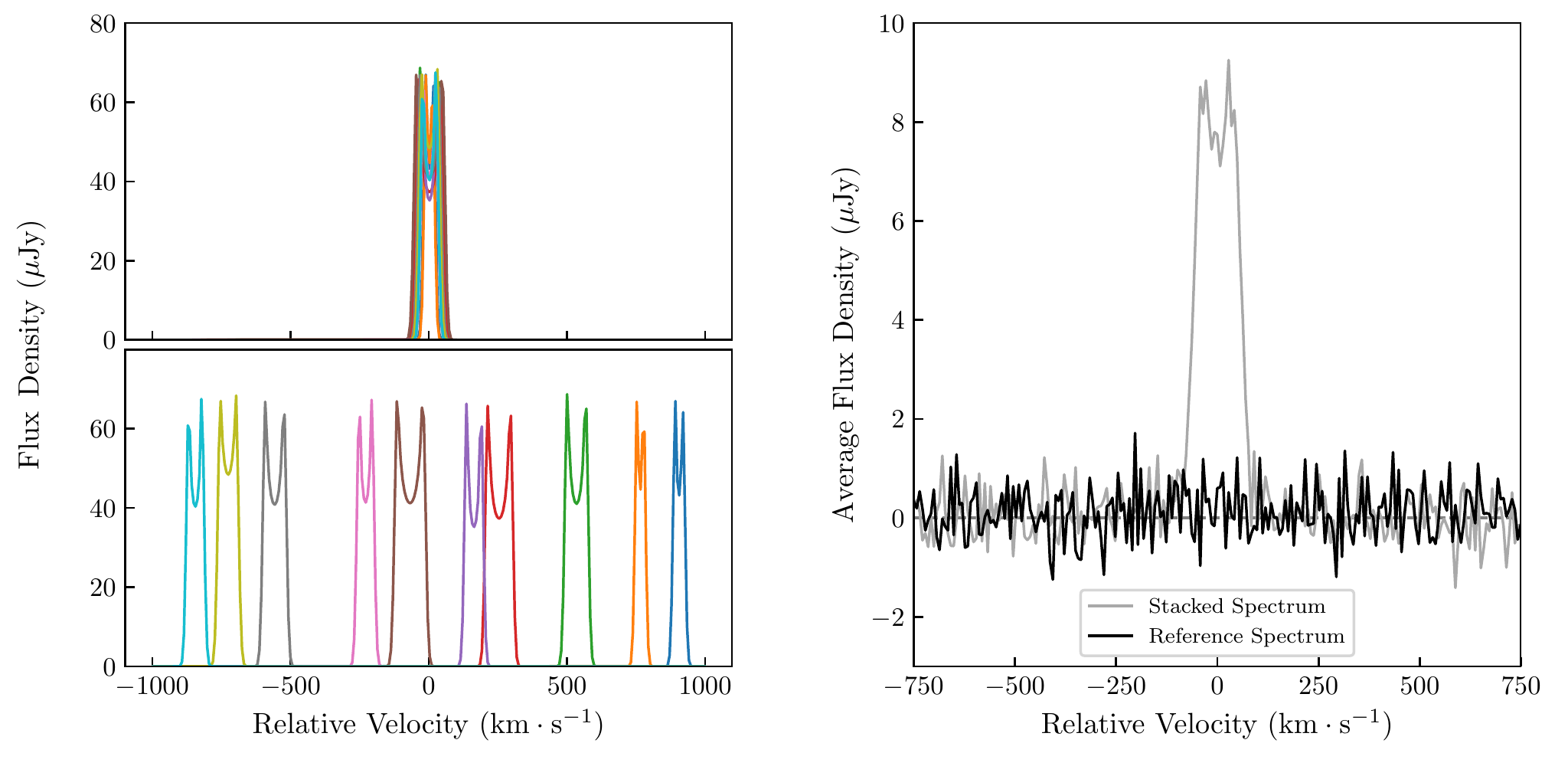}
				\caption{\textit{Top left:} the noise-free versions of the same ten simulated spectra that have been used previously. These spectra have been aligned using the correct redshifts. \textit{Bottom left:} each of the ten spectra has been shifted using the redshift from another galaxy. \textit{Right:} the reference spectrum for the 1000 noisy simulated spectra is plotted in black. The stacked spectrum for this sample is plotted in gray for comparison.}
				\label{fig:refspectrum}
			\end{figure*}

			The practice of creating a reference spectrum has been used since the first \hi stacking experiments. A reference spectrum is a stacked spectrum created by stacking the spectra in the target sample using the incorrect redshift to align the spectra. The various stacking studies have used two methods to create the reference spectrum. Some authors, who have had access to the radio data cube from which the \hi spectra were extracted, choose to create the reference spectrum by extracting spectra from random spatial locations in the data cube and using the target redshifts to align the spectra \citep[e.g.,][]{Zwaan2001}. Other authors use the spectra from their sample but randomize the sample redshifts \citep[e.g.,][]{Chengalur2001, Delhaize2013}. By randomising the sample redshifts, both the reference spectrum and the stacked spectrum have the same systematics arising from the shifting process \citep{Chengalur2001}. If the redshifts are randomly assigned to the sample spectra, the resulting average spectrum should take the appearance of a noise spectrum. If this is the case, it provides further confirmation that a detection in a stacked spectrum is legitimate.  \\

			The left panels of \autoref{fig:refspectrum} show how when the spectra are converted to the rest frame with the wrong redshift they end up completely misaligned. Each of the spectra in the bottom left panel of \autoref{fig:refspectrum} has been shifted using the redshift belonging to one of the other galaxies. The result of co-adding the misaligned spectra using \autoref{eqn:stack} is shown by the black line in the right panel of \autoref{fig:refspectrum}. The lack of any significant emission at \numunit{0}{\kms} supports the integrity of the detection of the corresponding stacked spectrum that has been created using the correct redshifts. For comparison, the thin grey line shows the stacked spectrum of the correctly aligned spectra.

	\subsection{Analysing the stacked spectrum}
		\label{sec:analysismodule}

		The Analysis Module provides the user with a collection of statistics, spectrum manipulation and profile characterisation routines without the need for third-party software. Upon initialisation of the Analysis Module, an initial statistical analysis is performed and the results are displayed either to a window or to the command line if running in suppress mode. At this point the first visual of the stacked spectrum is displayed in the form of Diagnostic Plot 1 (see \autoref{fig:dp1}.) The user is also asked if the integration window size should be changed from its setting from the input phase\footnote{The original integration window is specified by the user when setting \textit{Max galaxy velocity width of sample} (see bottom block of \autoref{fig:guimainwindow})}. The next steps are then displayed to the user:
        \begin{enumerate}
        \item Smooth the spectrum. (\autoref{sec:specmanip})
        \item Fit a number of other functions to the spectrum. (\autoref{sec:fitprofile})
        \item Rebin the spectrum. (\autoref{sec:specmanip})
        \item Do not fit anything to the spectrum, but continue with the analysis.
        \item Exit \stacker without doing anything else. (This option will not save data)
        \end{enumerate}

		\begin{figure*}
			\centering
			\begin{subfigure}{\textwidth}
				\centering
				\includegraphics[width=\textwidth]{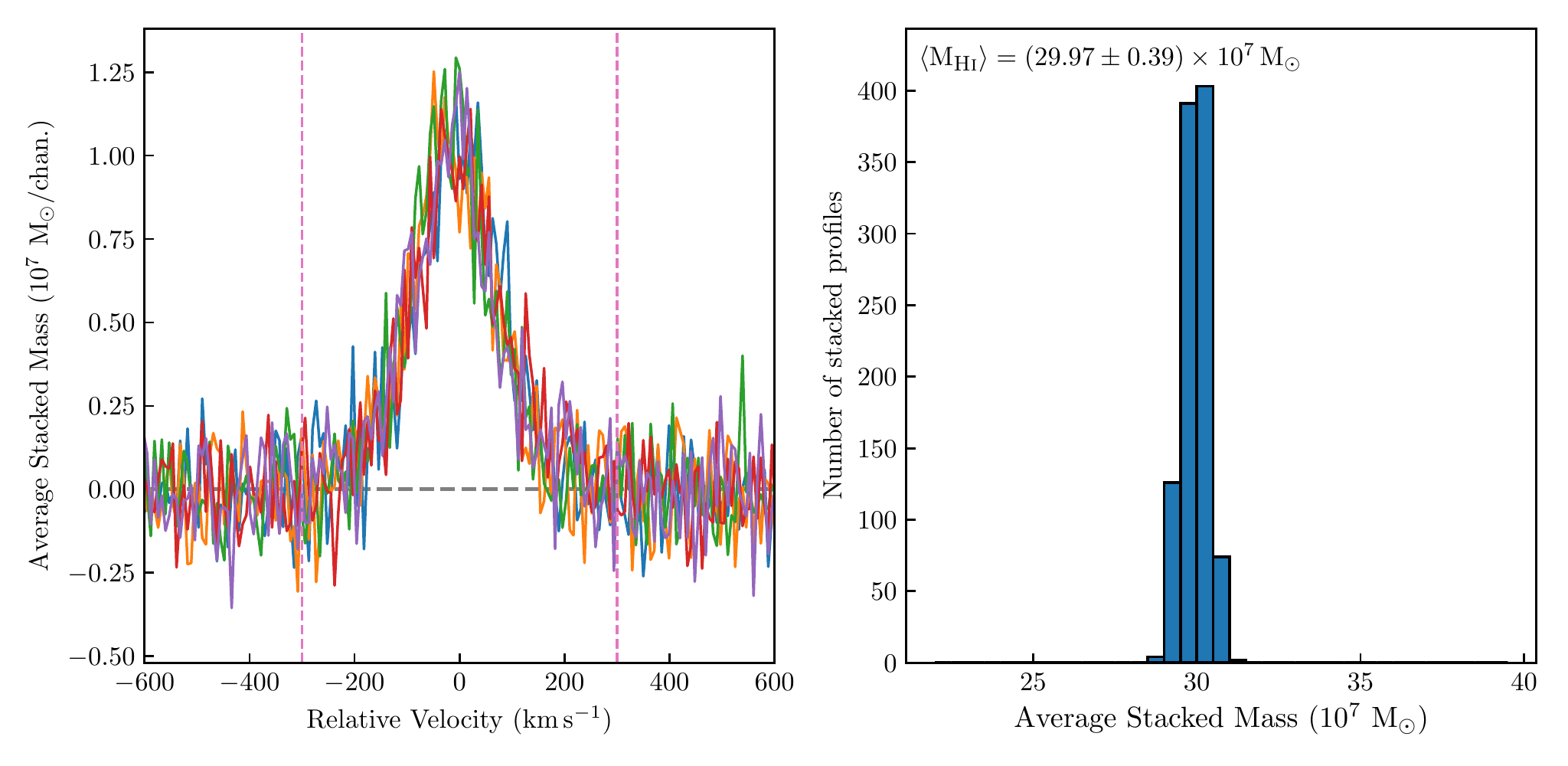}
				\caption[Redshift uncertainty example]{Redshift error analysis: The left panel shows five iterations of the stacked spectrum. Each of the stacked spectra has been created by varying the redshifts of the individual spectra. The errorbars for the final stacked spectrum are determined from the range of values in each channel after 1000 iterations. The right panel shows the \ave{\mhi} calculated from each iteration of the stacked spectrum by integrating the spectrum between the two vertical dashed lines.}
				\label{fig:redshift}
			\end{subfigure}
			\begin{subfigure}{\textwidth}
				\centering
				\includegraphics[width=\textwidth]{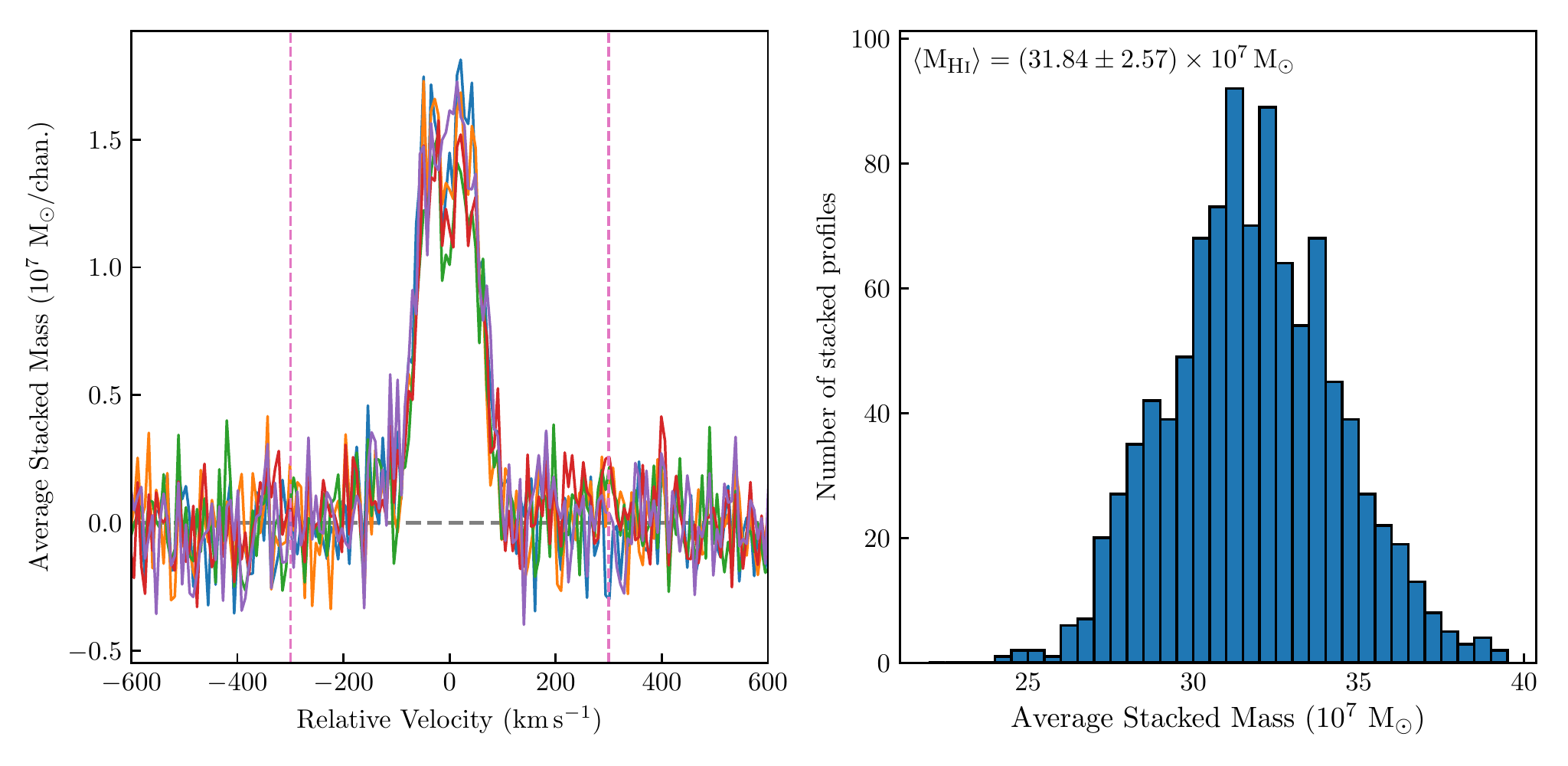}
				\caption[DAGJK uncertainty example]{Statistical error analysis: The left panel shows five iterations of the stacked spectrum. The errorbars for the final the stacked spectrum are determined from the range of values in each channel after 1000 iterations. Each of the stacked spectra has been created using a sub-sample of $75\%$ of the full sample. The right panel shows the \ave{\mhi} calculated from each iteration of the stacked spectrum by integrating the spectrum between the two vertical dashed lines.}
				\label{fig:dagjk}
			\end{subfigure}
			\caption{Uncertainty methods.}
		\end{figure*}

		\subsubsection{Profile detection statistics}
        	\label{sec:profilestats}

			The collection of statistics that are calculated by this module all characterise the significance of the possible detections in the stacked spectrum in some way. The statistics are displayed in the terminal after calculation, and are saved in the log file for perusal after \stacker has completed. The statistical analysis of the stacked spectrum includes three different methods of calculating the signal-to-noise ratio (S/N):
			\begin{itemize}
				\item peak flux density ($\mathrm{S}_\mathrm{peak}$) to noise ($\sigma_\mathrm{rms}$): 
					\begin{equation}
						\mathrm{S/N} = \mathrm{S}_\mathrm{peak} / \sigma_\mathrm{rms}
					\end{equation}
					where $\mathrm{S}_\mathrm{peak}$ is the maximum value of the spectrum within the integration window.
				\item integrated signal-to-noise: 
					\begin{equation}
						\mathrm{S/N}_{int} = \frac{\sum^{N_\mathrm{chan}}_i S_i\cdot \mathrm{dv}}{\sigma_\mathrm{rms} \mathrm{dv} \sqrt{N_\mathrm{chan}}},
					\end{equation}
					where $S_i$ is the flux density (in Jy) associated with channel $i$, $\sigma$ is the rms noise, dv is the channel width (in \kms), and $N_\text{chan}$ is the number of channels integrated over.
				\item ALFALFA S/N \citep[from][]{Haynes2011}: 
					\begin{equation}\label{eqn:snralf}
						\mathrm{S/N}_{ALFALFA} = \left(\left( \frac{\mathrm{F}_\hi}{W_{50} }  \right)\sqrt{ \frac{W_{50}}{2 R}  }\right) \Biggm/ \sigma_\mathrm{rms},
					\end{equation}
					where the integrated flux density of the detection is given by $\mathrm{F}_\hi$ (Jy \kms), $W_{50}$ (\kms) is the width of the spectrum at $50\%$ the peak value\footnote{The W$_{50}$ is obtained from the single Gaussian fit to the spectrum shown in \autoref{fig:dp1}.}, and $R$ is the velocity resolution of the spectrum (this is also called the channel width, unless the spectrum has been smoothed along the spectral axis).
			\end{itemize}

			The next statistic calculated is the probability that the signal in a stacked spectrum arises due to a random fluctuation -- this is known as the p-value, which will be as low as possible for a strong detection, from which the significance level of the detected signal can be calculated. This is done by fitting a single Gaussian to the spectrum and calculating the $\chi ^2$ value. The single Gaussian is used in favour of other models as it is the simplest model that can describe the shape of an \hi detection to first order; please note that in the case of a very clear detection which cannot be characterised by a single Gaussian, the p-value statistic provides no extra information. The null hypothesis is a horizontal line at zero.

		\subsubsection{Spectrum manipulation}
			\label{sec:specmanip}

			In order to boost the S/N of a stacked spectrum, the Analysis Module offers the user the choice of either re-binning the data \citep[lowering the data resolution, this method has been used by][]{Lah2007, Rhee2016}, or smoothing the spectrum using either a boxcar algorithm or a Hanning Window. \\

			All three manipulation routines allow the user to select either the window or the number of old bins to be rebinned into new bins. \stacker will not continue to the next step until the user is satisfied with the new version of the stacked spectrum. Diagnostic Plot 2 (which is identical to Diagnostic Plot 1, \autoref{fig:dp1}, but displays the updated spectrum) along with a recalculation of the initial statistics is produced every time a change is made to the spectrum.

		\subsubsection{Profile characterisation}
        \label{sec:fitprofile}

	Given that stacked spectra can have very low S/N, with a lot of noise spikes and dips that are unhelpful in determining the integrated flux density and line width, fitting a function to a stacked spectrum can assist in estimating the stacked profile properties. For example, determining the \hi line width ($W_{50}$) of stacked spectra can be useful for studies such as determining the \hi Tully-Fisher relation\footnote{An empirical relation which relates the luminosity of galaxy to its rotation velocity \citep{Tully1977}.} of non-detected galaxies \citep{Meyer2015}. The analysis module prompts the user to choose from a selection of seven different functions (between one and seven can be chosen simultaneously) that have been used in previous \hi studies to characterise \hi line profiles:
               
			\begin{enumerate}
				\item Single Gaussian
				\item Double Gaussian
				\item Lorentzian Distribution
				\item Voigt Profile
				\item 3$^{\rm rd}$ Order Gauss-Hermite Polynomial
				\item Busy Function \citep{Westmeier2013} -- with double-horn
                \item Busy Function -- without double-horn
			\end{enumerate}

			\autoref{fig:functionfit} shows the 7 functions fitted to the stack of the 1000 simulated profiles. Also shown is a table containing the integrated flux density, a measure of the goodness of fit, and the full-with at half-maximum (FWHM) of the fitted function. The user may select any number of the functions, and \stacker checks that the user is satisfied with the selection before continuing. A version of \autoref{fig:functionfit} is saved in PNG\footnote{Portable Network Graphics} format in the output location as Diagnostic Plot 3. The diagnostic plot is saved to file if \stacker is run in suppress mode (an option the user may select when initialising \stacker which suppresses all displayed output) so that the user may inspect the function fit before continuing with the analysis.

		\subsubsection{Uncertainty calculation}
			\label{sec:uncertcalc}

			The Uncertainty Module allows for two types of uncertainty calculations: a statistical error analysis and a redshift error analysis. The user specifies the type of uncertainty calculation (the two methods cannot be run simultaneously). The Uncertainty Module facilitates the storing of the final analysis options chosen by the user and uses those options to repeat the stacking process 1000 times, each time with slightly different parameters. \autoref{fig:outlineflowuncert} indicates the Uncertainty Module as a wrapper around the Stack and Analysis modules. \\

			1. Redshift error analysis: if the user chooses to engage the redshift uncertainty calculations, another loop is activated around the Stack and Analysis Modules which repeats the two processes one thousand times. In every iteration, the redshift associated with each spectrum in the input catalogue is changed by an amount that is selected from a normal distribution with a standard deviation equivalent to the particular spectrum's redshift uncertainty. Thus, the redshift for each spectrum becomes $z = z + \text{d}z$, where d\z has been sampled from a normal distribution defined by $N(\mu = z; \sigma = u(z) )$.\\

			The result of repeating the stack and analysis process means that there are 1000 slightly different versions of the stacked spectrum and their corresponding integrated flux values. Each of the 1000 versions of the analysis object (which each contain a selection of integrated fluxes and the stacked spectrum data) are stored by the Uncertainty Class. Upon completion, the Uncertainty Class processes the stored data: the error bars on the spectrum data are calculated from the minimum and maximum values per channel of the 1000 stored spectra, the data points are given by the mean values. The error-bars on the integrated fluxes are determined from the difference between the mean value (which serves as the quoted value) and the $25^{\rm th}$ and $75^{\rm th}$ percentile.\\
            
            An example of the stacked spectrum using this error calculation method and an average \numunit{\text{d}z = 60}{\kms} is shown in \autoref{fig:redshift}.  \\

			2. Statistical error analysis: the statistical error analysis implemented by this module is the \textit{Delete-A-Group Jackknife} \citep{Kott2001} error method. In the same way as mentioned above, another loop is activated around the Stack and Analysis Modules repeating the process a thousand times, however each iteration discards a percentage (as chosen by the user) of the total catalogue without replacement. The 1000 analysed stacked spectra are stored by the Uncertainty Module. The flux values for the final stacked spectrum are taken as the mean value of the 1000 versions in each channel. The error bars on the flux in each channel are calculated according to 
			\begin{equation}\label{eqn:dagjk}
				\sigma (s) = \sqrt{ \frac{R-1}{R} \sum^R_{n=0} \left( s_n - \bar{s} \right)^2 },
			\end{equation}
			where $R$ is the number of repeated estimates (which would be 4 if only $75\%$ of the population was used for each iteration), $s_n$ is the flux for the particular channel of the n$^{\rm th}$ subset, and $\bar{s}$ is the mean flux value for the particular channel \citep{Kott2001, Brown2015}. This is illustrated in \autoref{fig:dagjk}.\\

The shape of the redshift error analysis profile is smeared out into a more Gaussian shape than the statistical error analysis stacked profile due to each spectrum being allocated its correct redshift plus an offset value. This results in the individual line centres being smeared around the zero point. This was also observed by \citet{Maddox2013}.\\

The choice of preferred error analysis method is left to the user to decide based on their particular sample selection and optical catalogue redshift precision.

	\subsection{Save Results}
			\label{sec:outputmethods}

			There are nine possible files that can be saved to the output location -- 5 data files and 4 possible diagnostic plots (these inform user interactions at point 2):
			\begin{enumerate}
				\item Stacked Catalogue: this text file contains a list of all the spectra included in the stack. Along with the data provided by the user from the input catalogue, this table also contains the integrated flux density for each spectrum.
				\item Output Data: a FITS table file containing the spectrum data (stacked spectrum, spectral axis and reference spectrum), the fitted parameters of any fitted functions, the stacked noise, and if the spectrum has been smoothed then the original version of the spectrum data is also saved.
				\item Stacked Spectrum Plot: a PDF file of the stacked spectrum plotted with the fitted functions and the reference spectrum (a version of this is shown in the top left panel of \autoref{fig:fulldisplay}).
				\item Stacked Noise Plot: a PDF file of the stacked noise which has the expected $\sigma/\sqrt{N}$ line over-plotted (top right panel of \autoref{fig:fulldisplay}).
				\item Integrated Flux Data File: a table of the calculated integrated flux from the different functions as well as the flux integrated within the galaxy window. This file is saved in Encapsulated Comma Separated Value (\texttt{ecsv}) format using the \astropy ASCII module, which allows the units of the flux to be saved to file. Other columns in this file include the goodness of fit values from the function fits. A version of this table is shown in the bottom panel of \autoref{fig:fulldisplay}.
				\item Diagnostic Plot 1: this is the first plot of the stacked spectrum that is displayed and saved upon initialisation of the Analysis Module. (\autoref{fig:dp1})
				\item Diagnostic Plot 2: a PNG file that is only produced if the user decides to rebin or smooth the spectrum. This plot is almost identical to \autoref{fig:dp1}.
				\item Diagnostic Plot 3: a PNG file that is created when the user selects a function to fit to the stacked spectrum (\autoref{fig:functionfit}).
				\item Diagnostic Plot 4: this plot is only produced if the uncertainties have been calculated. The plot contains a series of histograms showing the spread of the integrated fluxes. Diagnostic Plot 4 is saved to file in PNG format (\autoref{fig:dp4}).
			\end{enumerate}

	\subsection{Displaying the results}
		\label{sec:displaymodule}

		Upon conclusion of the analysis of the stacked spectrum, all calculated quantities and plots are saved to the user-chosen output location. Amongst the saved files is a plot of the stacked spectrum which makes this final module an optional one. The Display module reads in the files saved to disk at the end of the analysis. The results are then displayed in an easy to read manner -- the stacked spectrum (along with the reference spectrum) is shown on the top left, to the right of the stacked spectrum is the stacked noise as a function the number of profiles. Below the two plots are the statistics associated with stacked spectrum, and a table containing the calculated integrated fluxes. A screenshot of the display window is shown in \autoref{fig:fulldisplay}.

\section{Application of \stacker to the NIBLES Survey}
	\label{sec:nibles}

In this second half of the paper, we will demonstrate some of the capabilities of \stacker by stacking different sub-samples of galaxies observed as part of the \nancay Interstellar Baryon Legacy Extragalactic Survey \citep[NIBLES;][]{VanDriel2016}, and re-visiting the well-known \hi scaling relations \citep[e.g.][]{Catinella2009,Catinella2012,Catinella2013,Brown2015}. \\

	NIBLES is a targeted \hi survey of galaxies selected by $M_\mathrm{z}$, a proxy for stellar mass, from the Sloan Digital Sky Survey \citep[SDSS;][]{York2000}. The \hi data were obtained using the \numunit{100}{\unit{m}} \nancay Radio Telescope (NRT) in the centre of France. NIBLES provides a unique opportunity to study the baryonic content in galaxies with a wide range of stellar mass ($10^6 < \mstar (\msol) < 10^{12}$). 

    \subsection{H~{\sc{i}} Data}
		\label{sec:galaxyselection}

		The NIBLES sample of 2600 galaxies was selected from SDSS Data Release 5 \citep[DR5;][]{AdelmanMcCarthy2007}. The selected galaxies were chosen to uniformly sample the stellar mass range of the ensemble of galaxies in the local Universe. The particular selection criteria \citep{VanDriel2016} were as follows:
		\begin{enumerate}
			\item The target must have both SDSS DR5 spectroscopy and photometry.
			\item NIBLES targets were limited to the Local Volume: $900 < cz < \numunit{12000}{\kms}$. 
			\item In order to study the \hi in galaxies as a function of stellar mass, $\sim$ 150 galaxies were selected per \numunit{0.5}{\text{mag}} bin for $ \numunit{-16.5}{\text{mag}} \leq M_\text{z} \leq \numunit{-24}{\text{mag}}$ ($H_0 = \numunit{70}{\kms\,\text{Mpc}^{-1}}$), if available. $M_\text{z}$ was used as a proxy for stellar mass\footnote{SDSS stellar mass catalogues were not yet publicly available at the time of the original NIBLES sample selection which was based on DR5 photometry.};
			\item In order to maintain a morphologically diverse sample, no colour selection was made.
		\end{enumerate}
		In addition to these criteria, \citet{VanDriel2016}, when selecting the target galaxies tried to avoid objects in and around the Virgo cluster as well as the ALFALFA $\alpha$.40 footprint \citep{Haynes2011}. The dense environment in the Virgo cluster is known to have noticeable effects on the \hi properties of the galaxies in the region. The distances to galaxies in the Virgo region also have large uncertainties.\\

		The \nancay Radio Telescope (see \citeauthor{VanDriel2016} for further details) is a \numunit{100}{\m} single dish radio telescope located in France. The telescope is a transit instrument with a collecting area of \numunit{6900}{\unit[2]{m}} and consists of two large reflectors. At \numunit{21}{\text{cm}}, the NRT beam spans $3.6'$ in right ascension and varies in declination from $22' - 33'$ depending on the declination of the source; the exact declination size of the beam can be determined using \autoref{eqn:beamsize}.  \\

		The spectra were collected from January 2007 to December 2010, for a total of 3500 hours of telescope time. Each target was initially observed for roughly \numunit{40}{\text{minutes}} using successive \numunit{40}{\text{s}} ON source and \numunit{40}{\text{s}} OFF source pointings, which was repeated as required and as observing time permitted. The OFF source pointing was chosen to be \numunit{20'}{\text{E}} of each target. The data reduction process is discussed in detail by \wim. Each published spectrum has a velocity resolution of \numunit{18}{\kms} and a typical noise level of \numunit{2.5}{\unit{mJy}}.

        \subsubsection{Higher-sensitivity Arecibo follow-up H~{\sc{i}} observations}
        
        An indication that stacking \nancay \hi spectra may lead to the detection of a significant signal can be found in the higher sensitivity follow-up observations that were obtained with the 305m Arecibo radio telescope, which resulted in a high detection rate of NRT non-detections \citep{Butcher2016}. Of the \nancay targets, 71 non-detections and 19 marginal detections were observed at Arecibo with a four times higher sensitivity (mean rms \numunit{0.57}{\text{mJy}}). This led to the detection of 85\% and 89\% of the NRT non-detections and marginal detections, respectively. It should be noted that most (59, or 66\%) of the re-observed NRT non-detections are blue, low mass galaxies ($\sim 10^{6.5} < \mstar\,(\msol) < 10^{8.5}$) which are expected to be relatively gas-rich (see \autoref{sec:stackall}). Within the same stellar mass range, the \ave{\mathrm{F}_\hi} of the new Arecibo detections is on average three times lower than for the \nancay detections. Alternatively, using the \stacker approach to reach the four times lower Arecibo rms level we could also have stacked 16 NRT spectra. The analysis of similar Arecibo data of a much larger sample of NIBLES \nancay non-detections of all colours has recently been published by \citet{Butcher2018}. The Arecibo data mentioned above were not used for the stacking study presented here.

 		\subsubsection{Comparison between NIBLES and other H~{\sc{i}} surveys}
 			\label{sec:compsurvey}

			Upon comparison with other single-dish \hi surveys (e.g. ALFALFA, HIPASS \citep{Barnes2001}; see Table A.4 in \citealt{VanDriel2016}), the ALFALFA $\alpha.40$ fluxes were found to be higher than the NIBLES fluxes by a factor \mbox{$\alpha.40$/NRT $= 1.45 \pm 0.17$}. The comparisons between NRT fluxes and the ALFALFA fluxes, as well as possible reasons for the difference in flux scales are discussed in detail in \S~4.6 in \citet{VanDriel2016}. In this work, the NIBLES fluxes are quoted. In the event of comparison between NIBLES measurements and ALFALFA measurements, the ALFALFA measurements are scaled to the NIBLES flux values using the ALFALFA/NRT flux ratio mentioned above. 

	\subsection{Optical Data}
	The optical photometric data used in this work is obtained from the SDSS Data Release 12 where the photometric quality flag was marked clean with no deblended ``child'' objects. In the DR12 database, the spectroscopic sources (from which the NIBLES sample is derived) are matched to the ``value-added'' JHU/MPA catalogues \citep{Brinchmann2004} which contain the stellar masses used in this work. The catalogue provides a number of different measures of the stellar mass, however for the purposes of the NIBLES analyses, the median values are used \citep{Butcher2018,VanDriel2016}.
   
     \subsection{NIBLES Stacking Sample}
 	\label{sec:stackingsample}
 
 	\subsubsection{Accounting for known contamination sources}
			\label{sec:contaminantemission}
			Studies have shown that galaxies which are nearby to the target galaxies in terms of both distance on the sky and systemic velocity can significantly contribute to the total emission in a target \hi galaxy spectrum \citep{Jones2015, Elson2016}. For example, \citet{Elson2016} showed that for a stacking experiment using simulated Parkes data at $15'$ resolution at $0.04 < \z < 0.13$, the average contaminant \hi mass per galaxy spectrum was \numunit{\sim 1.4 \times 10^{10}}{\msol}. Thus, before finalising the sample of spectra for stacking, it is necessary to remove spectra that are known to have nearby galaxies that could contribute significant contaminating emission to the spectrum. In other works \citep[e.g.,][]{Fabello2012}, a secondary source (irrespective of morphology) is considered to contaminate the target spectrum if it lies spatially within the half-power beam width and \numunit{\pm 300}{\kms} of the target. In this work, galaxies near the target source are considered as possible contaminators if
			\begin{itemize}
	        \item the secondary galaxy is located spatially within $1.2 \times$ N-S half-power beam width and $2 \times$ E-W half-power beam width, and
			\item the redshift of the secondary galaxy is within \numunit{\pm 300}{\kms} of the target redshift.
			\end{itemize}
			The above criteria were used to search for sources in both the SDSS Spectroscopic Database and the NASA/IPAC Extragalactic Database (NED). A total of 761 spectra were classified as potentially contaminated using these criteria. For more details see \appendref{sec:contamination}.

 \subsubsection{Defining H~{\sc{i}} detections vs. non-detections}
 	\label{sec:ndselect}
        
NIBLES galaxies were originally classified as `detected', `marginal' or `not detected' as explained in \wim by three independent adjudicators, who used statistical parameters describing the profile S/N and also inspected all spectra by eye. For this work we wanted a binary classification of `detected' or `not detected' for our stacking experiments and we wanted to use a method which did not rely on human intervention. Various methods exist for finding signals and classifying detections in \hi spectra (see for example \citealt{Saintonge2007,Verheijen2007,Haynes2011,Ramatsoku2016}). Using the original classifications in \wim as a guide, we used classification criteria, adapted from \citet{Ramatsoku2016} to reproduce the sample selection of \wim, while at the same time self-consistently and objectively classifying the marginal detections as either detections or non-detections. This method produced very similar detection classifications to \wim with over $96\%$ overlap in the classifications between the two schemes for both detections and non-detections, and resulted in the NIBLES marginal detections (52) being allocated as detections vs. non-detections in a ratio of one-third to two-thirds in our sample.\\

Following \citet{Ramatsoku2016} we classify galaxies as detected if there are a number of consecutive channels above a threshold value. The threshold values are set based on the noise level of the individual spectra which we estimate using the sub-interquartile range (IQR -- half the difference between the $75^\mathrm{th}$ and $25^\mathrm{th}$ flux percentiles). The IQR is less sensitive to outliers in a distribution than the standard deviation and therefore eliminates the need for manual masking of the target source and/or artefacts in the spectrum when estimating the noise (see \autoref{fig:detectioncriteria}). For classification of NIBLES spectra, we considered a window of \numunit{600}{\kms} centred on the target redshift. If any emission, satisfying the criteria below, was found in the window, the spectrum was classified as a detection:
\begin{itemize}
	\item 1 or more channels with flux density above $5 \times$ IQR (within a \numunit{400}{\kms} window)
	\item 2 or more consecutive channels above $4 \times$ IQR
	\item 4 or more consecutive channels above $3 \times$ IQR
	\item 5 or more consecutive channels above $2 \times$ IQR.
\end{itemize}

Visually inspecting the spectra after classification showed that some sources had an \hi detection in the off-source pointing which presents in the NIBLES spectra as an absorption-like feature, while other spectra had clear emission features from sources not previously identified in our possible contaminant source search (see \autoref{sec:contaminantemission}.) Examples of spectra with these features are presented in \autoref{fig:sketchyspec}. These features can bias the stacked spectrum and therefore in order to flag these spectra and eliminate them from our sample, we ran the classification algorithm a second time over a wider velocity range (\numunit{\pm 600}{\kms} centred on the target redshift and indicated by the vertical green dashed lines in \autoref{fig:sketchyspec}), to identify possible positive and negative emission not related to the target galaxy.
We rejected galaxy spectra that had either negative emission satisfying our classification criteria in any part of the larger search window (\numunit{cz \pm 600}{\kms}) or positive emission satisfying the criteria and falling outside a central range of \numunit{cz \pm 200}{\kms} but within \numunit{cz \pm 600}{\kms}. In total 24 spectra were removed from the stacking sample in this way.

	\subsubsection{Final stacking sample}
    	We excluded from our sample galaxies with \numunit{\mstar < 10^8}{\msol} due to unreliable SDSS photometry and therefore unreliable stellar masses. Our final catalogue consists of 1000 \hi spectra deemed free of nearby contaminant emission. Of the 1000 sources, 323 were classified as non-detections and 677 as detections (see \autoref{fig:nddetexampls} for examples). The catalogue for the non-detections can be found in \appendref{sec:ndcatalogue}. 

			\begin{figure}
				\centering
				\includegraphics[width=\textwidth]{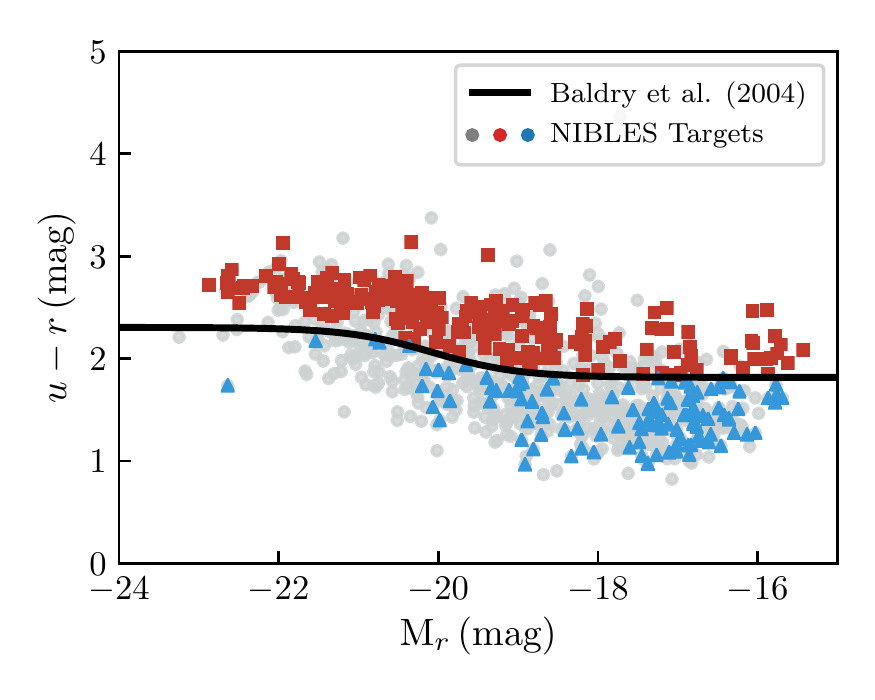}
				\caption{Optical colour-magnitude diagram for the NIBLES stacking sample galaxies: \ur colours derived from SDSS model magnitudes corrected for Galactic extinction as a function of absolute r-band magnitude. The \hi non-detected galaxies are plotted in either red or blue based on their position above or below the \citet{Baldry2004} red/blue colour divider which is indicated by the black line. The grey data points represent the \hi detected galaxies.}
				\label{fig:nibcolourmag}
			\end{figure}
	
	 \begin{figure*}
			\centering
			\includegraphics[width=\textwidth]{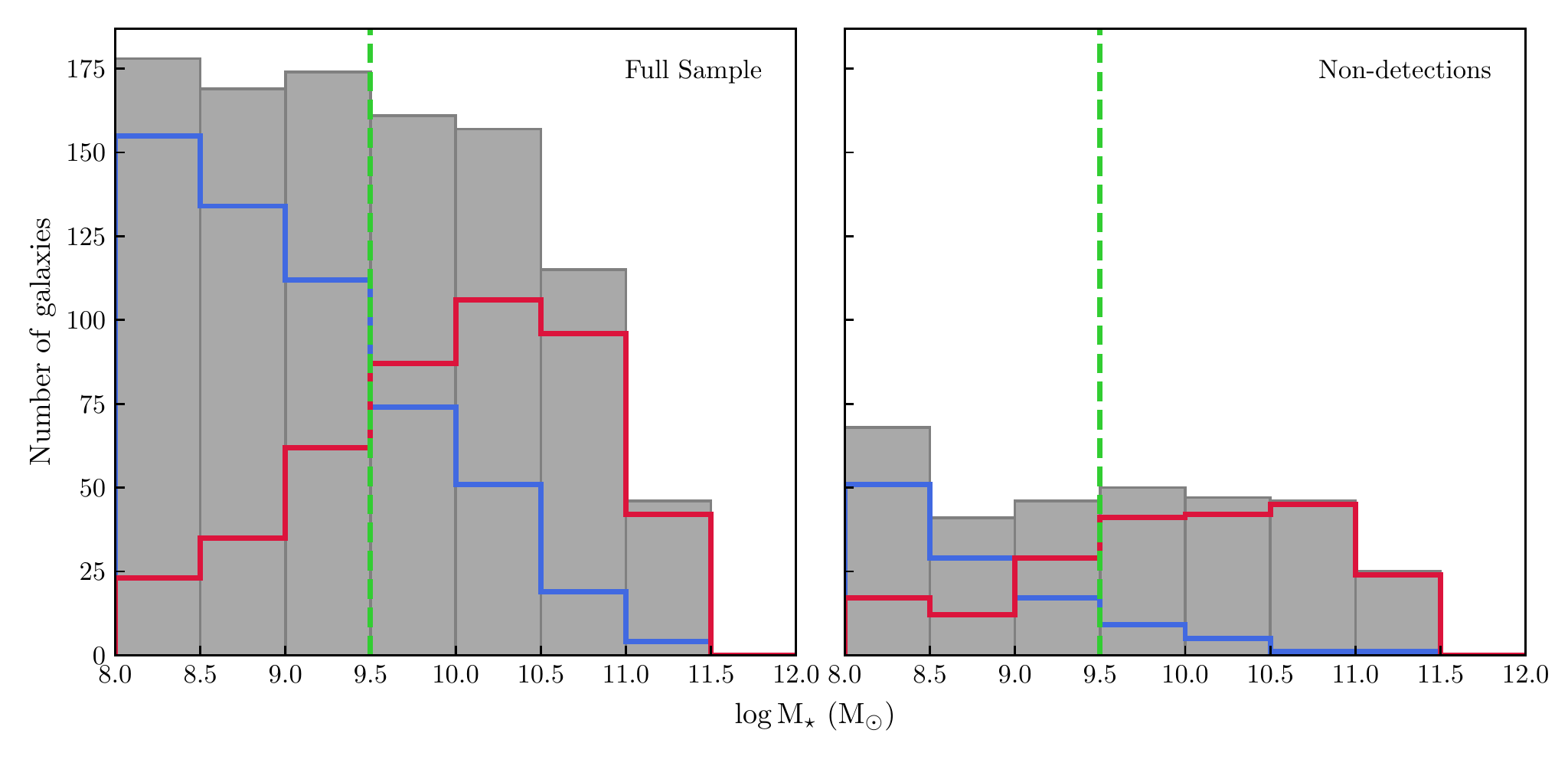}
			\caption{The stellar mass distribution for the full sample (\textit{left}) and the non-detections (\textit{right}). The grey bars indicate the total number of galaxies per bin, while the blue and red lines indicate the number of blue and red galaxies per mass bin. Each bin is \numunit{0.5}{\unit{dex}} wide. The dashed green line highlights the ``gas-richness'' threshold \citep{Kannappan2013}, where gas-rich galaxies are to the left of the line (lower \mstar) and gas-poor galaxies are to the right of the line (higher \mstar).}
			\label{fig:smbinsfc}
		\end{figure*}

\section{Gas scaling relations in NIBLES}
	\label{sec:stackall}

	    As has been previously stated, the NIBLES sample is morphologically diverse, and it ranges from gas-rich late-type galaxies to gas-poor early-type galaxies. In order to explore the contributions from the different morphological types to the stacked profiles, we separate the stacking sample into red and blue sub-samples using the \citet[Eq.~9]{Baldry2004} optimal colour divider as shown in \autoref{fig:nibcolourmag}. \\
     
    Having separated the full sample into blue and red sub-samples, there are 549 blue galaxies (436 detections, and 113 non-detections) and 451 red galaxies (241 detections, and 210 non-detections). The stellar mass distributions for the full sample and non-detections alone, and each broken down by colour, are shown in \autoref{fig:smbinsfc}. \\
 
 It is clear in both the full sample and the non-detected sample that the blue galaxies are predominantly in the low-\mstar regime, while the red sample contains mainly high-\mstar galaxies. The transition from predominantly blue galaxies to predominantly red galaxies occurs in our full sample around \numunit{\mstar = 10^{9.5}}{\msol} which also corresponds to the so-called ``gas-richness'' threshold \citep{Kannappan2013} which signifies the transition from \hi~gas-rich galaxies (at lower \mstar) to \hi~gas-poor galaxies (at higher \mstar).   
    \subsection{The average H~{\sc{i}} properties}
    
    \begin{figure*}
	\centering
    \includegraphics[width=\textwidth]{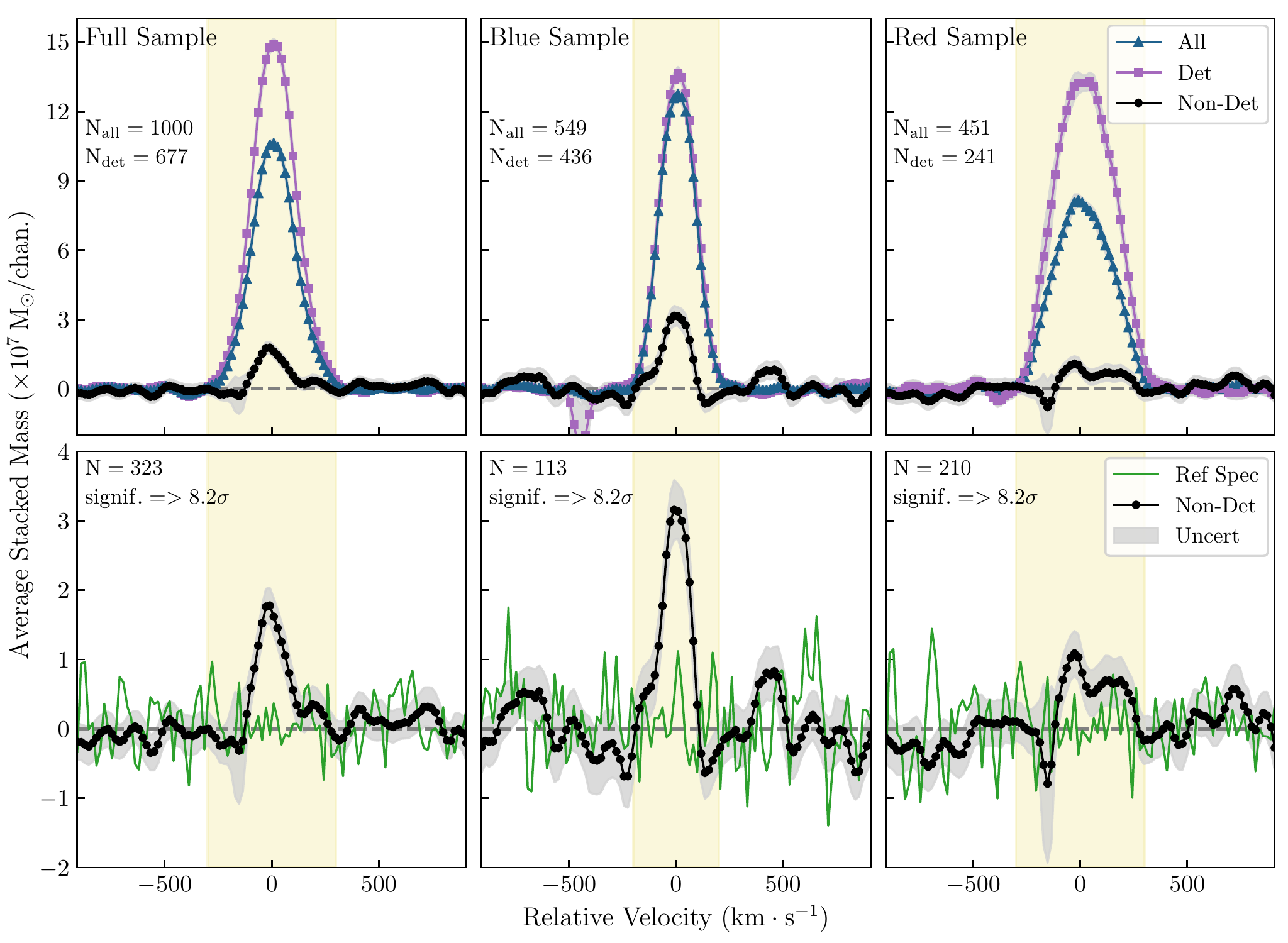}
    \caption{The three columns in the figure above show the stacked profiles for the full sample, blue sample, and red sample respectively. In the top row are the stacked profiles for all spectra (blue triangles), only detections (purple squares), and only non-detections (black circles). The bottom row shows the stacked profiles for the non-detections, the reference spectrum is shown in green, and the grey band represents the uncertainty of the stacked spectrum. The vertical yellow regions in each of the panels indicate the region over which the spectra are integrated to obtain the average \hi mass, \ave{\mhi}.}
    \label{fig:ninepanel}
\end{figure*}

	We separated the full sample into three samples: all spectra (full sample), detections only, and non-detections only. The spectra are stacked in units of \mhi and the resulting spectra are shown in the left column of \autoref{fig:ninepanel}. The top row of \autoref{fig:ninepanel}  shows the stacked profiles for the full sample, detections, and non-detections on the same set of axes; the bottom row shows only the stacked profiles for the non-detections, enlarged for clarity. The top left panel shows the comparison between the stacked spectra for the full sample (blue triangles), and the spectra for the sample of only detections (purple squares) and only non-detections (black filled circles). \\
    
    The middle and right panels of \autoref{fig:ninepanel} show the stacked spectra for the blue and red sub-samples respectively. As indicated by the vertical yellow bands, we integrate over a narrower velocity range for the blue sub-samples compared to the red sub-samples. The blue stacked spectra are expected to be narrower since the average stellar mass of the blue sample is lower than the red sample. \\
    
    The average \hi mass and \mhi to \mstar ratio (\fhi) measured from the stacked profiles for each of the nine sub-samples (all galaxies, all blue galaxies, all red galaxies; all \hi detections, blue \hi detections, red \hi detections; all non-detections, blue non-detections, and red non-detections) are shown in \autoref{tab:ninesampleplot}.  \\

\hi stacking is a powerful tool to probe the average \hi properties of various galaxy samples, however one should have a clear idea of the morphological properties of the galaxies that make up the sample. We have shown how the stacked \hi profiles can change by only looking at blue or red galaxies -- we find that the stacked profiles for the red samples are wider than the stacked profiles for the blue samples. This is to be expected based on the bimodal distribution of local galaxies where the red galaxies on average have higher stellar masses, and in line with the Tully-Fisher relation, higher rotation velocities \citep[see also][]{Fabello2011, Meyer2015}. \\
       
\begin{figure*}
			\centering
			\includegraphics[width=\textwidth]{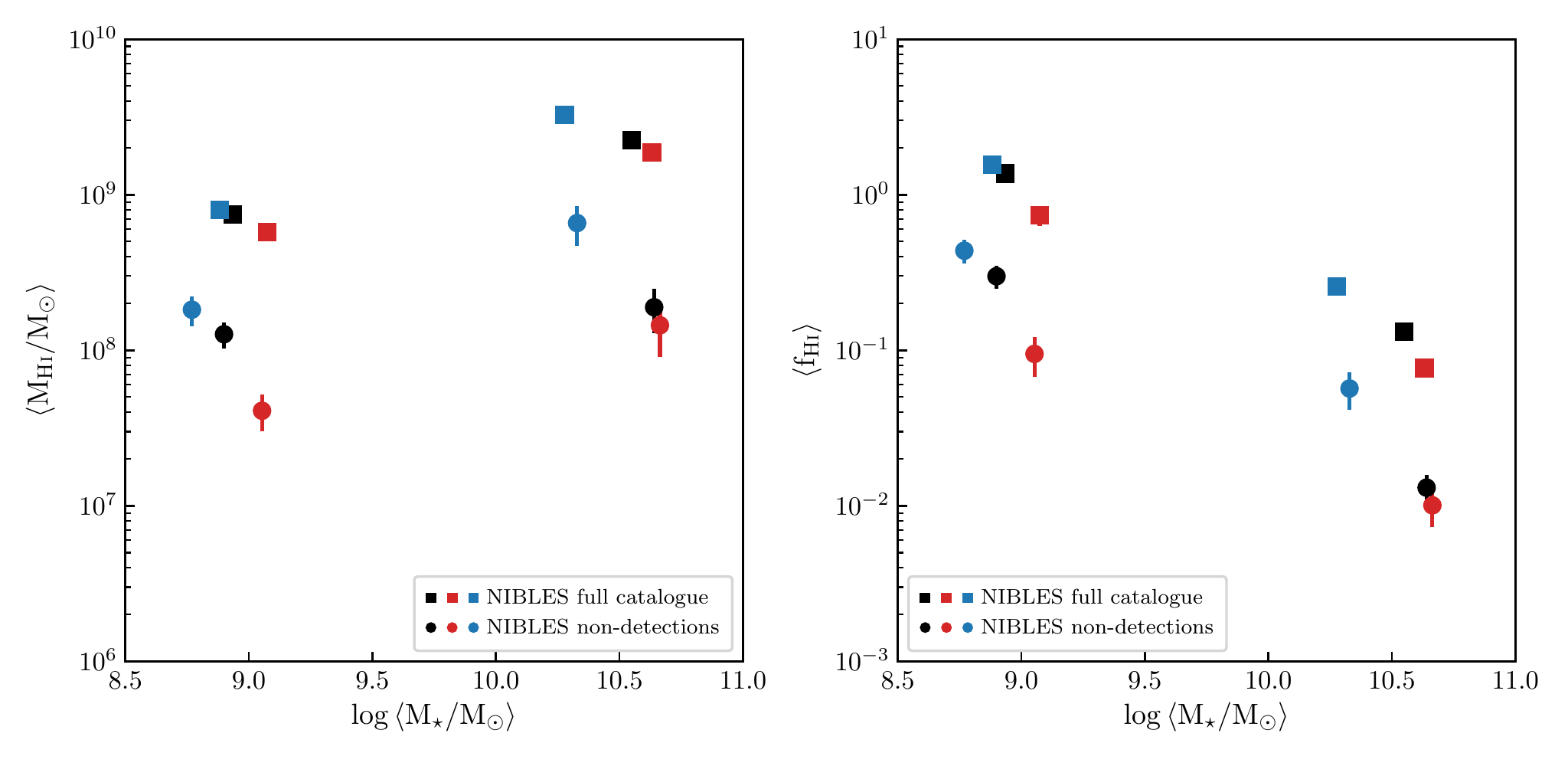}
			\caption{Average \mhi \ave{\mhi} (left), and the \ave{\mhi/\mstar} mass ratio, the average gas fraction \ave{\fhi} (right), as a function of \mstar for the full NIBLES sample as well as the sub-sample of \hi undetected galaxies. Each sample has been split into two stellar mass bins: $10^8 < \mstar\,(\msol) < 10^{9.5}$ and $10^{9.5} < \mstar\,(\msol) < 10^{12}$. Shown here are the mean \mstar of the galaxies in each of the two stellar mass bins, for each of the various samples. The round data points represent the non-detections, and the square points represent the full sample. The blue, red and black data points represent respectively the blue, red and total sub-samples. The error bars represent the statistical uncertainties, including on the stacks of non-detections. The stacked profiles which each data point represents, are presented in \appendref{sec:fig8profiles}.}
			\label{fig:2binsSM}
            \label{FIG:2BINSSM}
		\end{figure*}

        We divided our sample into two bins of low and high stellar mass (\numunit{\mstar < 10^{9.5}}{\msol} and \numunit{\mstar > 10^{9.5}}{\msol}) to explore the average \hi properties of blue and red galaxies in these 2 samples. As outlined earlier in this paper, \stacker is capable of directly determining the average \hi mass or \hi gas fraction from stacked spectra. The \hi-to-stellar mass fraction (gas fraction or \fhi) is a useful tool to probe how gas rich a galaxy or sample of galaxies is. Our results are summarized in \autoref{fig:2binsSM} and the stacked profiles are shown in \appendref{sec:fig8profiles}, with the measured quantities presented in \autoref{tab:fig9lowmass} and \autoref{tab:fig9highmass}. Note that all the stacked profiles have S/N$_\mathrm{ALFALFA} > 6.5$ and are also classified as detections according to the criteria in \autoref{sec:ndselect}.\\ 
        
        These results are consistent with work by \citet{Kannappan2004} which shows that while low mass galaxies tend to have lower \hi masses than their high mass counterparts, the low stellar mass galaxies tend to have higher \hi gas fractions. The right panel of \autoref{fig:2binsSM} highlights what was found by \citet{Brown2015} that the steep slope of the average \ave{\fhi} vs \mstar trend (indicated by the black squares and filled circles) from low to high stellar mass is driven by the predominance of high gas fraction blue galaxies at low stellar mass and the low gas fraction red galaxies at high stellar mass. This trend is seen for both the full stacked sample and when stacking non-detections alone.

\subsection{Gas fractions in NIBLES galaxies}

		\defcitealias{Brown2015}{B15}
        \defcitealias{Catinella2013}{C13}
        \defcitealias{Fabello2011}{F11}

\begin{figure}
			\centering
			\includegraphics[width=\textwidth]{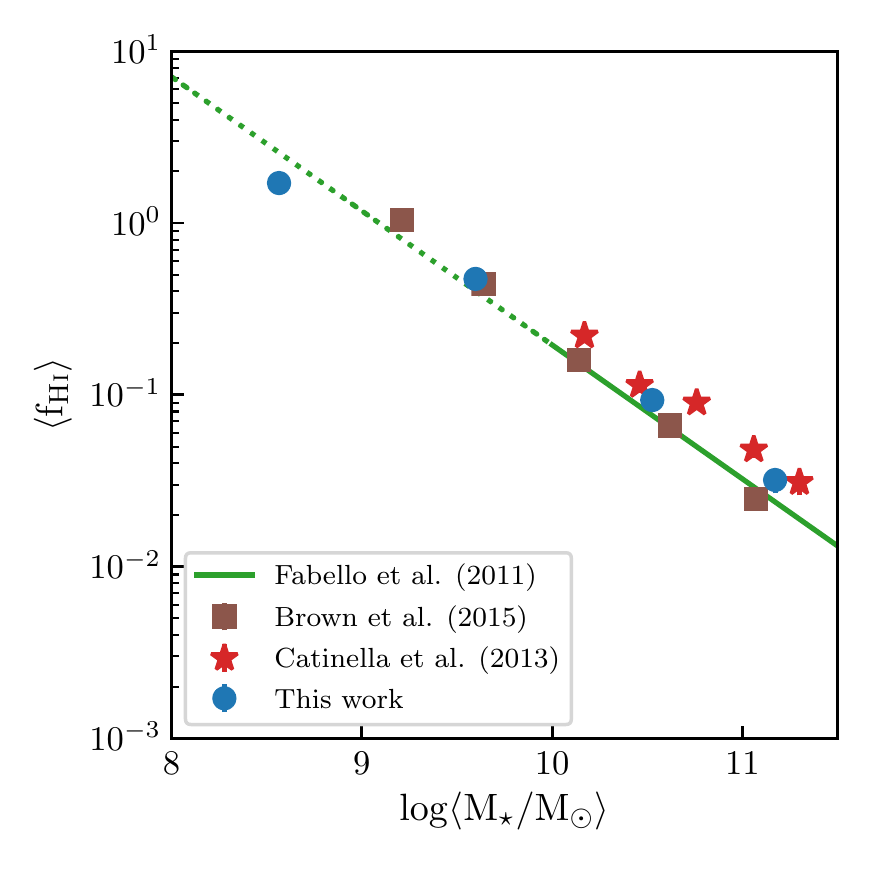}
			\caption{\hi gas fraction \ave{\fhi} as a function of \mstar for the full NIBLES stacking sample, including both detections and non-detections. Also shown, are \ave{\fhi} values from \citet{Brown2015} who stacked sub-samples of SDSS-selected galaxies with \hi data from ALFALFA, and \citet{Catinella2013} who calculated weighted averages for sub-samples of GASS galaxies. The green line is taken from \citet{Fabello2011} who fitted the slope of the gas scaling relation for a sub-sample of ALFALFA galaxies; the solid line represents the stellar mass range of their sample while the dotted part is a linear extrapolation. The errorbars on the NIBLES data points are the statistical uncertainties. The plotted values are given in \autoref{tab:avefhifig1112}, and the stacked profiles are shown in \autoref{fig:fig9profiles}.}
			\label{fig:fhismbins}
			\label{FIG:FHISMBINS}
		\end{figure}

Gas scaling relations have been used in conjunction with stacking (e.g., \citealt{Fabello2011}, hereafter F11; \citealt{Brown2015}, hereafter B15) to study various galaxy properties (star formation, stellar mass, etc.) and their influence on the \hi content of the galaxies. In \autoref{fig:fhismbins} we compare NIBLES gas fractions calculated using \stacker to gas fractions from \citet[hereafter C13]{Catinella2013}, \citetalias{Fabello2011} and \citetalias{Brown2015}. Adopting the method used by B15, we separated the full NIBLES sample (detections plus non-detections) into bins of $\log \mstar$ with widths of \numunit{1}{\text{dex}}. The sample stellar mass distribution is presented in \autoref{fig:smbinsfc}. For the NIBLES sample in \autoref{fig:fhismbins} (indicated by the round blue circles), the \mstar values represent the mean value in each bin. The NIBLES stacked profiles are presented in \autoref{fig:fig9profiles} and the measured quantities are listed in \autoref{tab:avefhifig1112}.\\

The \citetalias{Catinella2013} sample was selected from the GALEX Arecibo SDSS Survey \citep[GASS,][]{Catinella2009, Catinella2013}, a survey with the Arecibo telescope which targeted $\sim$~1000 massive galaxies randomly selected from the overlap between the SDSS DR6 spectroscopic survey and GALEX Medium Imaging Survey. The \citetalias{Fabello2011} and \citetalias{Brown2015} samples are selected from the overlap between ALFALFA, SDSS and GALEX. The \citetalias{Catinella2013} results show weighted average values (the \hi non-detections are set to upper limits) while the \citetalias{Fabello2011} and \citetalias{Brown2015} results are obtained from \hi stacking. To take into account the known flux offset between ALFALFA \hi spectra and NIBLES (see \autoref{sec:compsurvey}), the \citetalias{Brown2015} and \citetalias{Fabello2011} gas fractions have been scaled by the mean \mbox{$\alpha.40$/NRT flux ratio ($1.45$)}. The solid green line is the fit from \citetalias{Fabello2011} to their gas scaling relation and the dotted green line represents a linear extrapolation of this trend to lower stellar mass. \\

Due to the NIBLES sample selection, we are able to go down an order of magnitude lower in stellar mass than previous results by \citetalias{Brown2015}. For \numunit{\mstar > 10^9}{\msol} the NIBLES gas fractions agree well with the results by \citetalias{Catinella2013} and the adjusted gas fractions from \citetalias{Brown2015} and \citetalias{Fabello2011}. Below \numunit{\mstar < 10^9}{\msol}, the NIBLES \ave{\fhi} is lower than the trend set by the higher stellar mass galaxies. \\

\citet{Huang2012} and \citet{Maddox2015} have previously noted the change in slope of the \mhi vs \mstar relation between \numunit{10^8 < \mstar < 10^9}{\msol} in the \hi-selected population of ALFALFA galaxies where the relation is steeper at lower mass and flatter towards the high mass end. A change in slope is also seen in the ALFALFA \fhi-\mstar scaling relation by \citet{Huang2012} around a similar \mstar where the slope is flatter at lower \mstar and drops more steeply at higher \mstar. \citet{Huang2012} note that their average gas fraction relation has an overall offset to higher values as expected compared to the optically selected samples by \citet{Catinella2009} and \citet{Cortese2011} due to the ALFALFA sample bias towards more gas rich galaxies per stellar mass bin. Across the mass range investigated, per bin in stellar mass, they also observe a sharp cut-off in number density at the higher end of the gas fraction distribution, but a large dispersion towards lower gas fractions and note that for surveys with longer integration times, a lower average \fhi per stellar mass bin would be observed. A similar change in slope towards lower \mstar would also therefore be expected for stellar mass selected samples as for the ALFALFA sample. The divergence of the lowest stellar mass NIBLES gas fraction data point from the higher stellar mass trends from the extrapolated \citetalias{Fabello2011} fit is therefore consistent with expectations based on the \hi-selected ALFALFA sample \citet{Huang2012} gas fraction vs \mstar relation. \\

\begin{table*}
	\centering
    \caption{Average \hi gas fractions \ave{\fhi} for the scaling relation in Figure~\ref{fig:fhismbins}. The average \mhi/\mstar ratio \ave{\fhi} values for each data point are given along with the statistical uncertainties. The total number of galaxies per bin are given in the column titled N, with the number of non-detections per bin indicated by the number in brackets. The second-last column gives the Gaussian significance of the stacked profile obtained from \stacker (see \autoref{sec:profilestats} for details). The final column gives the S/N of the stacked spectrum calculated by \stacker using \autoref{eqn:snralf}.}
    \label{tab:avefhifig1112}
    \makegapedcells
	\begin{tabular}{ccccc}
    	\hline
        $\log \ave{\mstar/\msol}$  & \ave{\fhi} & N  & Significance & S/N$_\text{ALFALFA}$\\
     	 \hline
         8.56 & $ 1.709 \pm 0.089 $ & 347 (109) & $>8.2\sigma$ & 174.7 \\
         9.60 & $ 0.473 \pm 0.024 $ & 335 (96) & $>8.2\sigma$ & 231.8 \\
         10.53 & $ 0.093 \pm 0.005 $ & 272 (93) & $>8.2\sigma$ & 216.6 \\
         11.17 & $ 0.032 \pm 0.005 $ & 46 (25) & $>8.2\sigma$ & 49.3 \\
	\end{tabular}
\end{table*}

\section{Summary}

	This work has detailed the development of a new \numunit{21}{\text{cm}} \hi spectral stacking package in Python (\stackerfull: \stacker); while we have exclusively used \hi emission spectra to test \stacker, it is capable of stacking emission or absorption spectra of any characteristic line. \stacker takes the spectra for a sample of galaxies along with the accompanying redshifts recorded in the galaxy catalogue to produce an average spectrum from which a number of average properties (e.g., \hi mass \ave{\mhi}, \hi-to-stellar mass ratio \ave{\fhi}) for the sample may be extracted. \stacker also offers the user a choice of two built-in error analysis methods, and the ability to characterise the shape of the stacked spectrum through fitting a variety of functions.  \\ 

	We have applied \stacker to stacking of 1000 \hi spectra from the \nancay Interstellar Baryon Legacy Extragalactic Survey (NIBLES), which is a stellar mass-selected (with no colour-selection) \hi survey of SDSS galaxies in the nearby Universe. Due to the wide stellar mass range spanned by the NIBLES dataset ($10^8 < \mstar \,(\msol) < 10^{12}$), extending an order of magnitude lower than \citet{Brown2015}, we were able to extend the previously studied gas fraction vs. stellar mass gas scaling relation \citep{Brown2015, Fabello2011, Catinella2013} down to lower average stellar mass (\numunit{\ave{\mstar} = 10^{8.6}}{\msol}). With our stellar mass selected sample, we find good agreement with previous results at high stellar mass. At low stellar mass (\numunit{\mstar < 10^{9}}{\msol}) we observe a deviation from the extrapolated high mass relation which indicates a flattening of the slope in the scaling relation at low stellar mass qualitatively consistent with the trend seen in the \hi-blind, gas-rich ALFALFA sample \citep{Huang2012}. 

\section*{Acknowledgements}
JH acknowledges the bursary provided by South African Radio Astronomy Observatory; SLB, EE, and JH were supported by the South African National Research Foundation. We thank Barbara Catinella and Kelley Hess for their input and advice on the design of \stacker. JH wishes to thank S. Makhathini and L. Lindroos for helpful conversations during a trip to OSO, supported by MIDPREP. The \nancay Radio Astronomy Facility is operated as part of the Paris Observatory, in association with the Centre National de la Recherche Scientifique (CNRS) and partially supported by the R\'{e}gion Centre in France. This research has made use of the NASA/IPAC Database (NED) which is operated by the Jet Propulsion Laboratory, California Institute of Technology, under contract with the Nation Aeronautics and Space Administration, and data from Sloan Digital Sky Survey (SDSS-III). Funding for SDSS-III has been provided by the Alfred P. Sloan Foundation, the Participating Institutions, the National Science Foundation, and the U.S. Department of Energy Office of Science. The SDSS-III web site is http://www.sdss3.org/. SDSS-III is managed by the Astrophysical Research Consortium for the Participating Institutions of the SDSS-III Collaboration.


\bibliography{library} 


\appendix
\onecolumn
\section{Details of \stacker}
\subsection{Running \stacker} \label{sec:runhiss}

\stacker can be downloaded from \url{https://github.com/healytwin1/HISS}. Below is a summary of the README file which details the software requirements as well as how to install and run \stacker from the command line. \\

An example of \stacker run-time: to stack the 1000 NIBLES spectra with 1000 jackknife iterations on a standard laptop with 8GB of RAM, took $\sim 4-5$ hours.\\

FIRST TIME USERS: It is recommended to use the graphical interface to populate a config file. The graphical interface is called by the "hiss" executable.
To use \stacker from the command-line, enter the following command: \\ 

\texttt{python pipeline.py [-h] [-f <filepath+filename>] [-d] [-p] [-s] [-c] [-l]}\\

There are a number of different options that can be used to run this package: \\
\begin{minipage}{\textwidth}
\begin{verbatim}
optional arguments:
-h, --help            show this help message and exit.
-d, --display         Option to display progress window during the stacking process.
-p, --saveprogress    Option to save progress window during the stacking process.
-s, --suppress        Use this flag to suppress all output windows.
                      Note that [suppress] and [progress] cannot be used simultaneously.	
-c, --clean           Use [clean] for testing purposes and for stacking noiseless spectra as this option will 
                      bypass any noise-related functions and actions.	
-l, --latex           This option enables to the use of latex formatting in the plots. 
\end{verbatim}
\end{minipage}

\subsection{\stacker Configuration File}
\label{sec:configfile}
\onecolumn
\definecolor{light-gray}{gray}{0.85}
Below is the configuration file that is used by \stacker. This file is filled by the interface shown in \autoref{fig:guimainwindow}.
\setlength{\extrarowheight}{6pt}
{\footnotesize{
\begin{longtable}{l@{\hspace{2ex}}p{0.3\linewidth}p{0.45\linewidth}}

\hline
Config Parameter & GUI Parameter/Section & Description \\ \hline
\endfirsthead

\endhead

  \hline 
  \endfoot

  \hline \hline
  \endlastfoot
CatalogueFilename & Catalogue File & Path to catalogue file -- this should be the full path \\\arrayrulecolor{light-gray}\hline
CatalogueColumnNumbers & Catalogue Information & Column numbers for [Object ID, Filename, Redshift, Redshift Uncertainty, Stellar Mass, Other Data]. ``Other Data'' refers to data used to bin the sample - if one of the columns is not needed leave the column entry as an empty string. \\ \arrayrulecolor{light-gray}\hline
z\_max & \multirow{2}{*}{Redshift range} & Maximum redshift value of sample.\\ 
z\_min & & Minimum redshift value of sample. \\ \arrayrulecolor{light-gray}\hline
SpectrumFluxUnit &  \multirow{4}{*}{Input Spectra block} & Specify flux density units of input spectra. Values to use in config file: 1-Jy, 2-mJy, 3-uJy, 4-Jy/beam, 5-mJy/beam, 6-$\mu$Jy/beam \\
SpectralAxisUnit & & Values to use in config file: 1-Hz, 2-kHz, 3-MHz, 4-m/s, 5-km/s \\
VelocityType & & The choices here are optical/radio. The default is optical, and should be changed with care. This option will only appear in the GUI if a velocity spectral type is chosen. \\
ChannelWidth & & This should be entered in the same units as the spectral axis. This parameter is used as a diagnostic to flag spectra which have vastly different channel widths than the rest of the sample. \\
SpectrumLocation & Spectra location & Full path to location of spectra. \\\arrayrulecolor{light-gray}\hline
FirstRowInSpectrum & Input spectra file info & First row of data in spectrum file (numbering starts at 1). This option allows for header information in the spectrum file to be skipped. \\\arrayrulecolor{light-gray}\hline
RowDelimiter & Data delimiter & Spectrum file data delimiter - if the delimiter is whitespace, use "none". \\\arrayrulecolor{light-gray}\hline
SpectrumColumns & Spectrum Data Columns (Axis/Flux Density) & Column numbers for [freq/vel, flux density]. \\\arrayrulecolor{light-gray}\hline
StackFluxUnit & Stacked Spectrum Information & Choose unit in which to stack. Choices:1-Jy, 2-\msol, 3-\mhi/\mstar\\\arrayrulecolor{light-gray}\hline
H0 & \multirow{2}{*}{Cosmology} & Choice of H$_0$ -- default is \numunit{70}{\text{km}\cdot\text{s}^{-1}\cdot\text{Mpc}^{-1}} \\
Om0 & &  Choice of $\Omega_{m,0}$ -- default is 0.3 \\\arrayrulecolor{light-gray}\hline
OutputLocation & Output location & Full path to where results should be saved. \\\arrayrulecolor{light-gray}\hline
GalaxyWidth & Max galaxy velocity width of sample & This is the maximum velocity width in the sample. This value is used to mask the spectrum where emission is expected as well as uses to define the integration window. The value can be changed during the analysis process. \\\arrayrulecolor{light-gray}\hline
StackedSpectrumLength & Velocity width of spectrum & This defines the length of the spectrum. There is a hard limit that this value cannot be less than 3 times the maximum galaxy velocity width. Default is \numunit{2000}{\kms}.\\\arrayrulecolor{light-gray}\hline
WeightOption & Select Weighting Function & 1-$\text{w}=1$[default], 2-$\text{w}=\frac{1}{\text{rms}}$, 3-$\text{w}=\frac{1}{\text{rms}^2}$, 4-$\text{w}= \frac{1}{(D_L^2 \text{rms})^2}$\\\arrayrulecolor{light-gray}\hline
NumberCatalogueObjects & Stack entire catalogue & If `y' is chosen, need to entire the number of catalogue objects to stack. This will be set to the length of the catalogue if the chosen number is larger. \\\arrayrulecolor{light-gray}\hline
UncertaintyYN & Calculate Uncertainties & This is a [y]es/[n]o option. \\\arrayrulecolor{light-gray}\hline
UncertaintyMethod & Uncertainty method & redshift -- uses redshift uncertainties to determine uncertainty of stacked profile. dagjk -- Delete-a-group-jackknife, determines the statistical uncertainty of stacked profile. \\\arrayrulecolor{light-gray}\hline
BinYN & Stack in different bins? & This is a [y]es/[n]o option. \\\arrayrulecolor{light-gray}\hline
BinInfo & Information on bins & Column name of data to bin, Bin width, Start of Bin range, End of Bin range \\
\end{longtable}
}}

\subsection{Plots created by \stacker}
\label{sec:hissplots}
\begin{landscape}
\begin{figure}
	\centering
	\includegraphics[width=24cm]{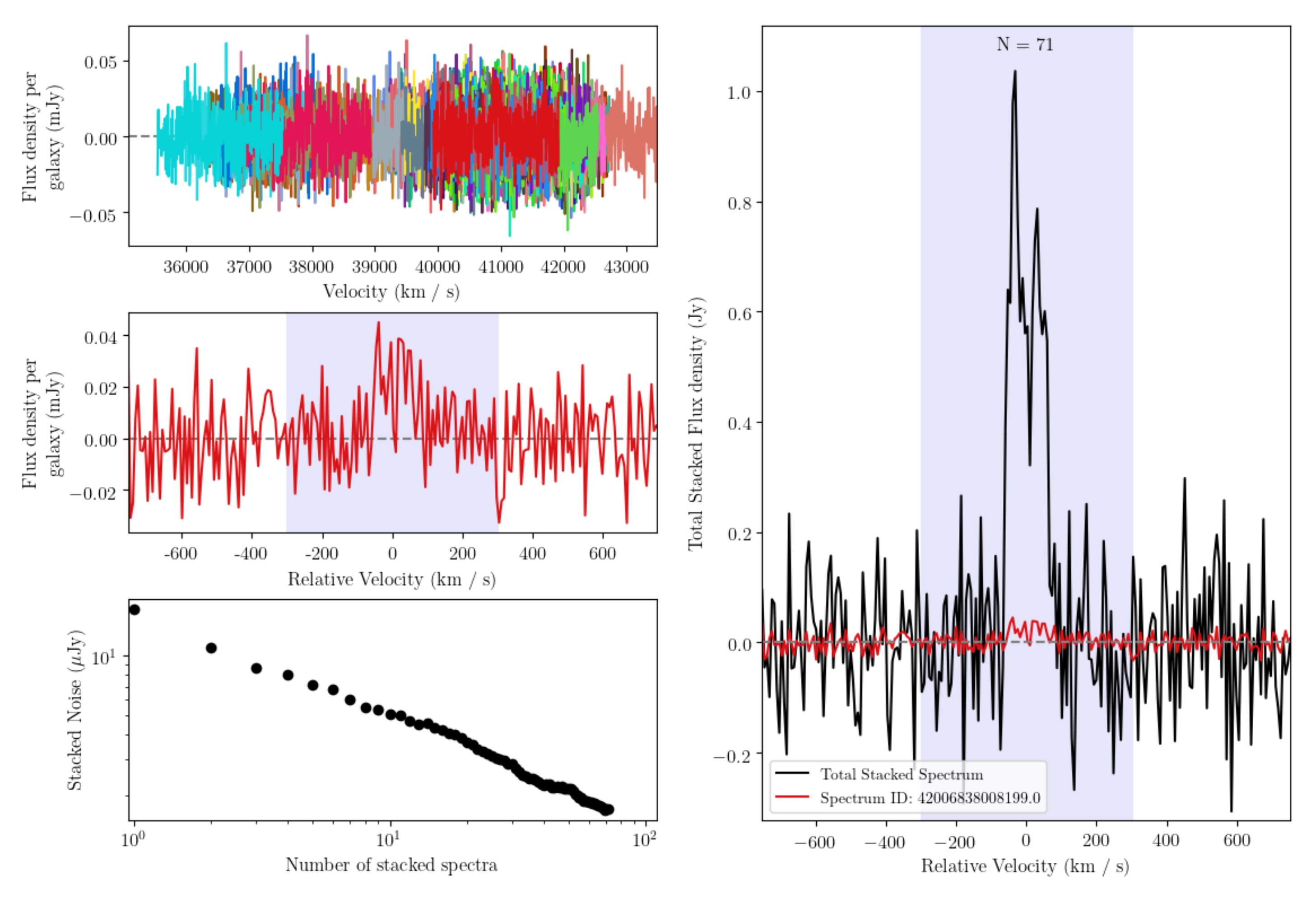}
	\caption{The top left panel shows where the input spectra lie in the observed frame. The middle left panel shows the now aligned spectrum in the velocity and flux density units in which it will be stacked. The bottom left panel shows how the average noise changes with the addition of each new spectrum. The large panel on the right shows the total stacked spectrum, in black, together with the incoming spectrum in blue.}
	\label{fig:progressplot}
\end{figure}
\end{landscape}

\begin{figure*}
	\centering
	\includegraphics[width=\textwidth]{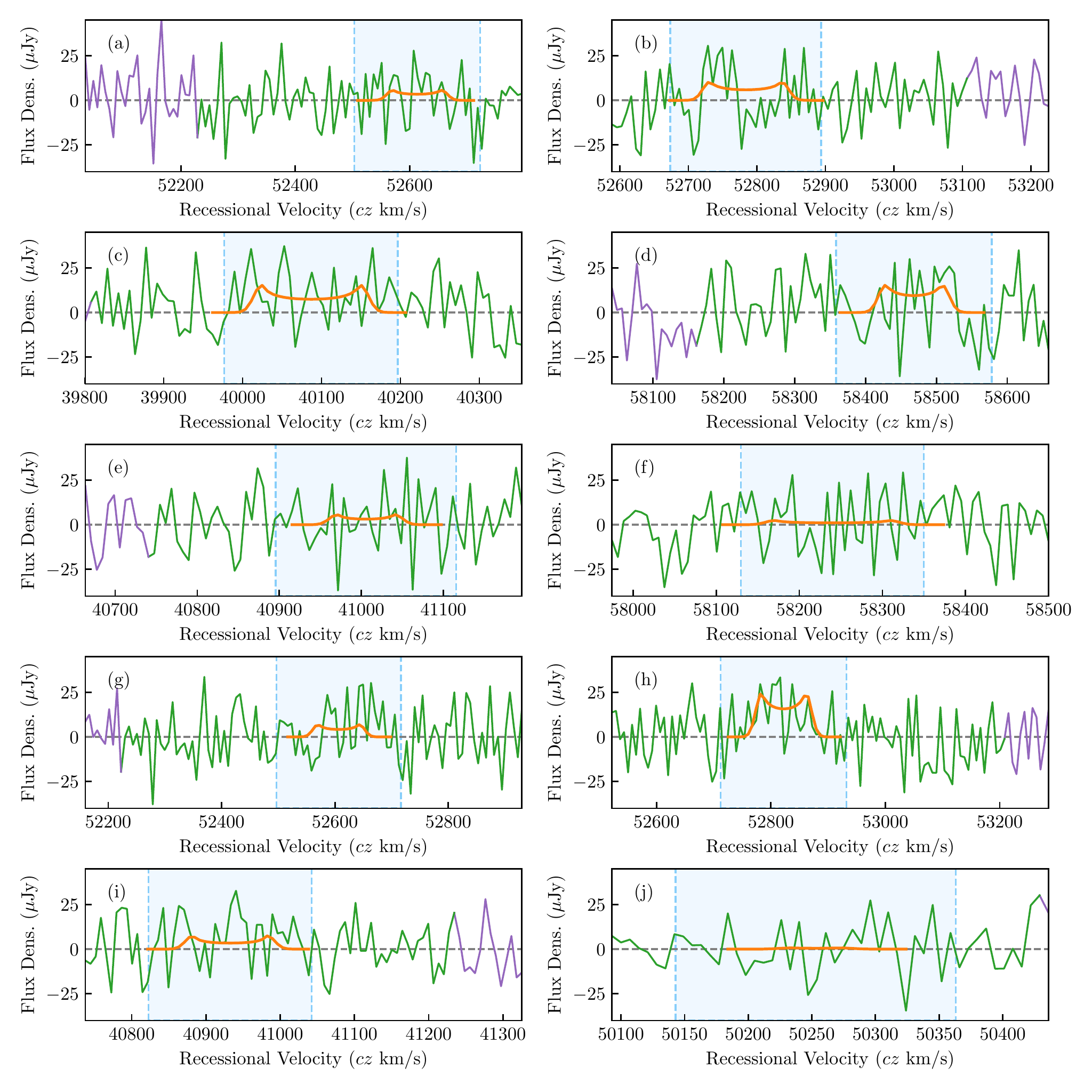}
	\caption{These ten spectra are a selection of the 1000 simulated spectra used in the stacking example throughout the paper. The spectra are plotted using their native flux density and spectral axes. The simulated, noiseless spectra are shown in orange. Shown in green and purple are the spectra after adding simulated Gaussian noise. The green section of the spectrum shows the part that will be shifted to the centre of the array while the purple section is the part that will be wrapped around and appended to the other end of the spectrum in the shifting process. The light blue region shows the section of the profile where the galaxy emission is expected to be located. The width of this region is defined by the user and forms part of the galaxy mask. Panel (j) is an example of a spectrum that will be excluded during the stacking process as there are not enough channels outside of the galaxy window.}
	\label{fig:inputspectra}
\end{figure*}

\begin{figure*}
	\centering
	\includegraphics[width=\textwidth]{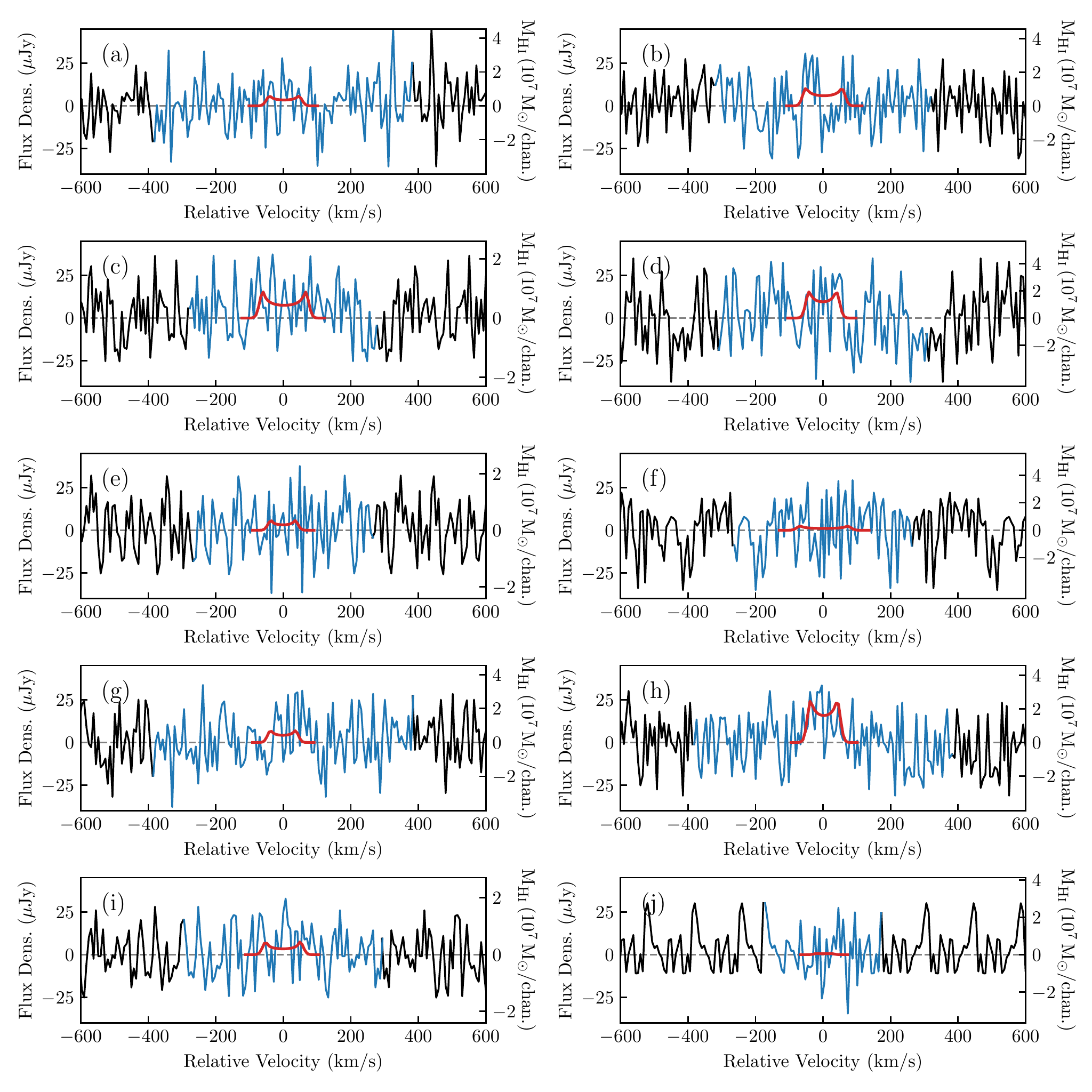}
	\caption[10 aligned simulated spectra.]{These are the same ten simulated spectra as those plotted in \autoref{fig:inputspectra}. Shown here in blue are the spectra that have been converted to their rest frame and centred. The original noise-free versions of each of the spectra are now shown in red. Shown in black are the parts of the extended spectrum that have been filled using the flux from outside the galaxy window. The right y-axis shows that each of these spectra have also been converted to mass spectra.}
	\label{fig:extendedspectra}
\end{figure*}

\begin{figure*}
    \centering
    \includegraphics[width=\textwidth]{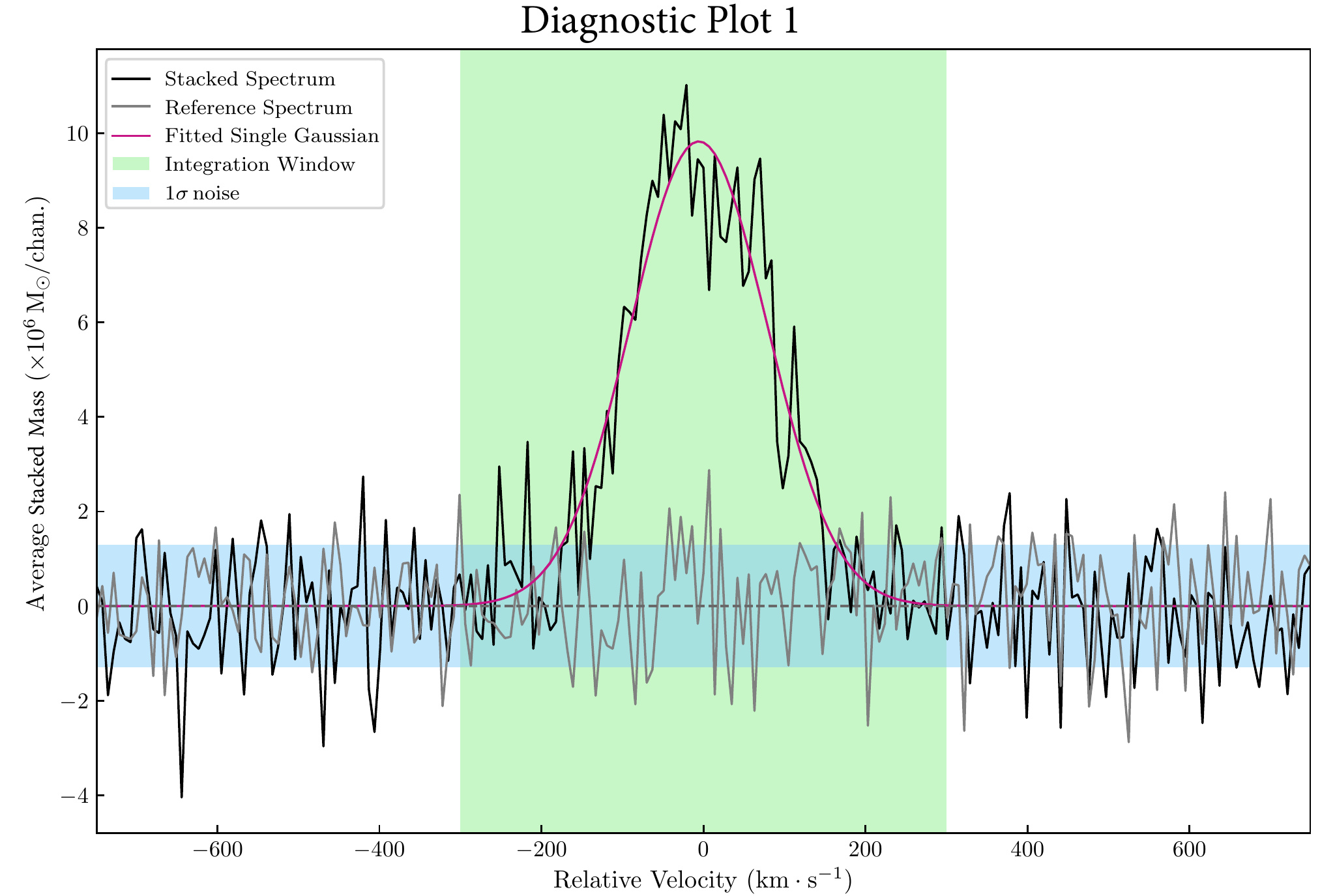}
    \caption[Diagnostic Plot 1]{Diagnostic Plot 1: this is the first plot of the stacked spectrum displayed to the user. The stacked spectrum is shown in black in the units the user chose as the stacking units. This plot is shown to the user before any spectrum manipulation routines have been applied to the spectrum. The magenta curve shows the Single Gaussian fit to the spectrum to determine the significance of the detection. The grey spectrum is the reference spectrum. The green vertical window indicates the part of the spectrum over which the average quantity will be measured. The $1\sigma$ noise level is shown by the horizontal blue band.}
    \label{fig:dp1}
\end{figure*}

\begin{landscape}
\begin{figure}
    \centering
    \includegraphics[width=0.9\textwidth]{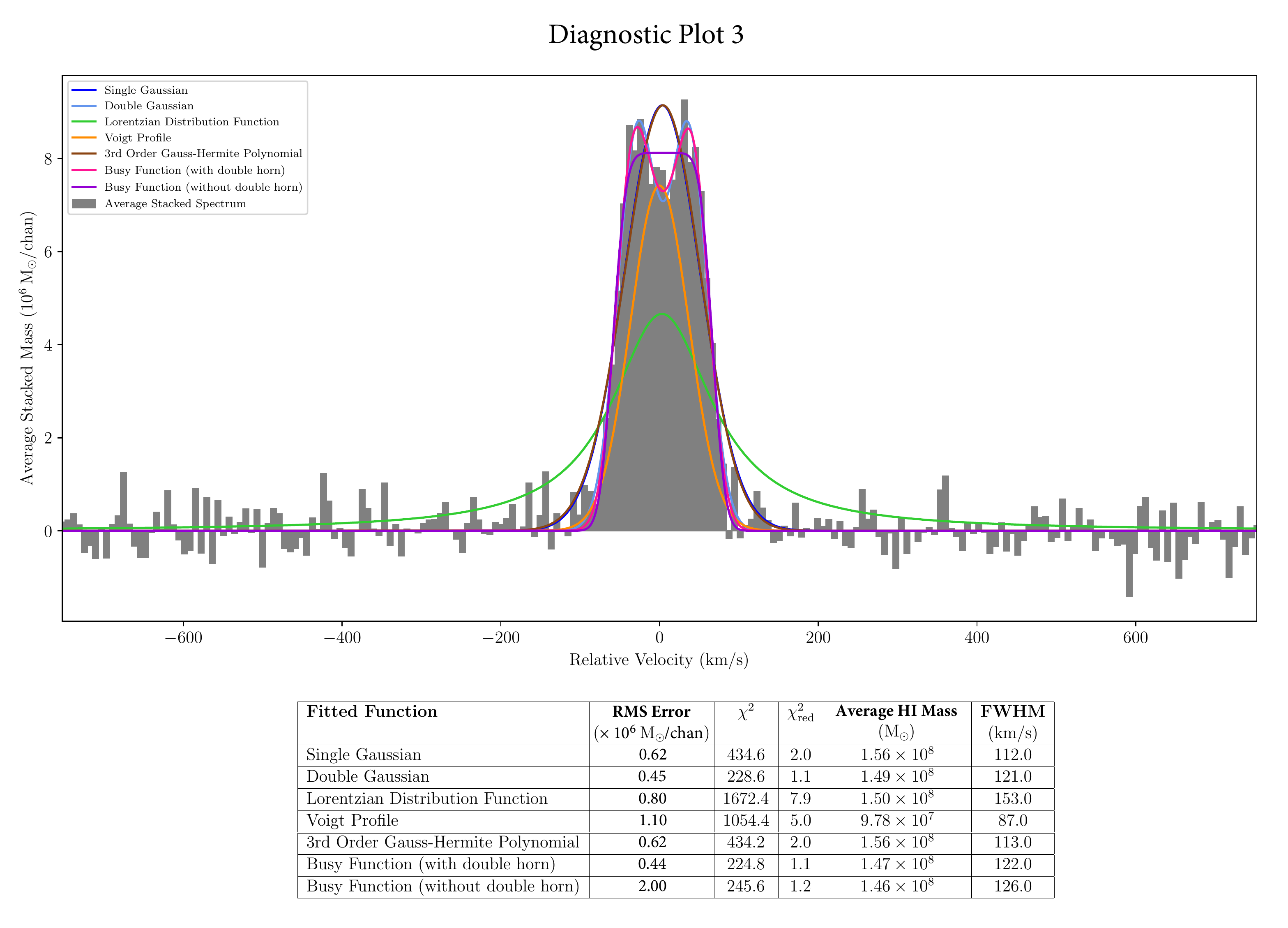}
    \caption[Diagnostic Plot 3.]{Diagnostic Plot 3: the top panel shows the seven different functions (in various colours) fitted to the stacked spectrum which is represented by the grey bars. The table in the bottom panel shows the measured average \hi mass as well as three measures of the goodness of fit -- here we use the standard $\chi^2$ goodness of fit parameter and the $\chi^2_\mathrm{red} = \chi^2/\mathrm{dof}$, where dof is the number of degrees of freedom, and the last measure is the RMS Error. This plot enables the user to inspect the quality of the fit to the data before committing to a selection of functions.}
    \label{fig:functionfit}
\end{figure}
\end{landscape}

\begin{landscape}
\begin{figure}
	\centering
	\includegraphics[width=0.98\textwidth]{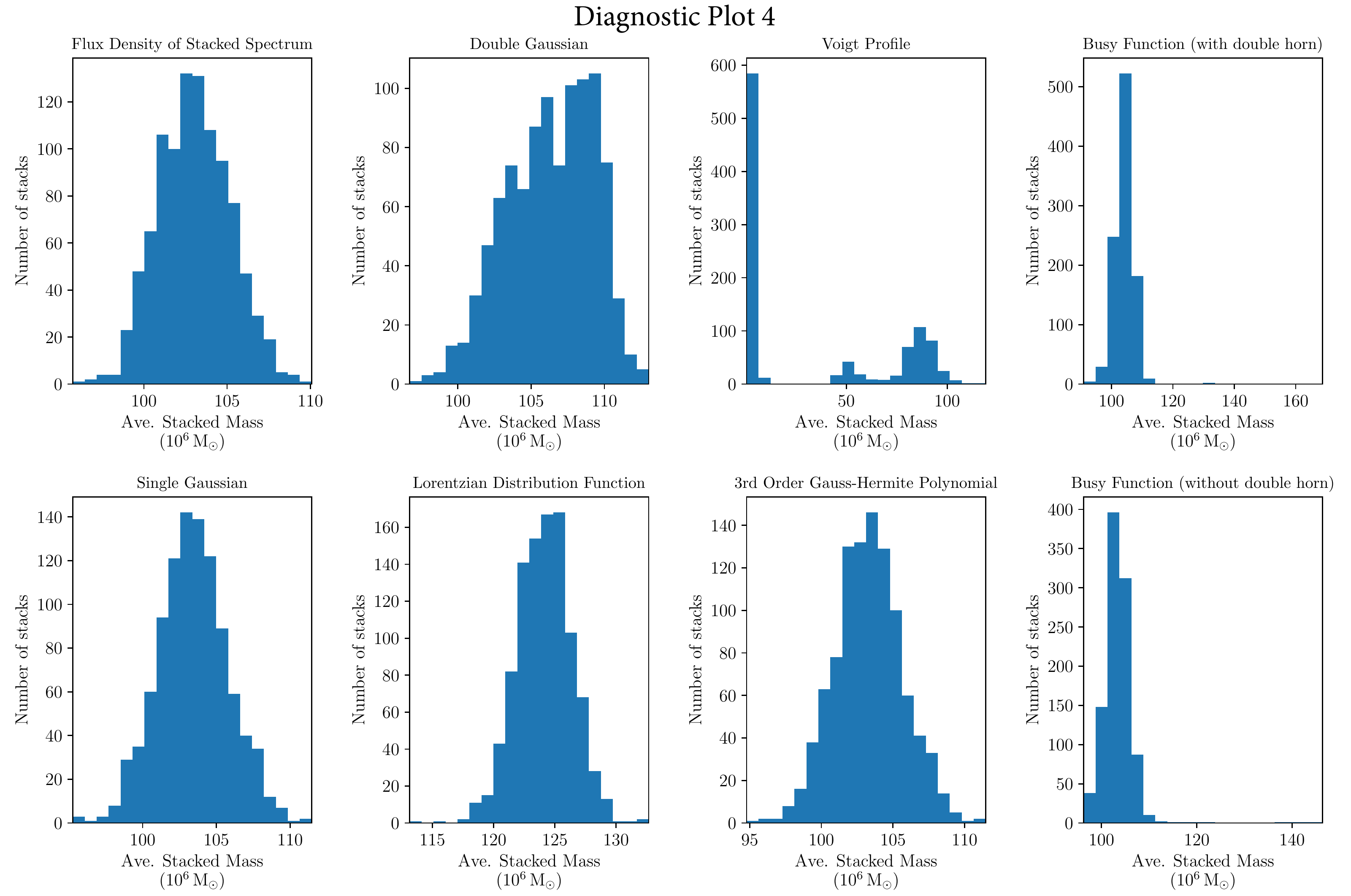}
	\caption{Diagnostic Plot 4: each panel shows a histogram of the average stacked mass measured from each profile fit for each of the 1000 stacking iterations (i.e. stacks). This plot provides a picture of how well each function was fit. It is from these histograms that the best estimate of the average quantity and associated error are calculated.}
	\label{fig:dp4}
\end{figure}
\end{landscape}

\begin{landscape}
\begin{figure}
	\centering
	\includegraphics[width=0.95\textwidth]{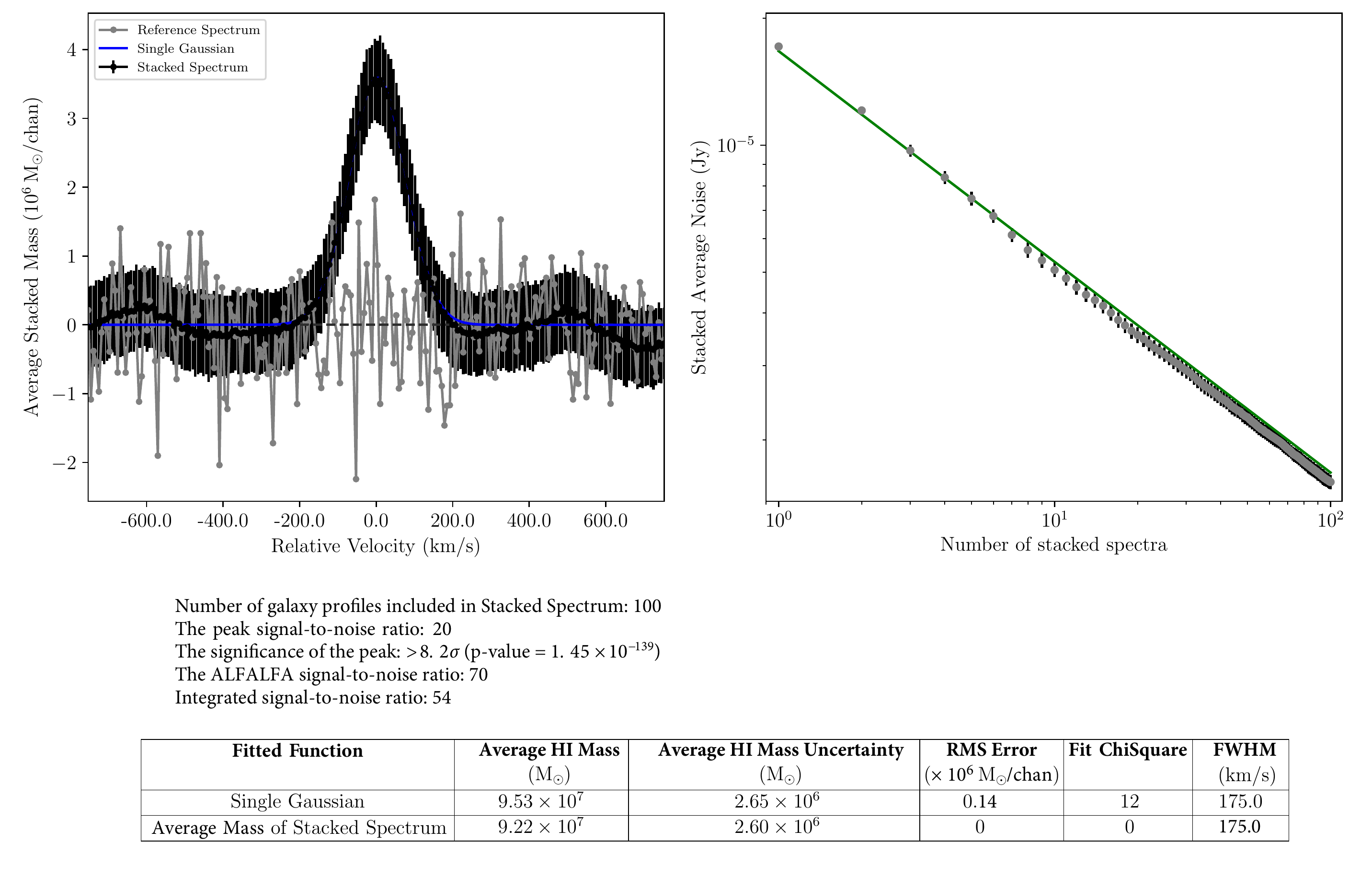}
	\caption{Screen shot of the output in full display mode. The top left panel shows the final average stacked spectrum with errorbars in black, and the chosen fitted Single Gaussian overlaid in blue. The top right panel shows the stacked noise as a function of the number of profiles in grey; overlaid in green is the theoretical response to stacked noise: $\sigma/\sqrt{N}$. The bottom half of the figure contains the table of measured quantities (average \hi mass and associated uncertainty), as well as the $\chi^2$ and RMS Error calculated from the function fit to the stacked spectrum. The second entry in the table (Average Mass of Stacked Spectrum) shows the quantities measured directly from the stacked spectrum over the range specified by the user. The S/N statistics, as discussed in \autoref{sec:profilestats}, for the stacked profile are given in the text above the table. }
	\label{fig:fulldisplay}
\end{figure}

\end{landscape}
\onecolumn
\section{Accounting for known nearby contamination}
\label{sec:contamination}
The \nancay telescope has a large beam which extends between $22' -  33'$ in the N-S direction depending on the declination of the source. The change in the N-S beam width can be approximated by the polynomial
		\begin{equation}\label{eqn:beamsize}
			y = (4.02 \times 10^{-9})\,x^5 - (6.90 \times 10^{-8})\,x^4 + (3.63 \times 10^{-6})\,x^2 + (5.45 \times 10^{-3})\,x + 21.95
		\end{equation}
		where $y$ is the beam width in arcminutes and $x$ is the declination in degrees \citep[Fig. 1.]{Matthews2000}.\\
        
In this work, galaxies near the target source are considered as possible contaminators if
			\begin{itemize}
	        \item the secondary galaxy is located spatially within $1.2 \times$ N-S half-power beam width and $2 \times$ E-W half-power beam width, and
				\item the redshift of the secondary galaxy is within \numunit{\pm 300}{\kms} of the target redshift.
			\end{itemize}
            
\begin{landscape}
\begin{figure}
	\centering
    \vspace{-0.5cm}
	\includegraphics[width=20cm]{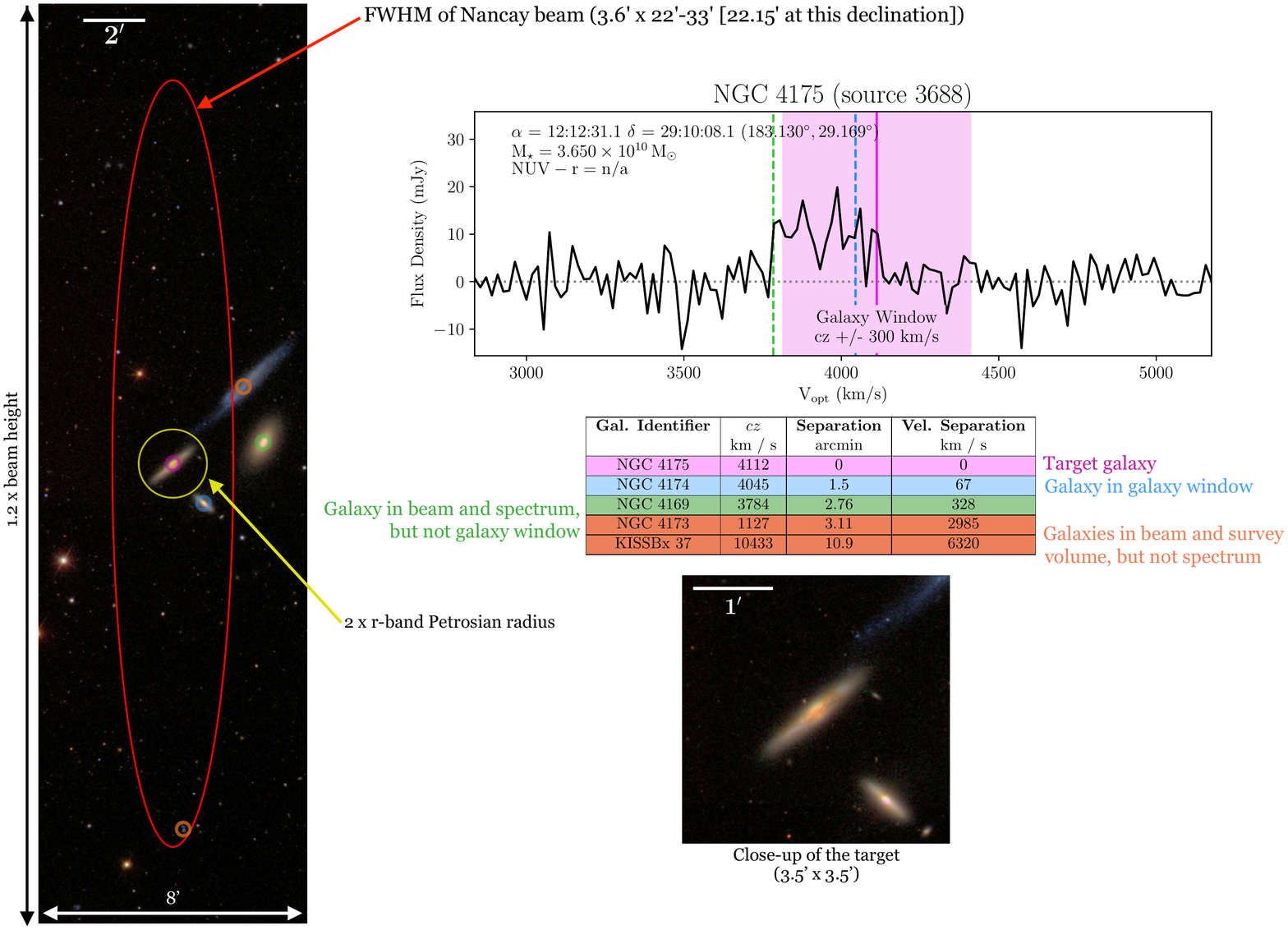}
 	\caption{Illustration of the method used to find any secondary contaminating sources. On the left of the figure is an SDSS optical image that is $8'$ wide and $1.2\times\text{beam in declination}$ centred on the coordinates of the target galaxy. Overlaid in red is the half-power size of the beam; the yellow circle has a radius of $2r_\text{P}$ which indicates the estimated size of the galaxy. The magenta open circular marker indicates the target source while the blue, green, and orange open circular markers indicate secondary targets in the entire survey velocity range (\numunit{900 < cz < 12000}{\kms}). Each of the targets highlighted in the large optical image are listed in the table to the right. The galaxies that are in the survey volume, but whose recessional velocities are not in the spectrum velocity range, are indicated by the orange circles. The green circles highlight galaxies that are within the velocity range covered by the spectrum -- these sources are also marked on the spectrum by the green dashed vertical lines. The blue circled sources are the ones that are considered contamination sources. These sources have a redshift (highlighted by the dashed blue line on the spectrum) that is within \numunit{\pm 300}{\kms} of the target redshift (indicated by the magenta line on the spectrum). The small optical image below the table is a $3.5' \times 3.5'$ close-up centred on the target source.}
	\label{fig:findingchart}
\end{figure}
\end{landscape}

 A number of the target galaxies were found to have possible contaminators with very small angular separation distances. Upon inspection of the optical images for these sources, it was discovered that the possible contaminators were, in fact, part of the target galaxy. The SDSS pipeline has a known issue of de-blending large sources, particularly those with knots of star formation. In order not to over-classify the contaminant sources, a new criterion was therefore introduced to the search: the contaminant source must lie outside the edge of the target source, as defined as twice the $r$-band Petrosian radius ($r_\mathrm{P}$), which according to \citet{Stoughton2002} is a good estimate of the size of the galaxy. \autoref{fig:findingchart} is an illustration of the method used to find any secondary contaminating sources.

\section{Defining a sample of \hi non-detections}
	\label{sec:ndsample}
	\begin{figure*}[!h]
	    \centering
	    \includegraphics[width=\textwidth]{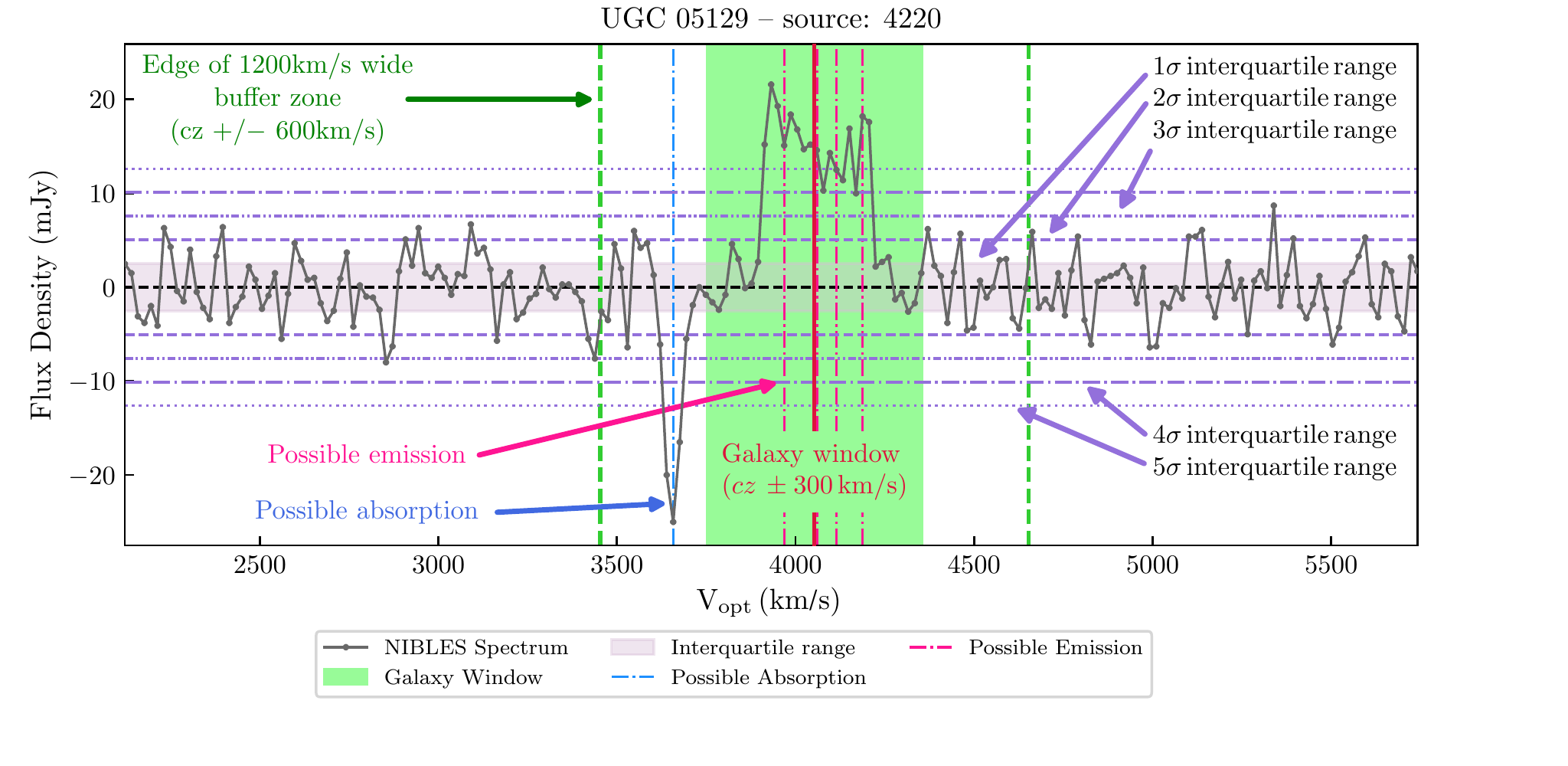}
	    \caption{Graphic example of how the detection algorithm flags different features in a spectrum. The spectrum used in this example is classified as a detection. The criteria we used for defining \hi detections is outlined in \autoref{sec:ndselect}. Using these criteria we reproduced with 96 percent overlap, the detection and non-detection classifications published in \wim. The marginal detections in \wim were classified in an approximately one-third/two-thirds ratio as detections vs. non-detections respectively.}
	    \label{fig:detectioncriteria}
	\end{figure*}
	
	\begin{figure}
	    \centering
	    \includegraphics[width=\textwidth]{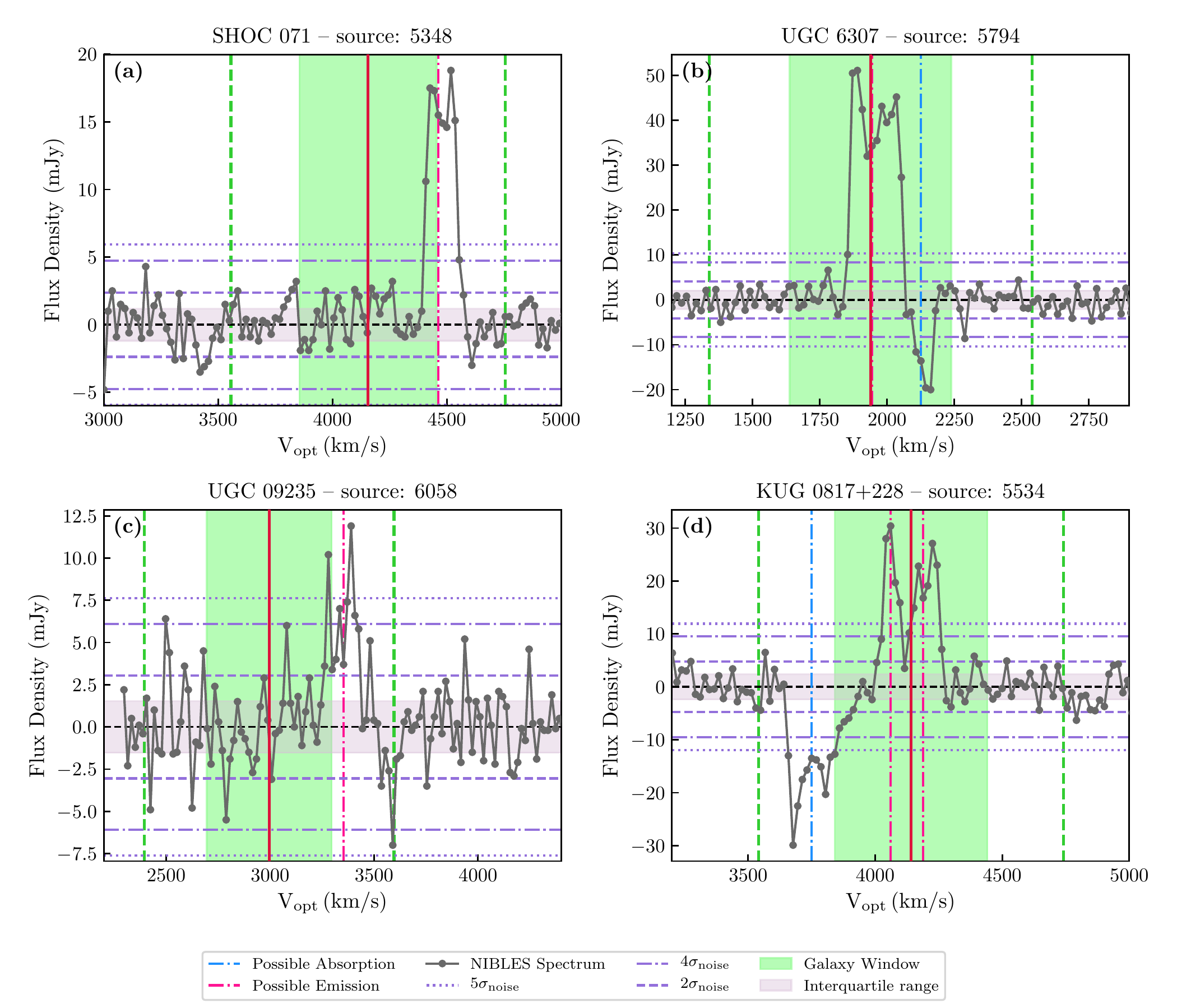}
	    \caption{Sources flagged by the second run of the detection algorithm as containing emission/absorption not related to the target galaxy. All four sources were removed from the sample. Panels (b) and (d) show evidence of an \hi detection in the off-source pointing. Panels (a) and (c) show clear \hi detections that are not associated with sources previously identified in our contaminant source search. The vertical red line in each panel indicates the optical redshift of the target source.}
	    \label{fig:sketchyspec}
	\end{figure}
	
	\begin{figure}
	    \centering
	    \includegraphics[width=\textwidth]{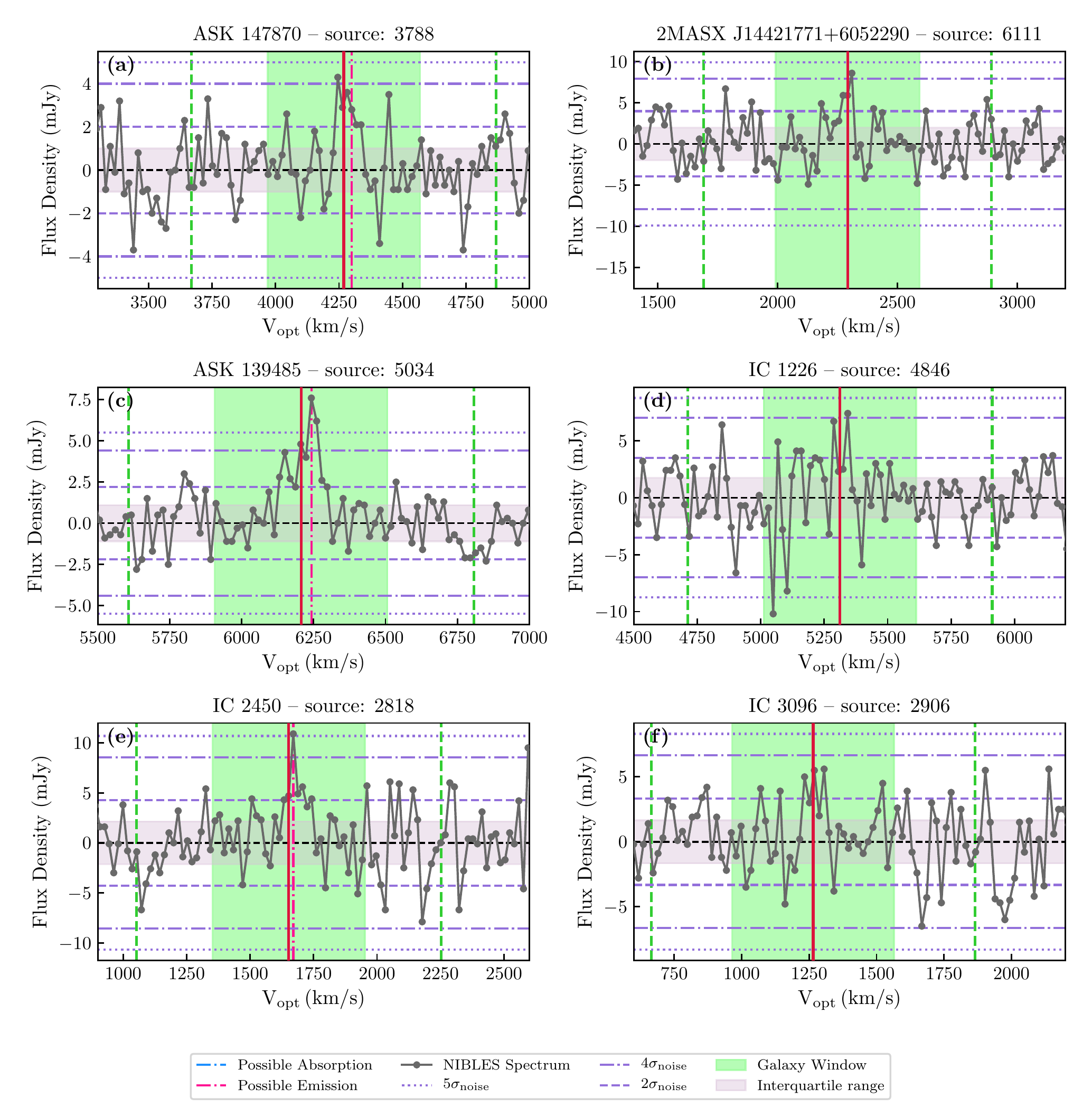}
	    \caption{Examples of NIBLES spectra classified as \hi detections (left column) and not classified as detections (right column) -- i.e. non-detections. Panel (a) Classified as a detection with $\geq 5$ consecutive resolution elements with flux density $>2 \times$ IQR. Panel (c) Classified as a detection with $\geq 1$ consecutive resolution elements with flux density $>5 \times$ IQR. Panel (e) Classified as a detection with 1 resolution element with flux density $>5 \times$ IQR. The vertical red line in each panel indicates the optical redshift of the target source. Panels (a), (c), (e) are also identified by \citet{VanDriel2016} as detections, however only panel (c) would be identified as a detection using the S/N criterion used by ALFALFA.}
	    \label{fig:nddetexampls}
	\end{figure}

\onecolumn

\section{Stacking NIBLES galaxies}
    \subsection{Quantities measured from the stacked spectra}

	\begin{table*}[!h]
    \centering
	\caption{Average quantities obtained from stacking the NIBLES galaxies in colour sub-samples and the full sample. The average \hi mass values are calculated from the stacked spectra shown in \protect\autoref{fig:ninepanel}. Gaussian significance values (i.e. $\sigma_{\mhi}/\sigma_{\fhi}$) of $>8.2\sigma$ have corresponding p-values that are too small to be numerically resolved by a computer, thus have been assigned the significance value of the smallest resolvable p-value.}
    \label{tab:ninesampleplot}
    \makegapedcells
	\begin{tabular}{|r|c|c|c|l|} \cline{2-4}
\multicolumn{1}{c|}{} & Full Sample & Blue Sample & Red Sample & \multicolumn{1}{c}{} \\ \hhline{-===-} 
                 & 1000 & 549 & 451 & N \\ \cline{2-4}
         & $>8.2/>8.2$ & $>8.2/>8.2$ & $>8.2/>8.2$ & $\sigma_{\mhi}$/$\sigma_{\fhi}$ \\ \cline{2-4}
         & $63.1/204.5 $ & $289.3/182.8 $  & $26.9/112.7 $  & S/N$_\mathrm{ALFALFA, \mhi}$/ S/N$_\mathrm{ALFALFA, \fhi}$ \\ \cline{2-4}
All & \resultmhi{4}{1.493929}{0.002390}{0.057224}{9} & \resultmhi{4}{1.471576}{0.002774}{0.071459}{9} & \resultmhi{4}{1.515453}{0.004225}{0.094565}{9} & \ave{\mhi} (\msol) \\ \cline{2-4}
        & $1.795 \times 10^{10}$ & $5.641 \times 10^{9}$ & $3.280 \times 10^{10}$ & \numunit{\ave{\mstar}}{(\msol)} \\ \cline{2-4}
        & \resultfhi{4}{0.777609}{0.002005}{0.035758} & \resultfhi{4}{1.209980}{0.003811}{0.058548} & \resultfhi{4}{0.253316}{0.001090}{0.032064} & \ave{\fhi} \\ \hhline{=====} 
        & 677 & 436 & 241 & N \\ \cline{2-4}
        & $>8.2/7$ & $>8.2/>8.2$ & $>8.2/>8.2$ & $\sigma_{\mhi}$/$\sigma_{\fhi}$ \\ \cline{2-4}
        & $61.2/339.4 $ & $194.9/242.6$ & $25.9/122.3 $  & S/N$_\mathrm{ALFALFA, \mhi}$/ S/N$_\mathrm{ALFALFA, \fhi}$ \\ \cline{2-4}
Detections & \resultmhi{4}{2.123094}{0.003028}{0.082458}{9} & \resultmhi{4}{1.592552}{0.002616}{0.086370}{9} & \resultmhi{4}{2.580637}{0.008046}{0.172777}{9} & \ave{\mhi} (\msol) \\ \cline{2-4}
        & $1.519 \times 10^{10}$ & $5.837 \times 10^{9}$ & $3.226 \times 10^{10}$ & \numunit{\ave{\mstar}}{(\msol)} \\ \cline{2-4}
        & \resultfhi{4}{1.080183}{0.002081}{0.048667} & \resultfhi{4}{1.322187}{0.003762}{0.073883} & \resultfhi{4}{0.340875}{0.001676}{0.040169} & \ave{\fhi} \\ \hhline{=====} 
          & 323 & 113 & 210 & N \\ \cline{2-4}
          & $>8.2/>8.2$ & $>8.2/>8.2$ & $>8.2/7.7$  &  $\sigma_{\mhi}/\sigma_{\fhi}$ \\ \cline{2-4}
          & $21.1/32.5 $ & $18.7/27.6 $ & $8.6/8.8 $ & S/N$_\mathrm{ALFALFA, \mhi}$/ S/N$_\mathrm{ALFALFA, \fhi}$ \\ \cline{2-4}
Non-detections & \resultmhi{4}{1.518298}{0.040888}{0.325137}{8} & \resultmhi{4}{2.488772}{0.058919}{0.467408}{8} & \resultmhi{4}{1.187917}{0.057730}{0.430991}{8} & \ave{\mhi} (\msol) \\ \cline{2-4}
        & $2.536 \times 10^{10}$ & $4.428 \times 10^{9}$ & $3.351 \times 10^{10}$ & \numunit{\ave{\mstar}}{(\msol)} \\ \cline{2-4}
        & \resultfhi{4}{0.140756}{0.003897}{0.027739} & \resultfhi{4}{0.377954}{0.009818}{0.071023} & \resultfhi{4}{0.034855}{0.001314}{0.007828} & \ave{\fhi} \\ \hline
	\end{tabular}
\end{table*}

\begin{table}
\centering
\caption{Average quantities and statistical errors obtained from \stacker for \protect\autoref{fig:2binsSM} for the low stellar mass sample \mbox{($10^8 < \mathrm{\mstar\,(\msol)} < 10^{9.5}$)}. The results with Gaussian significance values of $>8.2$ have corresponding p-values that are too small to be numerically resolved by a computer, thus have been assigned the significance value of the smallest resolvable p-value.}
\label{tab:fig9lowmass}
\makegapedcells
\begin{tabular}{|l|c|c|c|c|c|c|c|} 
\cline{2-8}
\multicolumn{1}{c|}{} & N & \numunit{\ave{\mhi}}{(\msol)} & $\sigma_{\mhi}$ & S/N$_\mathrm{ALFALFA, \mhi}$ & \ave{\fhi} & $\sigma_{\fhi}$  & S/N$_\mathrm{ALFALFA, \fhi}$ \\ \cline{2-8} \hline
Full Catalogue (all) & 521 & $7.46 \pm 0.33 \times 10^{8} $ & $>8.2$ & 277.7 & $1.372 \pm 0.069 $  &$ >8.2$ & 205.7 \\ \hline
Full Catalogue (blue) & 401 & $8.00 \pm 0.39 \times 10^{8} $ & $>8.2$ & 232.8 & $1.563 \pm 0.074 $ & $ >8.2$ & 190.5 \\ \hline
Full Catalogue (red) & 120 & $5.74 \pm 0.63 \times 10^{8} $ & $>8.2$ & 196.0 & $0.740 \pm 0.107 $ & $ >8.2$ & 127.3 \\ \hline
Non-detections (all) & 156 & $12.69 \pm 2.47 \times 10^{7} $ & $>8.2$ & 35.1 & $0.299 \pm 0.050 $ & $ >8.2$ & 33.7 \\ \hline
Non-detections (blue) & 97 & $18.25 \pm 3.95 \times 10^{7} $ & $>8.2$ & 29.5 & $0.437 \pm 0.077 $ & $ >8.2$ & 28.1 \\ \hline
Non-detections (red) & 59 & $4.09 \pm 1.08 \times 10^{7} $ & $6$ & 11.1 & $0.095 \pm 0.027 $ & $ 5.9$ & 7.5 \\ \hline
\end{tabular}
\end{table}

\begin{table}[!h]
\centering
\caption{Same as \autoref{tab:fig9lowmass} but for the high stellar mass sample \mbox{($10^{9.5} < \mathrm{\mstar\,(\msol)} < 10^{12}$)}.}
\label{tab:fig9highmass}
\makegapedcells
\begin{tabular}{|l|c|c|c|c|c|c|c|} 
\cline{2-8}
\multicolumn{1}{c|}{} & N & \numunit{\ave{\mhi}}{(\msol)} & $\sigma_{\mhi}$ & S/N$_\mathrm{ALFALFA, \mhi}$ & \ave{\fhi} & $\sigma_{\fhi}$  & S/N$_\mathrm{ALFALFA, \fhi}$\\ \cline{2-8} \hline
Full Catalogue (all) & 479 & $2.25 \pm 0.11 \times 10^{9} $ & $>8.2$ & 54.2 & $0.132 \pm 0.007 $ &$ >8.2$ & 172.6 \\ \hline
Full Catalogue (blue) & 148 & $3.26 \pm 0.22 \times 10^{9} $ & $>8.2$ & 171.6 & $0.256 \pm 0.014 $ &$ >8.2$ & 203.0 \\ \hline
Full Catalogue (red) & 331 & $1.87 \pm 0.13 \times 10^{9} $ & $>8.2$ & 33.3 & $0.077 \pm 0.005 $ &$ >8.2$ & 93.9 \\ \hline
Non-detections (all) & 174 & $1.89 \pm 0.60 \times 10^{8} $ & $>8.2$ & 16.6 & $0.013 \pm 0.003 $ &$ >8.2$ & 19.0 \\ \hline
Non-detections (blue) & 17 & $6.58 \pm 1.86 \times 10^{8} $ & $8$ & 12.1 & $0.057 \pm 0.015 $ &$ 4.6$ & 10.8 \\ \hline
Non-detections (red) & 157 & $1.45 \pm 0.54 \times 10^{8} $ & $7.9$ & 8.0 & $0.010 \pm 0.003 $ &$ >8.2$ & 16.0 \\ \hline
\end{tabular}
\end{table}

\subsection{Stacked Profiles for {\protect\autoref{FIG:2BINSSM}} and {\protect\autoref{FIG:FHISMBINS}}}
	\label{sec:fig8profiles}

\begin{figure*}
        \centering
        \includegraphics[width=\textwidth]{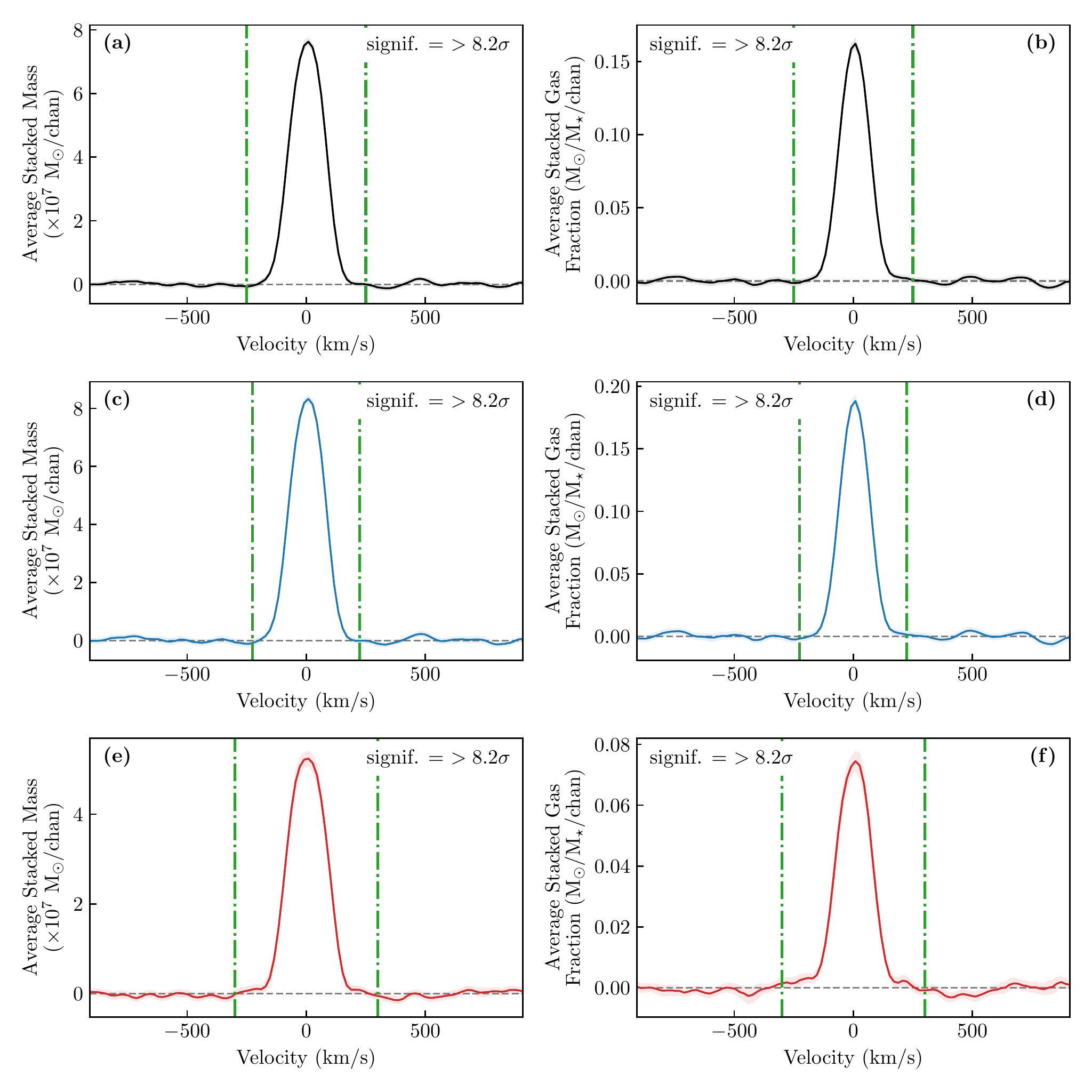}
        \caption{Stacked profiles for the full sample of \hi detections and non-detections with $10^{8} < \mstar (\msol) < 10^{9.5}$. The left column (panels a, c \& e) shows the stacked profiles from which the average \hi mass (\ave{\mhi}) is measured and the right column (panels b, d \& f) shows the stacked profiles from which the average gas fraction (\ave{\fhi}) is measured. The stacked profiles for all sources in this sample are shown in the top row. The middle row shows the red sub-sample, and the bottom row shows the blue sub-sample. The shading around each spectrum indicates the uncertainty in each channel calculated based on the redshift errors. The green dot-dashed lines in every panel mark the section of the stacked profiles used to measure the \ave{\mhi} and \ave{\fhi}. These spectra correspond to the measurements of \ave{\mhi} and \ave{\fhi} presented in \autoref{fig:2binsSM}. The measurements made from the above spectra can be found in \autoref{tab:fig9lowmass}.} \label{fig:fclowmass}
\end{figure*}
    
\begin{figure*}
    \centering
        \includegraphics[width=\textwidth]{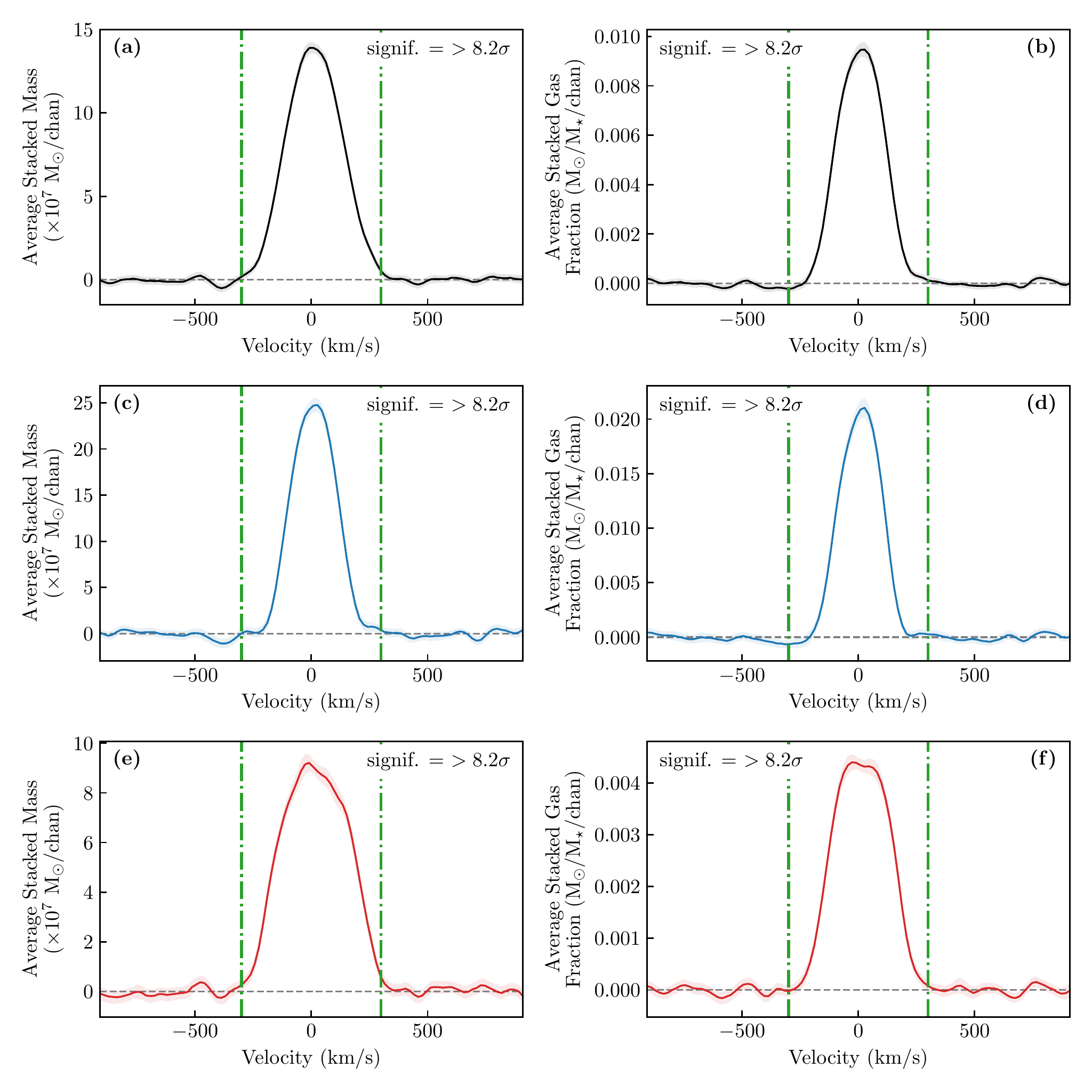}
        \caption{Same as \protect\autoref{fig:fclowmass}, but for the sample with  $10^{9.5} < \mstar (\msol) < 10^{12}$.  The measurements made from the above spectra can be found in \autoref{tab:fig9highmass}. }
        \label{fig:fchighmass}
    \end{figure*}
    
    \begin{figure*}
        \centering
        \includegraphics[width=\textwidth]{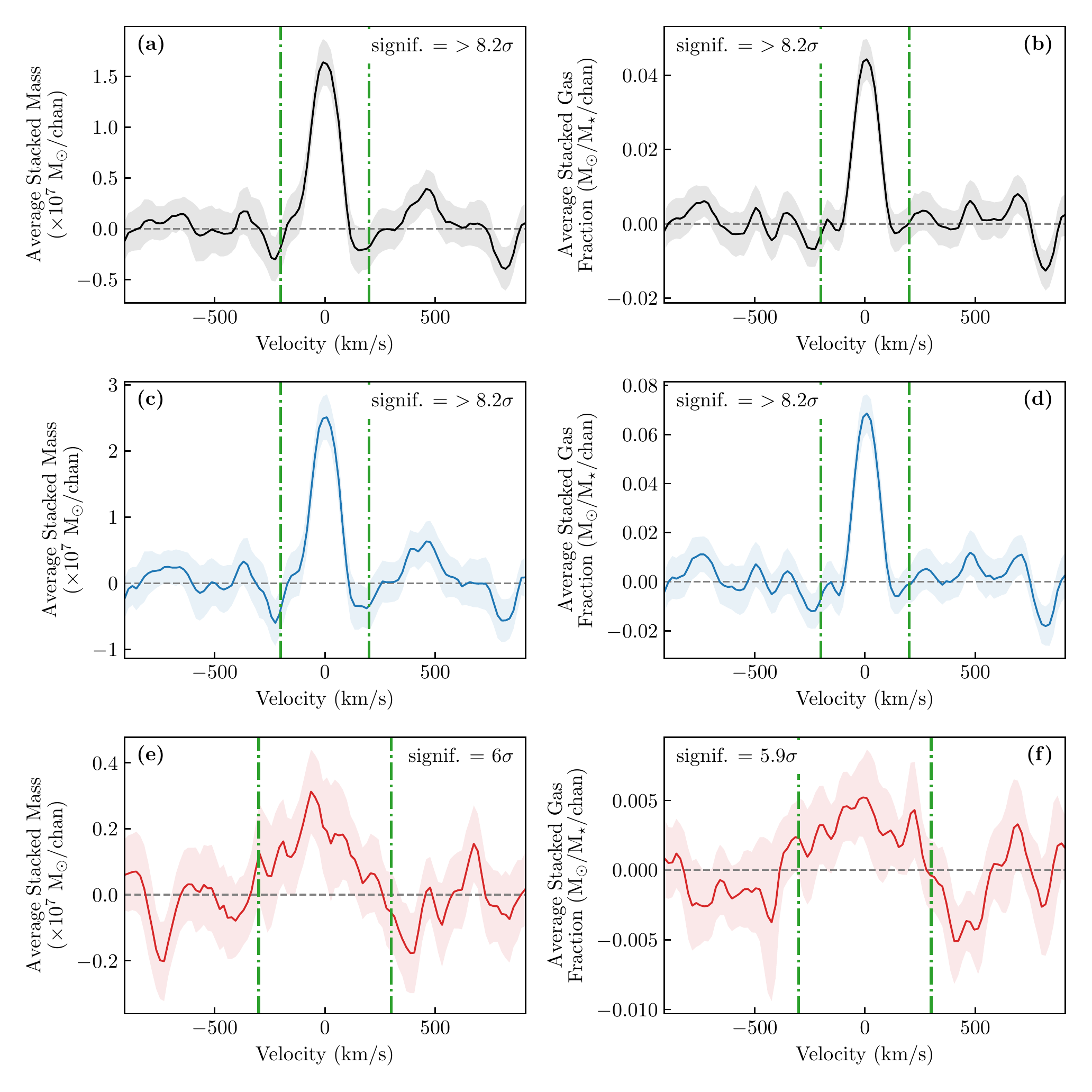}
        \caption{Same as \protect\autoref{fig:fclowmass}, but for the sample of only \hi non-detections with $10^{8} < \mstar (\msol) < 10^{9.5}$. \autoref{tab:fig9lowmass} contains the measurements made from the above spectra.}
        \label{fig:ndlowmass}
    \end{figure*}
    
    \begin{figure*}
        \centering
        \includegraphics[width=\textwidth]{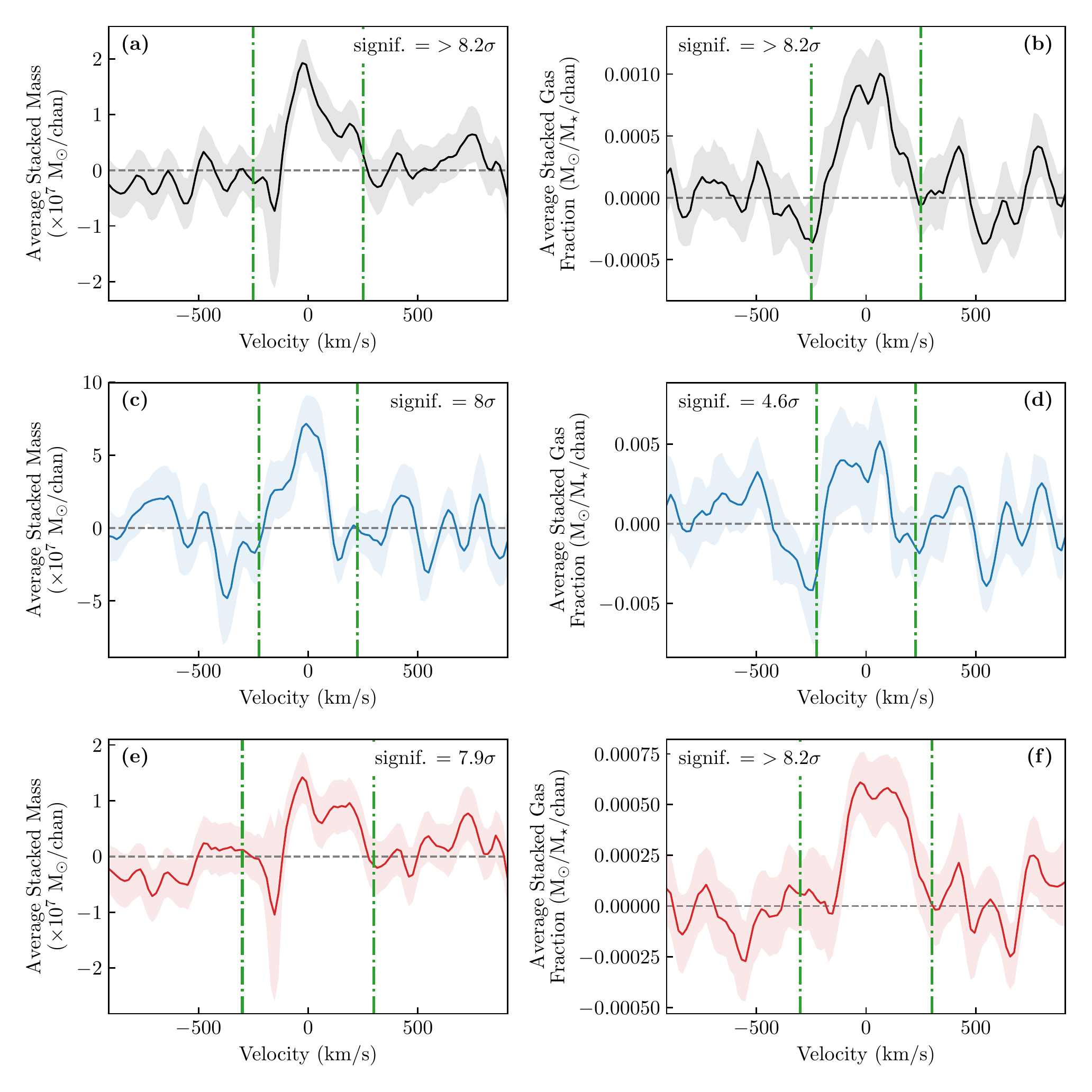}
        \caption{Same as \protect\autoref{fig:fclowmass}, but for the sample of \hi non-detections with  $10^{9.5} < \mstar (\msol) < 10^{12}$. \autoref{tab:fig9highmass} contains the measurements made from the above spectra. }
        \label{fig:ndhighmass}
    \end{figure*}

\begin{figure*}
	\centering
	 \includegraphics[width=\textwidth]{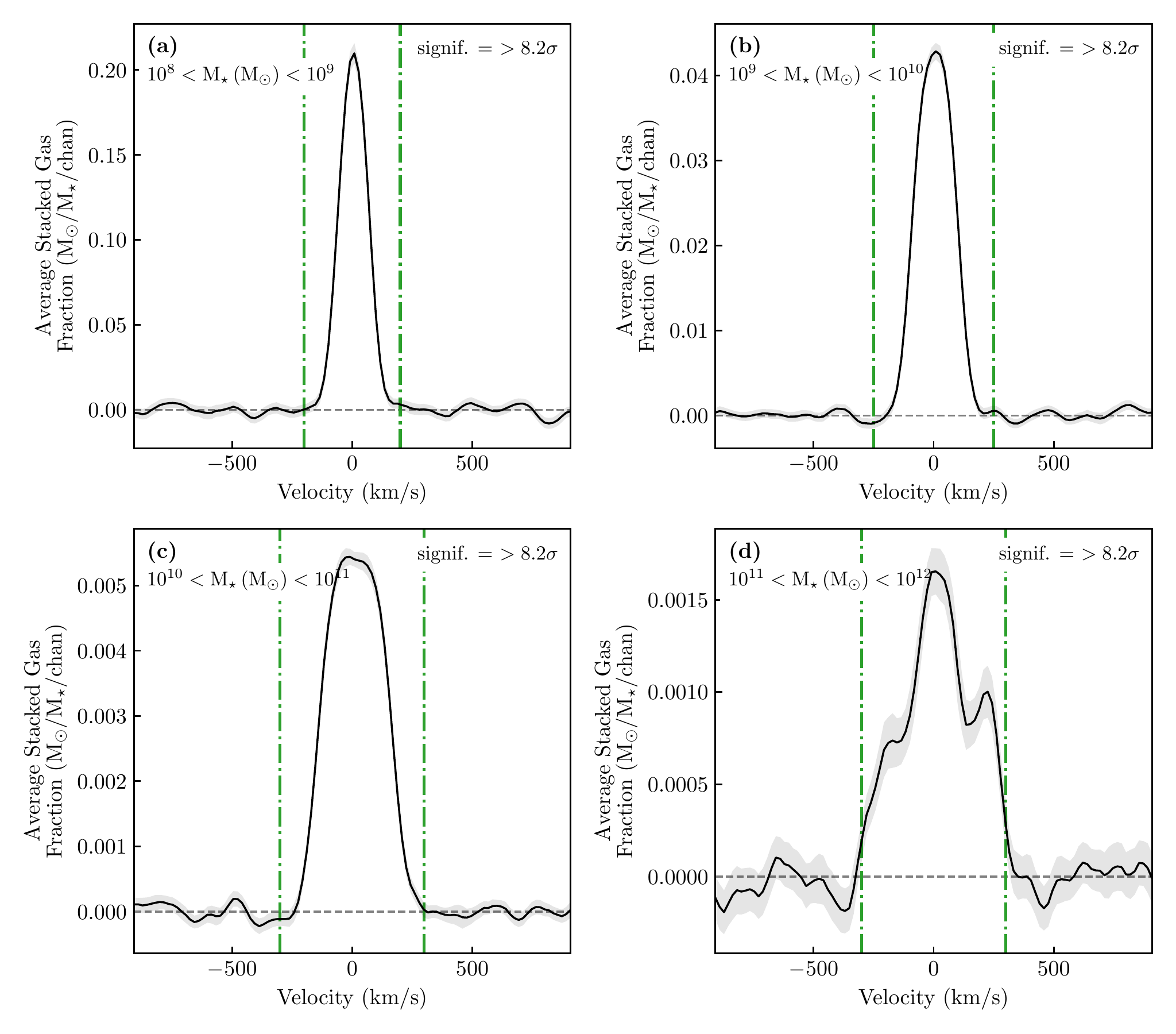}
	 \caption{All detections and non-detections stacked in bins of stellar mass. The bins increase in stellar mass from panel (a) to (d). The green dot-dashed lines in every panel mark the section of the stacked profiles used to measure the \ave{\fhi}. The gray band around the spectrum represents the uncertainty of the stacked spectrum. These spectra correspond to the \ave{\fhi} measurements in \autoref{fig:fhismbins}. See \autoref{tab:avefhifig1112} for measured quantities.}
	 \label{fig:fig9profiles}
\end{figure*}

\onecolumn
\section{Catalogue of \hi non-detections}
\label{sec:ndcatalogue}

        \footnotesize{
			\begin{longtable}{lrrlrrrr} 
				\caption{Data for the \hi non-detections in the stacking sample. The full table is available online.} \\\hline \hline
				\multicolumn{1}{c}{Source} & \multicolumn{1}{c}{RA} & \multicolumn{1}{c}{Dec} & \multicolumn{1}{c}{Name} & \multicolumn{1}{c}{$V_\text{opt}$} & \multicolumn{1}{c}{$u-r$} & \multicolumn{1}{c}{$M_r$} & \multicolumn{1}{c}{\mstar} \\
				 & & & & & & & $\left[ \log \right]$ \\
				 & \multicolumn{2}{c}{(J2000)} &  & \multicolumn{1}{c}{\kms} & \multicolumn{1}{c}{mag} & \multicolumn{1}{c}{mag} & \multicolumn{1}{c}{\msol} \\[0.5ex] \hline
				\endfirsthead

				\multicolumn{8}{c}{\normalsize{\tablename\ \thetable{}: continued from previous page}}\\[2ex]
				\hline \hline
				\multicolumn{1}{c}{Source} & \multicolumn{1}{c}{RA} & \multicolumn{1}{c}{Dec} & \multicolumn{1}{c}{Name} & \multicolumn{1}{c}{$V_\text{opt}$} & \multicolumn{1}{c}{$u-r$} & \multicolumn{1}{c}{$M_r$} & \multicolumn{1}{c}{\mstar} \\
				 & & & & & & & $\left[ \log \right]$ \\
				 & \multicolumn{2}{c}{(J2000)} &  & \multicolumn{1}{c}{\kms} & \multicolumn{1}{c}{mag} &  \multicolumn{1}{c}{mag} & \multicolumn{1}{c}{\msol} \\[0.5ex] \hline
				\endhead

				\hline 
				\endfoot

				\hline \hline
				\endlastfoot
					$1926$ & $11\, 23\, 18.84$ & $03\, 57\, 18.78$ & CGCG 039-161 & $1582 \pm 19 $ & $1.91$ & $-16.18$ & $8.23$\\
					$1954$ & $11\, 49\, 12.79$ & $-03\, 39\, 34.64$ & PGC 1068442 & $1572 \pm 3 $ & $1.57$ & $-15.78$ & $8.04$\\
					$2012$ & $12\, 19\, 21.12$ & $04\, 46\, 24.49$ & ASK 173829 & $2065 \pm 2 $ & $1.61$ & $-15.87$ & $8.09$\\
					$2244$ & $14\, 04\, 15.84$ & $04\, 06\, 43.85$ & UGC 08986 & $1239 \pm 3 $ & $2.19$ & $-17.79$ & $9.08$\\
					$2272$ & $14\, 25\, 08.99$ & $-01\, 06\, 48.52$ & CGCG 019-019 & $2643 \pm 3 $ & $2.11$ & $-17.94$ & $9.21$\\
					$2442$ & $15\, 17\, 29.45$ & $03\, 35\, 07.94$ & PGC 3124577 & $1882 \pm 2 $ & $1.62$ & $-15.69$ & $8.03$\\
					$2525$ & $09\, 35\, 44.04$ & $31\, 42\, 19.68$ & NGC 2918 & $6756 \pm 2 $ & $2.65$ & $-22.64$ & $11.28$\\
					$2533$ & $10\, 06\, 07.44$ & $47\, 15\, 45.53$ & NGC 3111 & $7437 \pm 2 $ & $2.68$ & $-22.52$ & $11.19$\\
					$2535$ & $14\, 30\, 25.58$ & $11\, 55\, 40.77$ & NGC 5644 & $7649 \pm 2 $ & $2.72$ & $-22.87$ & $11.33$\\
					$2541$ & $13\, 53\, 38.43$ & $36\, 08\, 02.57$ & NGC 5352 & $7956 \pm 3 $ & $2.81$ & $-22.64$ & $11.26$\\
					$2565$ & $00\, 35\, 59.83$ & $-10\, 07\, 18.07$ & NGC 0163 & $5917 \pm 4 $ & $2.70$ & $-22.05$ & $11.01$\\
					$2574$ & $11\, 00\, 35.40$ & $12\, 09\, 41.65$ & NGC 3491 & $6351 \pm 2 $ & $2.72$ & $-21.98$ & $11.06$\\
					$2589$ & $12\, 19\, 32.09$ & $49\, 48\, 56.71$ & UGC 07367 & $4093 \pm 2 $ & $2.60$ & $-21.45$ & $10.78$\\
					$2597$ & $12\, 19\, 32.59$ & $56\, 44\, 11.69$ & NGC 4271 & $4756 \pm 2 $ & $2.82$ & $-21.84$ & $10.93$\\
					$2606$ & $14\, 41\, 32.96$ & $38\, 51\, 04.95$ & UGC 09473 & $4669 \pm 2 $ & $2.17$ & $-21.54$ & $10.80$\\
					$2614$ & $12\, 12\, 17.27$ & $13\, 12\, 18.72$ & NGC 4168 & $2273 \pm 2 $ & $2.44$ & $-21.43$ & $10.63$\\
					$2615$ & $12\, 32\, 47.65$ & $63\, 56\, 21.18$ & NGC 4512 & $2516 \pm 1 $ & $2.72$ & $-20.75$ & $10.60$\\
					$2661$ & $10\, 36\, 38.44$ & $14\, 10\, 15.97$ & NGC 3300 & $3017 \pm 1 $ & $2.54$ & $-21.02$ & $10.53$\\
					$2675$ & $12\, 08\, 09.62$ & $10\, 22\, 43.96$ & NGC 4119 & $1651 \pm 2 $ & $2.28$ & $-19.74$ & $10.05$\\
					$2682$ & $12\, 27\, 13.34$ & $12\, 44\, 05.22$ & NGC 4425 & $1891 \pm 1 $ & $2.39$ & $-20.42$ & $10.22$\\
					$2688$ & $12\, 24\, 05.01$ & $11\, 13\, 05.04$ & NGC 4352 & $2083 \pm 1 $ & $2.36$ & $-20.12$ & $10.08$\\
					$2710$ & $12\, 27\, 02.54$ & $15\, 27\, 41.34$ & NGC 4421 & $1551 \pm 1 $ & $2.58$ & $-20.15$ & $10.04$\\
					$2715$ & $14\, 05\, 12.42$ & $55\, 44\, 30.67$ & NGC 5475 & $1647 \pm 1 $ & $2.54$ & $-19.58$ & $9.98$\\
					$2731$ & $12\, 37\, 48.38$ & $05\, 22\, 06.69$ & NGC 4580 & $1035 \pm 2 $ & $2.37$ & $-19.23$ & $9.75$\\
					$2742$ & $12\, 15\, 59.87$ & $66\, 13\, 51.00$ & NGC 4221 & $1314 \pm 1 $ & $2.51$ & $-19.48$ & $9.83$\\
					$2743$ & $10\, 12\, 41.25$ & $03\, 07\, 45.79$ & NGC 3156 & $1266 \pm 29 $ & $1.99$ & $-19.14$ & $9.42$\\
					$2744$ & $11\, 06\, 32.10$ & $11\, 23\, 07.51$ & NGC 3524 & $1357 \pm 1 $ & $2.36$ & $-19.08$ & $9.71$\\
					$2756$ & $11\, 08\, 40.38$ & $57\, 13\, 48.71$ & NGC 3530 & $1876 \pm 1 $ & $2.42$ & $-18.95$ & $9.69$\\
					$2769$ & $12\, 25\, 18.78$ & $64\, 56\, 00.52$ & NGC 4391 & $1320 \pm 1 $ & $2.52$ & $-18.73$ & $9.58$\\
					$2770$ & $13\, 00\, 10.57$ & $12\, 28\, 59.92$ & NGC 4880 & $1362 \pm 3 $ & $2.24$ & $-19.29$ & $9.51$\\
					$2775$ & $12\, 22\, 04.11$ & $12\, 47\, 14.95$ & NGC 4306 & $1520 \pm 3 $ & $2.21$ & $-18.70$ & $9.35$\\
					$2789$ & $11\, 21\, 24.99$ & $03\, 00\, 50.23$ & NGC 3643 & $1742 \pm 2 $ & $2.28$ & $-18.67$ & $9.53$\\
					$2790$ & $12\, 38\, 17.87$ & $13\, 06\, 35.64$ & NGC 4584 & $1715 \pm 3 $ & $2.01$ & $-18.96$ & $9.46$\\
					$2795$ & $12\, 41\, 59.35$ & $12\, 56\, 34.27$ & NGC 4620 & $1125 \pm 3 $ & $2.09$ & $-18.63$ & $9.24$\\
			\end{longtable}
		}
		
		\begin{figure}
			\centering
            \vspace{-0.5cm}
			\includegraphics[page=1, width=0.95\textwidth]{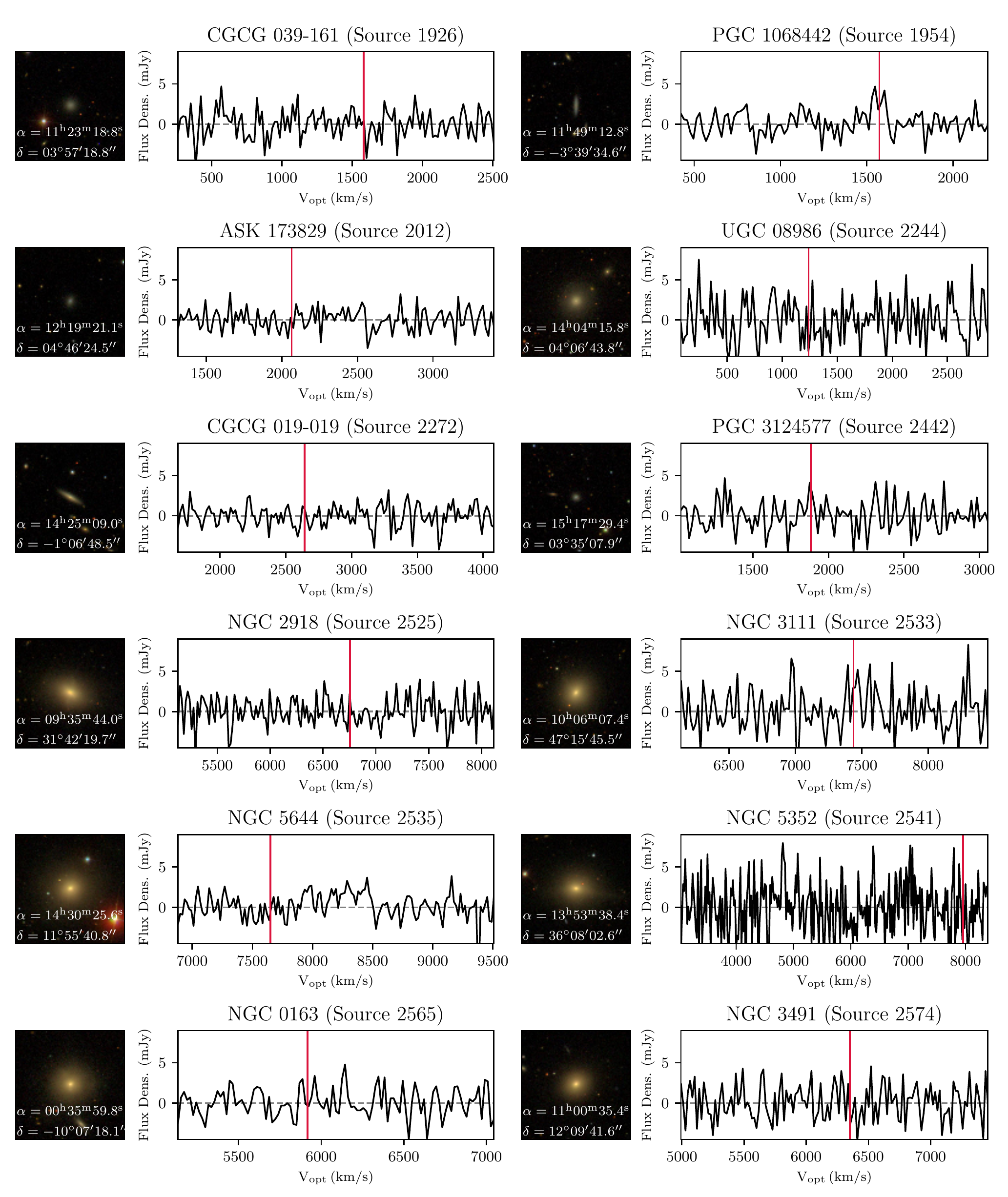}
			\caption{Optical images and \hi spectra for all 323 galaxies classified as \hi non-detections from the final NIBLES stacking sample (see \protect\autoref{sec:ndselect}). The SDSS redshift of each galaxy is represented by the red line in each spectrum. The rest of the images can be found in the online supplementary material.}
		\end{figure}


\bsp	
\label{lastpage}
\end{document}